\documentclass[
    aps,
    prb,
    twocolumn,
    nofootinbib
]{revtex4-2}

\newcommand{\psiB}{\psi}
\newcommand{\phiB}{\phi}
\newcommand{\omegaB}{\omega}

\newcommand{\bT}{{\cal T}^{(r)}}

\usepackage{array}
\usepackage{titlesec}
\usepackage{amsbsy}

\titleformat{\paragraph}
{\normalfont\normalsize\bfseries}{\theparagraph}{1em}{}
\titlespacing*{\paragraph}
{0pt}{3.25ex plus 1ex minus .2ex}{1.5ex plus .2ex}
\renewcommand*{\theparagraph}{\roman{paragraph})}

\usepackage{enumitem} 

\usepackage{amssymb, amsmath,amsthm,bm,bbm}    
\usepackage{graphicx}   
\usepackage{verbatim}   
\usepackage{color}      
\usepackage[labelformat=simple]{subcaption}
\usepackage{mathrsfs} 
\usepackage{dsfont}
\usepackage[linktocpage = true]{hyperref}  
\usepackage{cancel}
\usepackage{graphicx}
\usepackage{caption} 
\usepackage{tcolorbox}
\usepackage{yfonts}
\usepackage{multirow}
\definecolor{green}{RGB}{35,142,35}
\usepackage{float}
\usepackage{upgreek}

\usepackage{empheq} 
 
\usepackage{ulem}

\usepackage[framemethod=tikz]{mdframed}

 \usepackage{physics}
 
 \usepackage{bigints}

 \usepackage{braket}

\hypersetup{
    colorlinks   = true,
    linkcolor = blue,
    citecolor    = red
}  
\makeatletter
\renewcommand*{\p@subsection}{\thesection.}
\makeatother

\usepackage{cleveref}
\usepackage{multirow}
\usepackage{xcolor}
\allowdisplaybreaks

\newcommand{\eq}[1]{Eq. \eqref{#1}}
\newcommand{\fig}[1]{Fig. \ref{#1}}

\newcommand{\dslash}{\not{\hbox{\kern-2pt $\partial$}}}
\newcommand{\bqa}{\begin{eqnarray}} 
\newcommand{\eqa}{\end{eqnarray}}
\newcommand{\nn}{\nonumber \\}

\newcommand{\beq}{\begin{equation}}
\newcommand{\eeq}{\end{equation}}

\def\be{\begin{eqnarray}}
\def\ee{\end{eqnarray}}

\newcommand\thetaq{\Theta (\theta,\vec q)}
\newcommand\thetasq{\Theta (\theta,\vec q)}
\newcommand\deltaq{\Delta (\delta,\theta,\vec q)}

\newcommand\KFthetadim{{\bf K}_{F,\theta}}

\newcommand{\nge}{\tilde {G}_e}

\newcommand{\tdelta}{{\bm \delta}}

\NewDocumentCommand{\aV}{e{_^}}{%
  \ensuremath{%
    V%
    \IfValueT{#1}{_{#1}}
    \IfValueTF{#2}{^{\prime\,#2}}{^{\prime}}
  }%
}

\NewDocumentCommand{\av}{e{_^}}{%
  \ensuremath{%
    {\mathfrak V}%
    \IfValueT{#1}{_{#1}}
    \IfValueTF{#2}{^{\prime\,#2}}{^{\prime}}
  }%
}

\NewDocumentCommand{\avn}{e{_^}}{%
  \ensuremath{%
    {\mathfrak V}%
    \IfValueT{#1}{_{#1}}
    \IfValueT{#2}{^{#2}}
  }%
}

\newcommand\KFAVdim{ \mathbf k_F}

\newcommand\discinf{\eta_{P;-\infty}}

\newcommand{\halfominus}{
  \begin{tikzpicture}[baseline=(base)]
    \node[inner sep=0pt] (base) at (0,0) {};
    \draw[thick] (0,0) circle (0.06);
    \fill[black] (0.06,0) arc (0:180:0.06) -- cycle; 
    \draw[thick] (-0.06,0) -- (0.06,0); 
  \end{tikzpicture}
}

\newcolumntype{P}[1]{>{\centering\arraybackslash}p{#1}}

\newcommand\mK{{\mathcal K}}
\newcommand\mH{{\mathcal H}}
\newcommand\mS{{\mathcal S}}

\newcommand{\Gc}[1]{#1_c^{(4)}}
\newcommand{\Gce}[1]{#1_{c,0}^{(4)}}
\newcommand{\Gcr}{G_c^{(4,r)}}
\newcommand{\Gcre}{G_{c,0}^{(4,r)}}
\newcommand{\Gammar}{\Gamma^{(r)}}

\newcommand{\shubham}[1]{ { \color{red} [(\textsf{SHUBHAM}) \textsf{\textsl{#1}} ]}}

\newcommand{\tv}{\tilde{v}}
\newcommand{\te}{\tilde{\eta}}

\newcommand{\Vcrs}{\av^\star_{S,n}(\varepsilon)}
\newcommand{\Vcrl}{\av^\star_{L,n}}

\begin{document}

\title{\textbf{
Categorical classification of non-Fermi liquids in 2+1 dimensions
}}

\author{Shubham Kukreja$^\dagger$}
\author{Dawson M. Willerton$^\dagger$}
\author{Sung-Sik Lee}
\affiliation{$^{1}$Department of Physics \& Astronomy, McMaster University, Hamilton ON L8S 4M1, Canada}
\affiliation{$^{2}$Perimeter Institute for Theoretical Physics, Waterloo ON N2L 2Y5, Canada}
\affiliation{$^{\dagger}$ 
These are the co-first authors.
}

        \date{\today}

\begin{abstract}

A defining feature of non-Fermi liquids is a lack of the quasiparticle peak 
on the Fermi surface
in the spectral function,
which is usually realized by coupling 
a finite density of fermions with critical collective fluctuations.
However, these features of the two- and three-point vertex functions are qualitatively the same across many non-Fermi liquids, 
which has made it difficult to classify them beyond the definition by negation.
We point out that the four-point function in the pairing channel
contains rich information that positively differentiates distinct types of non-Fermi liquids.
Using the generalized notion of fixed points that captures scale-dependent universal pairing interaction, we provide a non-perturbative
classification of $2+1$-dimensional non-Fermi liquids.
In the large Fermi momentum limit with fixed dimensionless couplings,
metals coupled with long-wavelength critical fluctuations are classified into seven superuniversality classes. 
Each of them exhibits a unique phase diagram in the space of temperature and bare four-fermion coupling.
The present work generalizes the earlier perturbative classification to strongly interacting non-Fermi liquids by leveraging kinematic constraints.

\end{abstract}

\maketitle
\tableofcontents




\newcommand{\dawson}[1]{ { \color{red} [(\textsf{DAWSON}) \textsf{\textsl{#1}} ]}}

\section{Introduction}

Non-Fermi liquids are unconventional metals 
that do not fit within the quasiparticle paradigm of Fermi liquids\cite{LANDAU,LANDAU2}.
Despite the progress made over the years\cite{
HERTZ,
MILLIS,
VARMALI,
POLCHINSKI2,
PLEE1,
ALTSHULER,
YBKIM,
ABANOV2,
SSLEE,
MAX0,
MAX2,
NAYAK2,
SCHOFIELD,
DENNIS,
SENTHIL,
MROSS,
FITZPATRICK,
SHOUVIK2,
SCHLIEF,
PhysRevX.11.021005,
PhysRevB.107.165152,
annurev:/content/journals/10.1146/annurev-conmatphys-031218-013339},
a systematic classification of non-Fermi liquids is still lacking, as the indefinite name may suggest.
While most phases of matter are classified by scale invariant fixed points of the renormalization group (RG) flow\cite{WILSONRG1,WILSONRG2}, in metals,
Fermi momentum prevents scale invariance from emerging at low energies \cite{BORGES2023169221,PhysRevB.110.155142}. 
Fermi momentum $\KFAVdim$ is a large momentum scale reciprocal to the interparticle distance; 
yet, it does not decouple from low-energy physics because it controls the size of the Fermi surface on which zero-energy modes reside.

\begin{figure}[ht]
\includegraphics[width=0.35\textwidth]{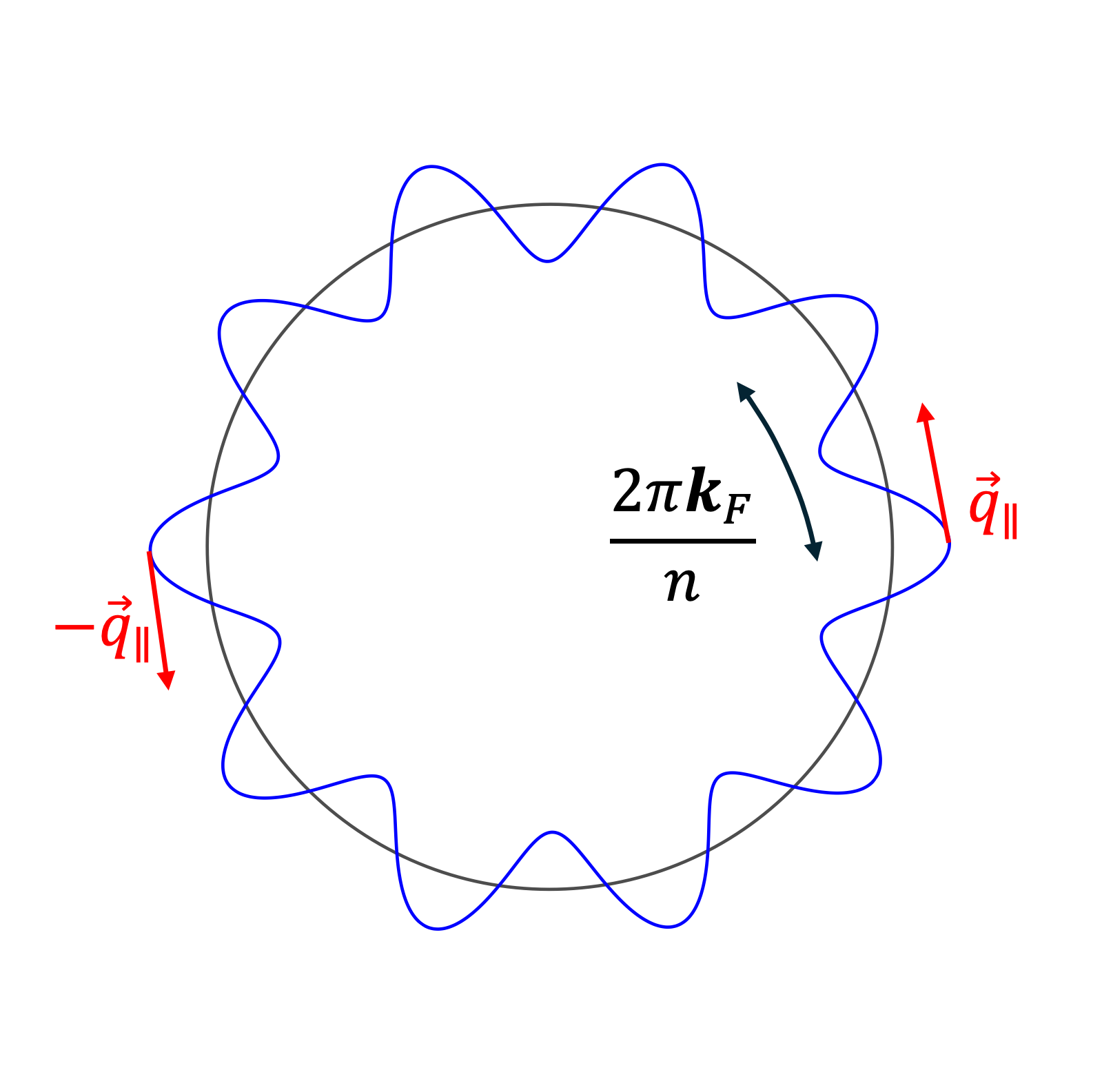}
\caption{
For a convex Fermi surface, frequency scales with the momentum perpendicular and parallel to the Fermi surface as
$\omegaB^{1/z} \sim q_\perp^z \sim q_\parallel^{2}/\KFAVdim$, respectively,
where $z$ is the dynamical critical exponent defined with respect to $q_\perp$.
Perturbatively, $z=3/2$ for many non-Fermi liquids, resulting in $\omega \sim q_\parallel^3$.
Low-energy bosons, which primarily scatter fermions tangentially, carry typical momentum $q_\parallel \sim \omegaB^{1/2z}$ at energy $\omegaB$.
The universal pairing interaction generated in the angular momentum channel $n$ 
at energy $\omega$ 
is determined by the nature of the critical boson, and the ratio between $q_\parallel$ and the pitch of the pair wavefunction.
For example, the transverse gauge field generates a positive scattering vertex through Amperean law.
This gives an repulsive pairing interaction
for $q_\parallel \ll \pi\KFAVdim/n$ 
as the critical boson connects Cooper pairs of the same phase.
The interaction becomes attractive for $q_\parallel \sim \pi\KFAVdim/n$ that connects Cooper pairs out of phase.
The interaction becomes negligible for $q_\parallel \gg \pi\KFAVdim/n$ due to the destructive interference.
}
\label{fig:pitch_vs_q}
\end{figure}

In quantum critical metals,  the lack of scale invariance originates from the indefinite growth of the ratio between $\KFAVdim$ and the momentum of critical fluctuations  under coarse graining.
The dimensionless parameter that grows indefinitely spoils scale invariance by creating a series of crossovers that persist to arbitrarily low energies\cite{kukreja2026classificationnonfermiliquidsuniversal}.
Such crossovers are particularly prominent in the universal pairing interactions generated by critical fluctuations. 
Unlike single-particle excitations,
Cooper pairs with zero net momentum are not kinematically confined within a small patch of the Fermi surface at low energies.
Since they can be `delocalized' across the entire Fermi surface, their dynamics are sensitive to 
$\KFAVdim$ 
as well as the momentum transferred by critical fluctuations.
Specifically, the renormalization of the pairing interaction in the angular momentum channel $n$\footnote{
Even when there is no rotational symmetry, a Cooper pair wavefunction on a simply connected Fermi surface can be labeled by the generalized angular momentum $n \in \mathbb{Z}$, which represents the winding number of the wavefunction around the Fermi surface.
For simplicity, we refer to it as angular momentum.
}
at each energy scale is determined by the ratio between
the pitch of the Cooper pair wavefunction $\sim 2\pi\KFAVdim/n$ 
and the momentum $q_\parallel$ carried by the critical 
fluctuations at that energy.
This causes a crossover when $q_\parallel$
becomes comparable to $2\pi \KFAVdim/n$
(see \fig{fig:pitch_vs_q}).
While the crossover momentum scale is proportional to $\KFAVdim$,
it becomes arbitrarily small in the large $n$ limit.
This means that the pairing interaction in the angular momentum channel $n$ undergoes a crossover at a very low energy in the large $n$ limit. 
Such low-energy crossovers also arise in the kinetic equation that governs the dynamics of shape fluctuations of the Fermi surface\cite{PhysRevB.52.17275}.
This inseparability of large momentum scales from low-energy physics is what causes the lack of scale invariance.

\begin{figure}[ht]
\includegraphics[width=0.45\textwidth]{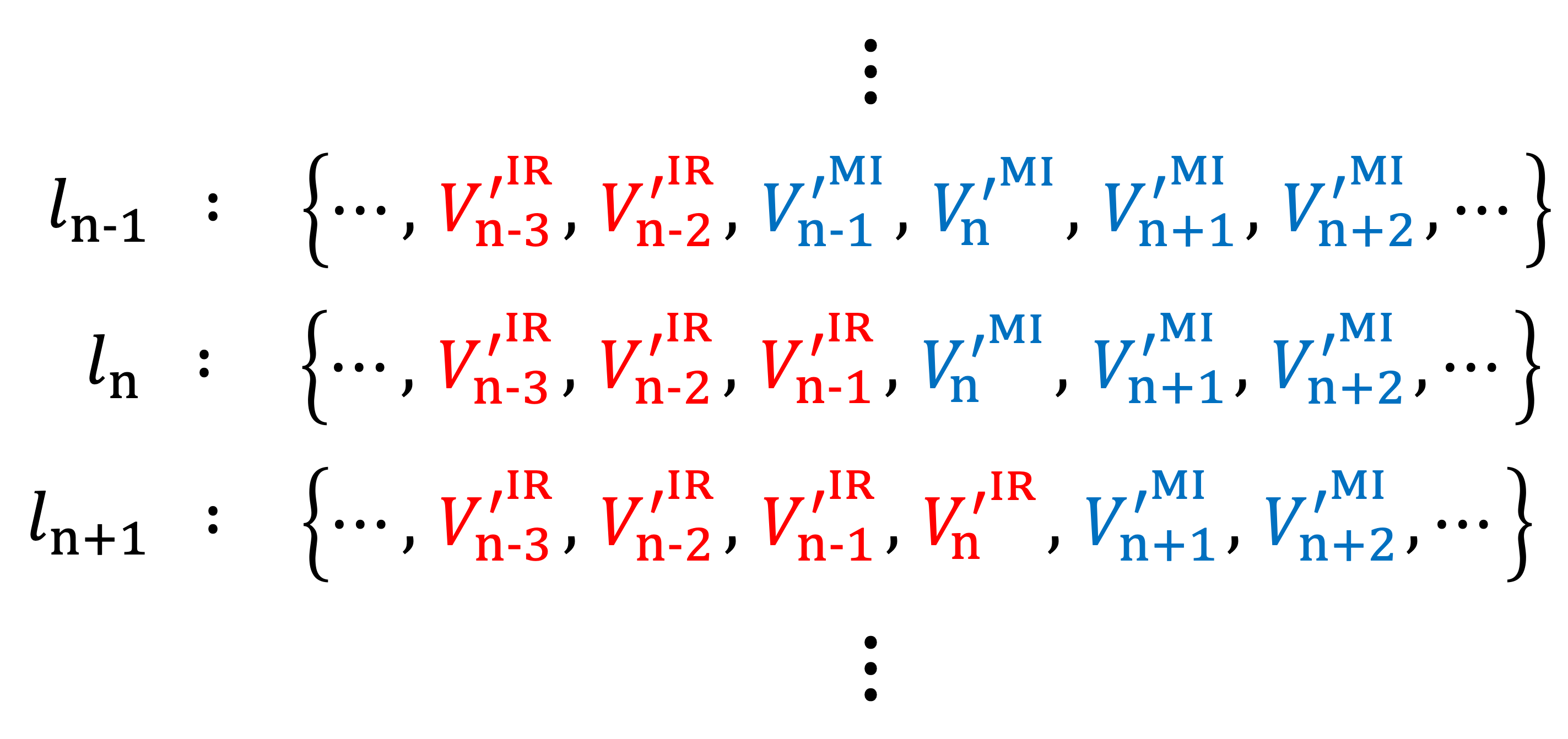}
\caption{
The set $\{ \aV_n | n \in \mathbb{Z} \}$, which consists of
the pairing interaction in every angular momentum channel,
flows as the logarithmic length scale $l =\frac{1}{z} \log \Lambda/\omega$ increases from top to bottom.
The coupling in each angular momentum channel $n$ crossovers from 
a mid-infrared scale value 
$\aV^{MI}_n$
to 
an infrared asymptotic value 
$V^{IR}_n$
below a crossover energy scale.
Because the crossover energy scale approaches zero with increasing angular momentum,
there are infinitely many couplings that are yet to undergo crossovers
at any finite energy scale.
}
\label{fig:crossovers}
\end{figure}

Since angular momentum is unbounded, there exists an infinite series of crossovers extending down to the zero energy limit.
In the first place, such incessant crossovers are possible because
the space of marginal or relevant coupling functions is infinite dimensional.
The renormalization group (RG) flow maps couplings at one energy scale to those at another.
In non-metallic phases of matter, 
the low-energy physics is determined by a few marginal and relevant couplings, modulo small corrections.
Once they flow to fixed points at low energies below a finite crossover energy scale, RG trajectories are not important.
However, this is not the case for metals.
The infinite set of couplings never reaches the low-energy `fixed point set' at any finite energy scale because the crossover scales for some elements are arbitrarily small.
No matter how low the energy scale is, there are couplings that change significantly as the energy is further lowered.
This is illustrated in \fig{fig:crossovers}.
Therefore,  one cannot use scale invariant fixed points to classify non-Fermi liquid states.

Nonetheless, the notion of universality remains intact:
different microscopic theories exhibit the same crossovers at low energies if they are in the basin of attraction of one universality class.
For example, the coupling in channel $n$ first flows to a mid-infrared coupling $\aV^{MI}_n$ at low but finite energies and then flows to an IR coupling $\aV^{IR}_n$ 
below a lower crossover energy scale.
For large $n$, the crossover from the mid-infrared to low-energy scales occurs at energies far below the UV cutoff scale. 
Consequently, $\aV_n^{MI}$ and $\aV_n^{IR}$, and the crossover between them become universal at large $n$.
Therefore, metallic universality classes are captured by RG `fixed trajectories' that describe the universal crossovers.
The universal RG trajectories
are extended indefinitely in the infrared direction, along which the coupling function undergoes incessant crossovers as\cite{kukreja2026classificationnonfermiliquidsuniversal}
\bqa
\cdots
\leftarrow
\{ \aV_n(l_2) \}
\xleftarrow{ \mathcal{R}_{l_2,l_1} }
\{ \aV_n(l_1) \}
\xleftarrow{ \mathcal{R}_{l_1,l_0} }
\{ \aV_n(l_0) \},
\label{eq:category}
\eqa
where $l$ represents the logarithmic length scale.
Such universal RG trajectories are referred to as projective fixed points (PFPs) because a sense of fixed points arises only after quotienting the rescaling of Fermi momentum\cite{PhysRevB.110.155142}.
In other words, there is scale invariance only when physical observables are compared between two metals with different Fermi momenta at two different energy scales\cite{BORGES2023169221}.

\begin{figure}[ht]
\begin{subfigure}{0.35 \textwidth}
\includegraphics[width=\textwidth]{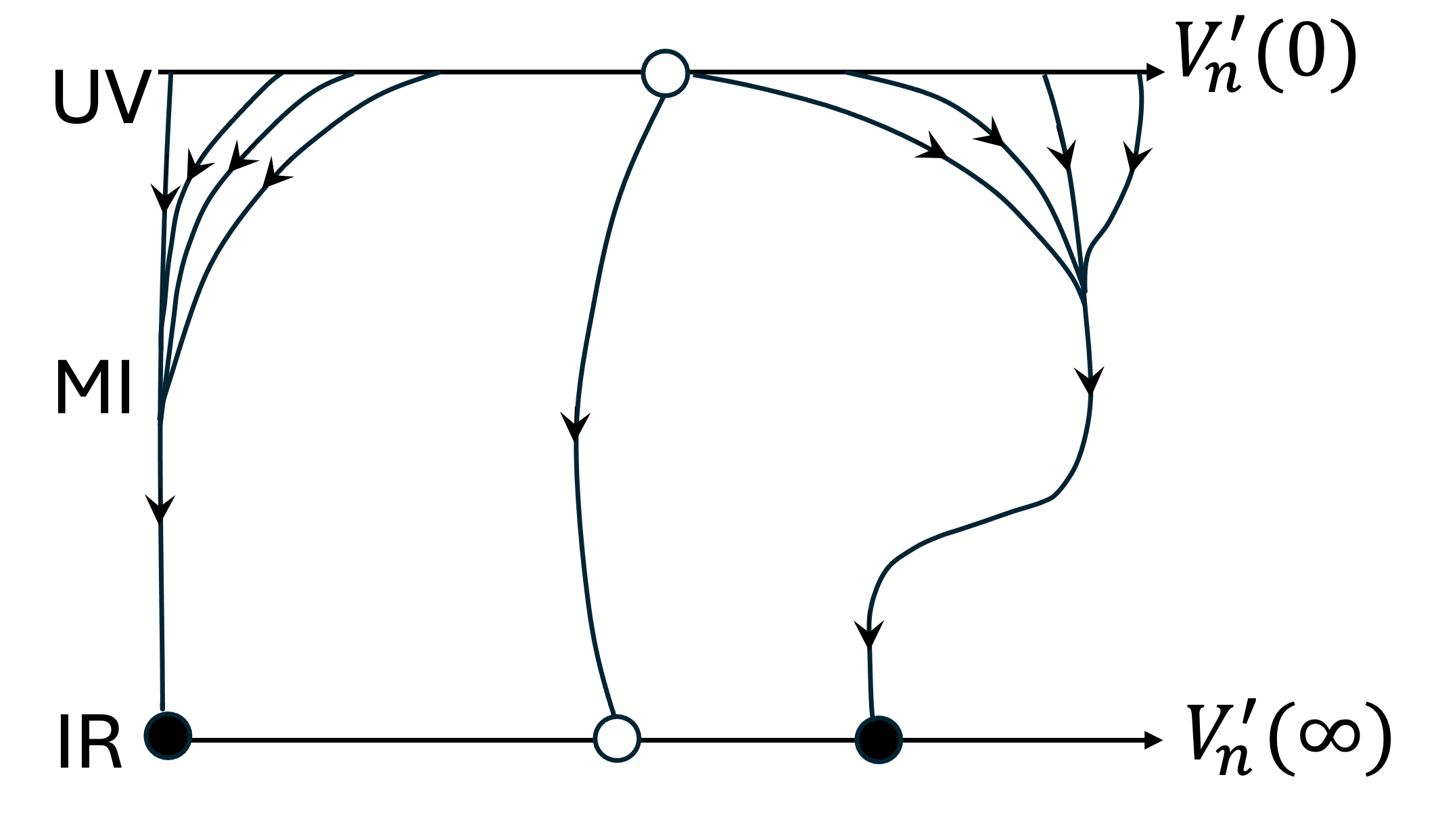}
\caption{}
\label{fig:toplogy1}
    \end{subfigure}
\begin{subfigure}{0.35\textwidth}
\includegraphics[width=\textwidth]{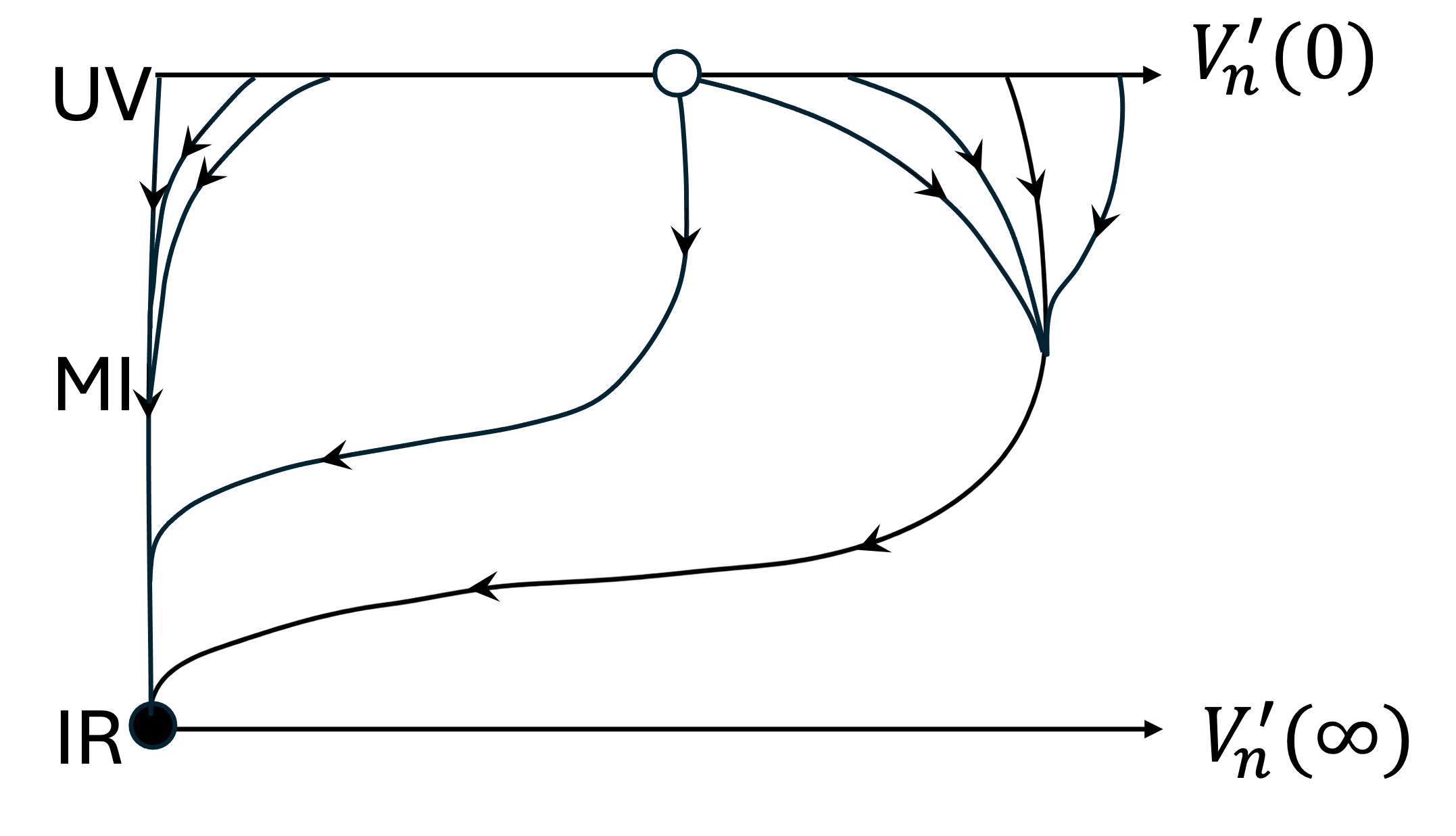}
        \caption{}
        \label{fig:toplogy2}
\end{subfigure}
\caption{
The RG flow generates the mapping
$R_{l^*,0}$
from couplings at a short-distance scale ($l=0$) to those at a mid-infrared scale ($l=l^*$),  and $R_{\infty,l^*}$ from $l^*$ to  the infrared limit ($l=\infty$).
In the large angular momentum limit, $l^*$ becomes arbitrarily large.
The topological data of the RG trajectories includes the number of disjoint images of 
$R_{l^*,0}$ and $R_{\infty,l^*}$ in the large $l^*$ limit,
and the mappings between them.
In (a), each of the three disjoint images of  $R_{l^*,0}$ flow to different images of $R_{\infty,l^*}$.
In (b), all images of $R_{l^*,0}$ flow to only one disjoint image of $R_{\infty,l^*}$.
(a) can not be smoothly connected to (b) without a topological phase transition.
}
\label{fig:topology}
\end{figure}

The RG trajectories correspond to mappings $\mathcal{R}_{l_2,l_1}$
from couplings at scale $l_1$ to those at $l_2>l_1$. 
%
The universal RG trajectories contain detailed dynamical information, such as the strength and scale-dependent profile of the couplings.
But they also contain topological data that captures global patterns of crossovers (see \fig{fig:topology}).
The topological data does not uniquely specify a universality class.
However, it draws boundaries between qualitatively distinct types of non-Fermi liquids:
each topologically distinct group defines a superuniversality class that contains multiple universality classes whose universal crossovers can be smoothly deformed into one another.
Hence, classifying the topology of projective fixed points is the first step towards the full classification of non-Fermi liquids. 
Recently, it was shown that perturbative non-Fermi liquids can be grouped into seven superuniversality classes\cite{kukreja2026classificationnonfermiliquidsuniversal}.
However, it is a priori not clear whether the classification based on the perturbative flow equation 
is applicable to general non-Fermi liquids with strong coupling.
The goal of this paper is to non-perturbatively extend the classification of superuniversality classes to strongly interacting non-Fermi liquids in 2+1 dimensions and make further physical predictions that can be tested experimentally.

In general, it is not easy to solve $2+1$-dimensional non-Fermi liquid theories unless there are extensive kinematic constraints or small parameters that either emerge or are introduced by hand\footnote{
Examples include chiral non-Fermi liquids with extensive kinematic constraints\cite{ SHOUVIK2} and the antiferromagnetic quantum critical metals in the near nesting limit\cite{
SCHLIEF}.
For recent progress on large $N$ theories with random couplings and non-local theories, 
see Ref. \cite{doi:10.1126/science.abq6011} 
and
Ref.  \cite{Phillips:2020aa}, respectively.}.
Nonetheless, we are able to make progress in classifying non-Fermi liquids in $2+1$ dimensions for the following two reasons.
Firstly, quantum critical metals naturally come with a small parameter at low energies: $q_\parallel/\KFAVdim$.
A large set of quantum corrections is suppressed in the limit where the momenta of critical fluctuations become much smaller than the Fermi momentum.
This is because the phase space for virtual particles is suppressed by $q_\parallel/\KFAVdim$ unless they can stay close to the Fermi surface while being able to exchange large momenta.
Henceforth, we refer to 
the large $\KFAVdim/q_\parallel$ limit with fixed dimensionless coupling as the large $\KFAVdim$ limit, for short.
Secondly, the topological features of RG trajectories are insensitive to the dynamical details such as the strength of couplings. 
This allows one to reach general conclusions based on locality and kinematic constraints that hold non-perturbatively.
In the end, we show that non-Fermi liquids in $2+1$ dimensions are still classified by the seven superuniversality classes obtained in Ref. \cite{kukreja2026classificationnonfermiliquidsuniversal} 
under one assumption, that is, high-energy modes decouple from the low-energy physics.
Although this is a foundational statement believed to hold generally, we state it as an `assumption' because it has not yet been explicitly tested for all non-Fermi liquids.
In the following section, 
we will be more specific about where this assumption is used beyond the validity of low-energy effective theories constructed from the usual derivative expansion.

Our work is built upon previous works\cite{
PhysRevD.59.094019,
PhysRevB.91.115111,
PhysRevD.59.094019,
PhysRevB.91.115111,
PhysRevB.102.024524,
PhysRevB.102.024525,
PhysRevB.102.094516,
PhysRevB.103.024522,
PhysRevB.103.184508,
PhysRevB.104.144509,
PhysRevB.95.165137,
PhysRevB.92.205104,
CHUBUKOVSC,
PhysRevB.91.115111,
ABANOV3,
PhysRevB.95.174520,
LEDERERREV,
LEDERER,
PhysRev.135.A550,
PhysRevB.69.020505,
PhysRevB.95.165137,
HAUCK2020168120,
PhysRevB.107.L041103}, but differs from them in scope and method.
Firstly, our goal is to classify non-Fermi liquids in $2+1$ dimensions based on their universal superconducting fluctuations.  
For this, we use the RG flow equation of the pairing interaction, which is exact in the large $\KFAVdim$ limit.
Hence, our classification holds for strongly interacting non-Fermi liquids at low energies.
Secondly, we consider superconducting fluctuations in all angular momentum channels beyond low angular momentum channels\cite{ PhysRevD.59.094019, PhysRevB.91.115111}.
This is necessary for our purpose because low-energy crossovers that occur in high angular momentum channels serve as the fingerprint that classifies non-Fermi liquids.
To understand universal superconducting fluctuations in high angular momentum channels, it is crucial to take into account the scale dependence of the flow equation, since the crossover energy scale below which such scale dependence disappears approaches zero as the angular momentum increases.
Finally, superconductivity in non-Fermi liquids has often been studied 
at a specific bare four-fermion coupling, such as $V=0$\cite{
Wang_Chubukov_First_Matsubara,
Wu_Chubukov_Matsubara,
Pairing_Wu_Chubukov,
CHUBUKOV2020168142}.
Here, we study superconducting fluctuations and instabilities of non-Fermi liquids for general four-fermion couplings.
Non-Fermi liquids can, in principle, remain stable down to zero temperature for a repulsive bare coupling, even if there are superconducting instabilities at $V=0$, as will be shown later.
Since $V$ can be tuned, e.g., by changing the screened Coulomb repulsion, it is desirable to understand how superconducting fluctuations vary as a function of $V$.
Furthermore, the way superconducting fluctuations and instabilities `respond' to tuning the bare coupling contains useful information that can be used to classify non-Fermi liquids.

Here is the outline of the paper.
Since we use the universal pairing interaction as the main diagnostic to differentiate non-Fermi liquids, the four-fermion coupling plays a crucial role.
In Sec. \ref{sec:theory}, we discuss why the four-fermion coupling can change the fate of non-Fermi liquids at low energies, although it is superficially `irrelevant'.
In Sec. \ref{sec:flow}, the exact flow equation for the four-point vertex function
is derived in the large $\KFAVdim$ limit.
This functional RG equation is quadratic, thanks to the factorization of the four-fermion vertices in the large $\KFAVdim$ limit.
We then take advantage of the emergent rotational symmetry in non-Fermi liquids with convex Fermi surfaces to diagonalize the flow equation in the generalized angular momentum basis.
In Sec. \ref{sec:classification}, we discuss non-perturbative kinematic constraints and use them to classify the topology of the solutions of the flow equation.
Sec . \ref{sec:predictions} discusses experimental predictions that can uniquely differentiate distinct superuniversality classes.
We conclude by summarizing our findings and discussing future directions
in Sec . \ref{sec:summary}.

\section{
Pairing interaction 
in the large $\KFAVdim$ limit 
}
\label{sec:theory}

We consider theories that describe a Fermi surface coupled with a critical boson in 2+1 dimensions.
We consider theories with $U(1)$ charge conservation and $SU(2)$ spin rotational symmetry, but 
it is straightforward to generalize the following discussion to other internal symmetries.
The action reads
\begin{align}
 S
&=
\int 
d \mathbf k \,
\psiB^\dagger_j(\mathbf k
)
\Bigl[
i k_0
+
v_{F,\theta}
\delta
\Bigr]
\psiB^j(\mathbf k)
\nonumber 
+
\frac12
\int 
d \mathbf q\,
{\bf q}^2
|\phi_a(\mathbf q)|^2
\nonumber \\ &
+
\frac{1}{\sqrt 2}
\int 
d \mathbf k d \mathbf q \,
~
e_{\theta}(\varphi)
\phiB_a( \mathbf q)
\psiB^\dagger_i( \mathbf k 
+ \mathbf q)
T^{a,i}_{j}
\psiB^j(\mathbf k)
\nonumber
\\
&
+
\int 
d \mathbf k  d \mathbf k' d \mathbf q \,
V^{i'j'}_{ij} (\theta',\theta;\vec q)
\times \nn & ~~ 
\psiB_{i'}^{\dagger}\left(\frac{\mathbf q}{2}-\mathbf k'\right)
\psiB_{j'}^{\dagger}\left(
\frac{\mathbf q}{2}
+
\mathbf k'
\right)
\psiB^j\left(
\frac{\mathbf q}{2}
+\mathbf k
\right)
\psiB^i\left(\frac{\mathbf q}{2}-\mathbf k\right) \nn
& + \frac{u}{4} \int  d \mathbf k  d \mathbf k' d \mathbf q \, 
\phiB_a( - \mathbf q-\mathbf k')
\phiB_a(\mathbf k')
\phiB_b(\mathbf k)
\phiB_b(\mathbf q-\mathbf k). 
\label{action_low_energy}
\end{align}
Here, $\psiB^j(\mathbf k)$ is the fermion field of 
spin $j=\uparrow, \downarrow$
with three momentum 
$\mathbf k = (k_0,\delta, \theta)$. 
For fermions, we use the `fermionic polar coordinate' for spatial momentum: $\theta$ is the polar angle of $\vec k$ and
$\delta = (|\vec k| - \KFthetadim)$ is the radial deviation of momentum from the angle-dependent Fermi momentum $\KFthetadim$. 
In this paper, we focus on convex Fermi surfaces.
The overall size of the Fermi surface is denoted by one scale $\KFAVdim = 
\int 
\frac{ d \theta  }{2\pi}
\KFthetadim$.
$v_{F,\theta}$ is the angle-dependent Fermi velocity in the radial direction. 
We assume that there is $C_2$ symmetry and
$
{\bf K}_{F,\theta} = {\bf K}_{F,\theta+\pi}$
and
$v_{F,\theta} = v_{F,\theta+\pi}$.
$\phiB_a(\mathbf q)$ represents a critical boson with 
$\mathbf q= (q_0, q, \varphi)$.
The spatial momentum of boson is written in the usual polar coordinate: $\vec  q = q\left(\cos(\varphi),\sin(\varphi)\right)$.
If the fermion-boson coupling is in the spin singlet channel, $T^{a,i}_{j} = \delta^{i}_{j}$ with a singlet boson.
In the triplet channel, 
$T^{a,i}_{j}$ with 
$a=1,2,3$ are Pauli matrices and $\phi_a$ carries spin $1$.
$e_{\theta}(\varphi)$ is the fermion-boson coupling function.
The angle-dependent form factor of the coupling function depends on the nature of the critical mode.
For example, $e_{\theta}(\varphi) \sim \cos (n\theta)$ for the Ising-nematic critical fluctuation for the 
$C_{2n} \rightarrow C_{n}$ symmetry breaking phase transition.
For the U(1) gauge theory, $e_{\theta}(\varphi) \sim i \sin (\varphi-\theta)$.
The critical fluctuations can  either be in the spin singlet or triplet channel.
The integration measures of the fermion and boson momenta are written as $d {\bf k} =
\frac{dk_0 d\delta d \theta \KFthetadim} {(2\pi)^3}$ and $d {\bf q} =
\frac{dq_0 dq  d\varphi q}{(2\pi)^3}$, respectively, which uses $|\vec k| \approx \KFthetadim$.
$V_{ij}^{i'j'} (\theta',\theta;\vec q)$
is the pairing interaction that describes scatterings of Cooper pairs
from angles 
$(\theta, \theta+\pi)$
to
$(\theta', \theta'+\pi)$ 
with center of mass momentum $\vec q$.
It can be decomposed into the singlet ($-$) and triplet ($+$) channels as $V^{i'j'}_{ij} (\theta',\theta;\vec q) = \sum_{r=\pm} V^{(r)}_{\theta',\theta}(\vec q) \left(P^{(r)}\right)^{i',j'}_{i,j}$, where
$\left(P^{(\pm)}\right)^{i',j'}_{i,j}  = \frac{ \delta_i^{i'}\delta_j^{j'} \pm  \delta_i^{j'}\delta_j^{i'} }{2}$
is the projector to each channel.
$u$ is the quartic boson coupling.

For the tree-level scaling analysis, we adopt a scale transformation in which the fermion momentum is scaled towards the Fermi surface while the boson momentum is scaled towards zero:
$(k_0,\delta,\theta)  \rightarrow (s k_0,s \delta,\theta)$,
$(q_0,q,\varphi)  \rightarrow (s q_0, s q, \varphi)$
with $s<1$.
At first, it may appear that the Fermi momentum should be kept invariant because $\delta$ is reduced for the fermion.
However, the Fermi momentum is forced to be scaled to keep the rule of momentum conservation invariant under the transformation.
For example, a fermion near the Fermi surface changes its angle by $\delta \theta = q
\sin(\varphi-\theta)
/\KFthetadim
$
 when it absorbs a boson with momentum $\vec q$. 
With the angles fixed under the scaling, the Fermi momentum changes as $\KFthetadim \rightarrow \KFthetadim/s$.\footnote{Alternatively, one should assign a positive scaling dimension to angles\cite{PhysRevB.78.085129}.}
This captures the fact that the circumference of the Fermi surface, measured in units of $q$, increases as $q$ is reduced. 
Both  $d {\bf k}$ and $d {\bf q}$ have dimension $3$, and the fields must be rescaled as
$\psiB \rightarrow s^{-2} \psiB$ and $\phiB_a \rightarrow s^{-5/2} \phiB_a $.
Under this Gaussian scaling that keeps the kinetic term invariant,
the dimensionless Fermi momentum grows to $\infty$ as the energy scale is lowered.
$e_\theta(\varphi)$, $u$ and  $V_{\theta',\theta}^{(r)}$ have tree-level scaling dimensions $1/2$, $1$ and $-1$, respectively.
\eq{action_low_energy} is valid below a high-energy cutoff energy scale and the couplings in the action are defined at that energy scale.
However, the dimensionful fermion-boson coupling sets a dynamically important energy scale  $\Lambda \equiv \max_{\theta,\varphi} |e_\theta(\varphi)|^2$. 
It plays the role of a crossover scale that separates the non-universal high-energy physics and the universal non-Fermi liquid physics within the low-energy effective theory.
In this paper, we will focus on physics below $\Lambda$.

At energy scales below $\Lambda$, the fermion-boson coupling becomes strong, and the non-Fermi liquid effects are important.\footnote{
On the other hand, the quartic boson coupling becomes irrelevant once the boson is dressed by Landau damping in a large class of theories, at least perturbatively\cite{PhysRevB.110.155142}.
}
In \eq{action_low_energy}, the seemingly irrelevant four-fermion coupling is also included 
because it can give rise to infrared singularities 
in the pairing channel 
despite the negative scaling dimension. 
This is possible because
the phase space available for virtual Cooper pairs with zero center of mass momentum is proportional to $\KFAVdim$,
and those quantum corrections are controlled by the dimensionless coupling $\mathfrak{V} \sim  \KFAVdim V$.
In effect, the Fermi momentum shifts the scaling dimension of the quartic coupling in the pairing channel by $1$.
In Fermi liquids, this is marginally relevant for attractive interactions, which cause the BCS pairing instability\cite{BCS1,BCS2,SHANKAR}.
In non-Fermi liquids, the fermion-boson coupling can introduce an additional anomalous dimension, which can make the effective quartic coupling relevant, marginal, or irrelevant, depending on the nature of the critical fluctuations.
Even if the four-fermion coupling is irrelevant in one non-Fermi liquid, one can access a new (critical) non-Fermi liquid by tuning its strength.
Since we aim to classify various non-Fermi liquids that can be accessed through tuning the bare coupling, we include the four-fermion coupling in the pairing channel.

\begin{figure}[t]
\includegraphics[width=0.5\textwidth]{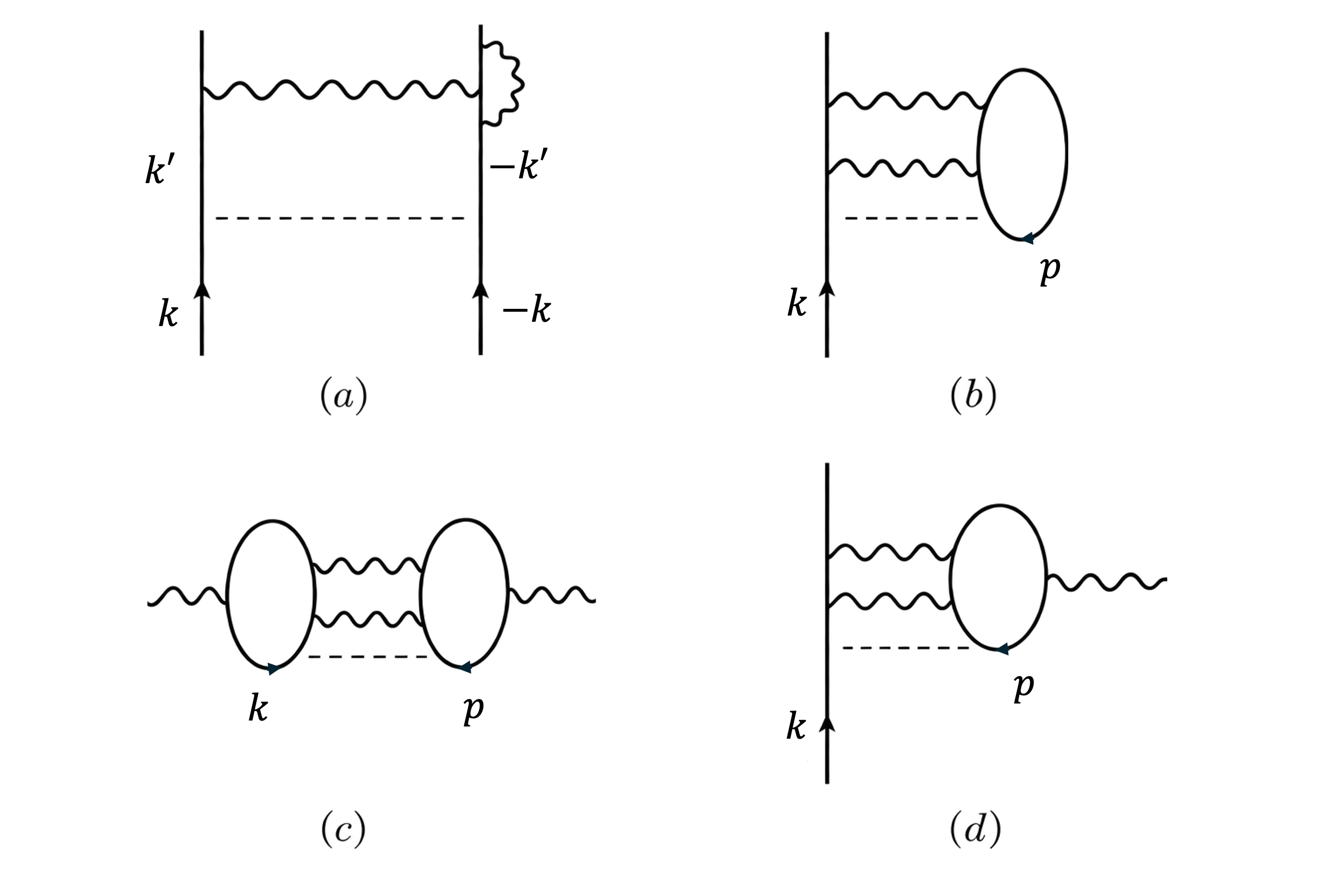}
\caption{
Quantum corrections that include the four-fermion coupling denoted by the dashed line.
The solid and wiggly lines are the fermion and boson propagators, respectively.
(a) For the four-point function in the pairing channel with zero net momentum, the virtual particle can go to any $\vec k'$ on the Fermi surface without costing energy. 
As a result, every vertex $V$ is enhanced by a factor of $\KFAVdim$. 
(b)-(d) For the two- and three-point functions, $V$ is not enhanced for generic internal momenta, $\vec p$ and $\vec k$.
These corrections can be ignored in the large $\KFAVdim$ limit 
in \eq{eq:largekF}.
}
\label{fig:largekFlimit}
\end{figure}

On the other hand,
the four-fermion coupling in the particle-hole channel can be dropped as it is not enhanced by $\KFAVdim$ in the absence of nesting\cite{PhysRevB.110.155142}.
Therefore, \eq{action_low_energy} is the minimal action in 
\bqa
\mbox{
the large $\KFAVdim/\Lambda$ limit with fixed $\mathfrak{V} \sim \KFAVdim V$
},
\label{eq:largekF}
\eqa
where $\Lambda$ is the energy scale associated with the fermion-boson coupling.
In the {\it large $\KFAVdim$ limit},
quantum corrections generated from $V$ survive only if the low-energy phase space of the virtual particles 
is proportional to $\KFAVdim$.
For example, the quantum corrections to the fermionic four-point function in the pairing channel have such an enhancement of $\KFAVdim$ because virtual Cooper pairs with zero center of mass momentum can be scattered anywhere on the Fermi surface without an energy penalty.
On the other hand, quantum corrections in the particle-hole channel and
the contributions of $V$ to the 
two-, three-, and bosonic four-point functions are sub-leading because the phase space of virtual excitations is limited 
(see  \fig{fig:largekFlimit}).

\section{Non-perturbative flow equation}
\label{sec:flow}

In metals, the information on the universal low-energy dynamics is encoded in the vertex function evaluated
near zero momentum for bosons and as a function of angles around the Fermi surface for fermions.
At low energies, all vertex functions are fixed by a few low order vertex functions, up to corrections that vanish in positive powers of the energy scale.
Here, we focus on non-Fermi liquids whose low-energy physics is fixed at most by the two-, three-, and four-point functions.
These low order vertex functions are, in turn, determined by the set of coupling functions,
$\left\{ \KFthetadim, v_{F,\theta}, e_\theta(\varphi), V^{(r)}_{\theta',\theta}, u \right\}$ 
in \eq{action_low_energy}.

\begin{figure}[ht]
\includegraphics[width=0.3\textwidth]{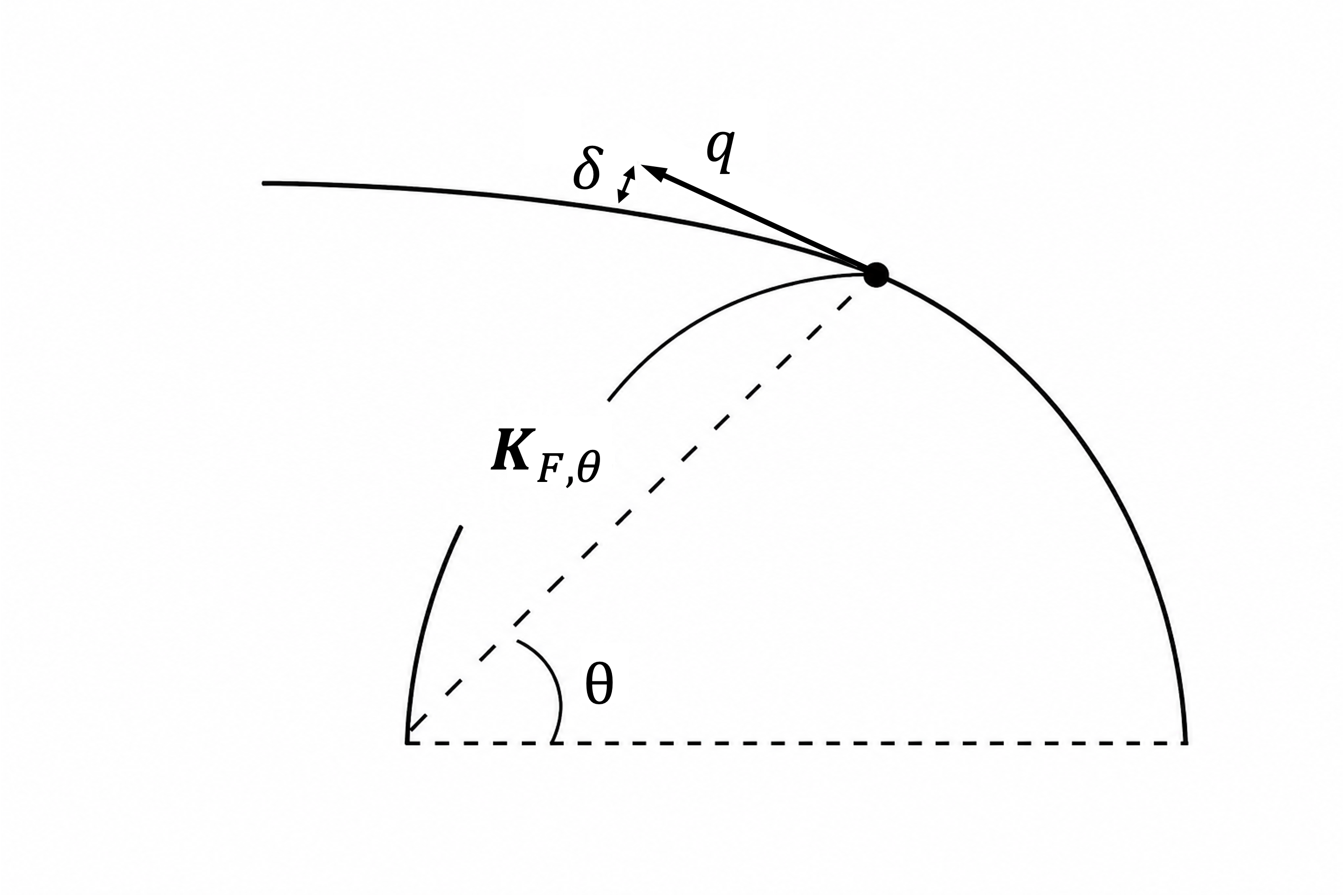}
\caption{
Bosons decay primarily into particle-hole pairs near the Fermi surface tangential to its momentum.
}
\label{fig:tangential_scattering}
\end{figure}

In non-Fermi liquids, the two-, three-, and bosonic four-point functions are rather featureless.
In $2+1$ dimensions, the dimensionless fermion-boson vertex 
and quartic boson vertex typically acquire $O(1)$ values, which give rise to a correction to the dynamical critical exponent ($z-1$) and anomalous dimensions for the fermion 
($\eta_{\psiB,\theta}$) 
and boson ($\eta_\phiB$). 
For convex Fermi surfaces,
the fermion and boson propagators take the form of
\bqa
&& G(\omegaB, \delta, \theta) \sim 
\frac{1}{\delta^{
2-z-2\eta_{\psiB,\theta}
}}
~
f_\psi \left( \frac{\delta}{ \omegaB^{1/z}} \right),  
\nn &&
D(\omegaB,q,\varphi) \sim  \frac{1}{q^{ 3-z-2\eta_\phiB }}
~ f_\phi \left( 
\frac{q^2}{
\KFAVdim \omegaB^{1/z}} \right) 
\label{eq:GD}
\eqa
at low energies.
Here, we define the dynamical critical exponent $z$ as the scaling dimension of frequency relative to momentum perpendicular to the Fermi surface.
In this convention,
$\omegaB \sim \delta^z$ for the fermion
and 
$\omegaB \sim q^{2z}$ for the boson.\footnote{ Perturbatively, $z=3/2$ for non-Fermi liquids with hot Fermi surfaces,
which results in 
$\omegaB \sim q^{3}$ for the boson.} 
The discrepancy in the scaling forms of the fermion and the boson originates from the fact that
$\KFAVdim$, which controls the curvature of the Fermi surface and determines the decay rate of the boson into the particle-hole continuum,
has the scaling dimension $1$.
Namely, a boson with momentum $\vec q$ primarily mixes with particle-hole excitations created near the Fermi surface tangential to $\vec q$, and
 the energy of the particle-hole pair scales as
$\delta \sim |\vec q|^2/\KFAVdim$
(see \fig{fig:tangential_scattering}).


In the two- and three-point vertex functions, virtual excitations created by $V$ have a limited low-energy phase space.
Since the four-fermion coupling is not enhanced by $\KFAVdim$ in such scattering processes, those vertex functions do not receive singular corrections from $V$ in the large $\KFAVdim$ limit.
Therefore, the two- and three-point functions are only renormalized by critical bosons. 
Well below the cutoff energy scale $\Lambda$, the dimensionless couplings that determine the anomalous dimensions become independent of scale, as do $\eta_{\psi,\theta}$
and $\eta_\phi$.
Furthermore, we assume that the high-energy bosons decouple from the low-energy physics,
namely, bosons with large momenta/energies do not create quantum corrections that are singular at low energies.
Perturbative theories that have been studied explicitly\cite{kukreja2026classificationnonfermiliquidsuniversal} satisfy this condition.
With the decoupling of high-energy modes at low energies, the anomalous dimensions are captured within the patch theory that describes small angle scatterings around the momentum of the external fermion.
For a convex Fermi surface, every patch is identical to each other up to a finite rescaling of fields and momentum, 
as will be explicitly shown in  \ref{sec:decoupling}.
As a result, $\eta_{\psiB,\theta}$ becomes independent of $\theta$ at low energies, except for cold spots with vanishing measure (if present).

Because the two- and three-point functions are determined by small-angle scatterings confined within a small patch of the Fermi surface, they are qualitatively similar for many non-Fermi liquids.\footnote{
The same is true for the quartic boson vertex function.}
On the other hand, the fermionic four-point function captures the dynamics of Cooper pairs, which are not kinematically confined within one patch of the Fermi surface, even at low energies.
It describes the interpatch couplings mediated by universal pairing interactions and contains the global information about the Fermi surface.
Therefore, the fermionic four-point function serves as a unique fingerprint that differentiates non-Fermi liquids beyond what individual patches can capture.

\subsection{Factorization of the four-fermion vertices}
\label{sec:factorization}

\begin{figure}[t]
\begin{subfigure}{0.4 \textwidth}
\includegraphics[width=\textwidth]{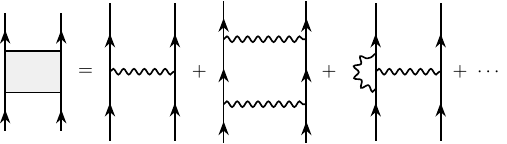}
\caption{}
\label{fig:Ge}
    \end{subfigure}
\begin{subfigure}{0.29\textwidth}
\includegraphics[width=\textwidth]{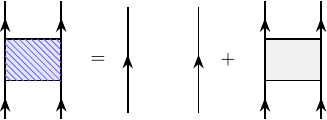}
        \caption{}
        \label{fig:G40}
\end{subfigure}
\begin{subfigure}{0.45\textwidth}
\includegraphics[width=\textwidth]{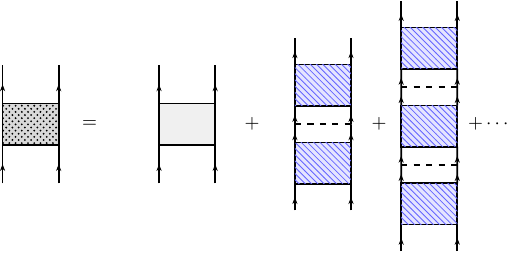}
        \caption{}
        \label{fig:V}
\end{subfigure}
    \caption{
(a) The exact connected four-point function 
for  $V=0$. 
The solid and wiggly lines denote the exact fermion and boson propagators, respectively, evaluated at $V=0$.
(b) The four-point function  
for $V=0$ that includes the disconnected parts.
(c) 
For $V \neq 0$, the connected four-point function is given by the sum of ladders made by alternating $V$ (dashed line) and the four-point function evaluated at $V=0$.
All other diagrams are suppressed by powers of 
$\omegaB/\KFAVdim$ in the large $\KFAVdim$ limit.
}
\label{fig:1}
\end{figure}

The universal superconducting fluctuations are encoded in the four-point function in the pairing channel.
Let 
$G^{(4)(j_1, j_1');(\nu_1,\delta_1,\theta_1)}_{c~~
(j_2, j_2');
(\nu_2,\delta_2,\theta_2)
}(\omegaB)$ 
be the connected four-point correlation function 
for a pair of incoming fermions with three momenta
${\bf k}_2 = (\omegaB/2+\nu_2, \delta_2, \theta_2)$,
${\bf k}_2' = (\omegaB/2-\nu_2, \delta_2, \theta_2
+\pi)$
and spins $j_2, j_2'$,
and outgoing fermions 
with 
${\bf k}_1 = (\omegaB/2+\nu_1, \delta_1, \theta_1)$,
${\bf k}_1' = (\omegaB/2-\nu_1, \delta_1, \theta_1+\pi)$
and $j_1, j_1'$.
$\omegaB$ is the center of mass frequency of Cooper pairs.
$\nu_1$ and $\nu_2$ are the relative frequencies.
$\delta_1$ and $\delta_2$ are deviations of the fermion momentum away from the Fermi surface in the radial direction.
The pairing vertex is obtained by removing the external legs: 
\bqa
\Gamma(\omegaB)=-K^{-1} \bullet G^{(4)}_c(\omegaB) \bullet K^{-1}.
\label{eq:Gamma}
\eqa
Here,
$K^{(j_1,j_1');(\nu_1,\delta_1,\theta_1)}_{(j_2,j_2');(\nu_2,\delta_2,\theta_2)}
=  2
\frac{
(2\pi)^3
}{\mathbf{K}_{F,\theta_1}}
\tdelta^{j_1}_{j_2}\delta^{j_1'}_{j_2'}\tdelta(\nu_1-\nu_2)\tdelta(\delta_1-\delta_2)\tdelta(\theta_1-\theta_2)G(\omegaB/2+\nu_1,\delta_1,\theta_1)
G(\omegaB/2-\nu_1,\delta_1,\theta_1+\pi)$
is the disconnected four-point function,
where  $G(\omegaB,\delta,\theta)$
is the exact Green's function.
$G^{(4)}_c$, $\Gamma$, and $K$ are viewed as matrices whose rows and columns run over $\{(j,j');(\nu,\delta,\theta)\}$.
The matrix product is defined as $(A\bullet B)^{(j_1,j_1');(\nu_1,\delta_1,\theta_1)}_{(j_2,j_2');(\nu_2,\delta_2,\theta_2)}=\sum_{i,j}\int \frac{d\nu\,d\delta}{(2\pi )^2}\int_{-\pi}^{\pi}\frac{ \KFthetadim\, d\theta}{2\pi}A^{(j_1,j_1');(\nu_1,\delta_1,\theta_1)}_{(i,j);(\nu,\delta,\theta)}B^{(i,j);(\nu,\delta,\theta)}_{(j_2,j_2');(\nu_2,\delta_2,\theta_2)}$.
The inverse is defined with respect to the identity matrix,
${\bf I}^{(j_1,j_1');(\nu_1,\delta_1,\theta_1)}_{(j_2,j_2');(\nu_2,\delta_2,\theta_2)}
=
\frac{
(2\pi)^3
}{\mathbf{K}_{F,\theta_1}}
\delta^{j_1}_{j_2}
\delta^{j_1'}_{j_2'}
\tdelta(\nu_1-\nu_2)
\tdelta(\delta_1-\delta_2)
\tdelta(\theta_1-\theta_2)
$.

At low energies, the dependence of the vertex function on $\nu_i$ and $\delta_i$ does not carry information independent of its dependence on $\omega$.
Hence, we define the vertex function that only depends on the angles and the center of frequency in the spin channel $r$ as
\bqa
&&\avn^{(r)}_{\theta_1,\theta_2} =
\frac{\sqrt{{\bf K}_{F,\theta_1}{\bf K}_{F,\theta_2}}}{ \mathcal N^{(r)} }
\left(P^{(r)}\right)^{j_2,j_2'}_{j_1,j_1'}
\bar \Gamma^{
(j_1,j_1');
\theta_1
}_{
(j_2,j_2');
\theta_2
}(\omegaB).
\label{eq:dimensionless4pt}
\eqa
Here,
$
\bar \Gamma^{
(j_1,j_1');
\theta_1
}_{
(j_2,j_2');
\theta_2
}(\omegaB)
=
\int 
d \nu_1 d \delta_1
d \nu_2 d \delta_2 ~
W^*_{\nu_1,\delta_1,\theta_1}(\omega)
\times$
$
\Gamma^{
(j_1,j_1');
(\nu_1,\delta_1,\theta_1)
}_{
(j_2,j_2');
(\nu_2,\delta_2,\theta_2)
}(\omegaB)
W_{\nu_2,\delta_2,\theta_2}(\omega)$ 
represents the vertex function averaged over relative frequencies and radial momenta with a distribution function
$W_{\nu,\delta,\theta}(\omega)$.
The distribution is peaked around $\nu =\delta=0$ with widths $\omega$ and $\omega^{1/z}$ in the $\nu$ and $\delta$ directions, respectively.
The smooth distribution function regularizes singularities that arise when the momentum in the particle-hole 
($t$ and $u$) channels vanishes at 
$\theta_1=\theta_2$ and 
$\theta_1=\theta_2+\pi$.
Note that the infrared singularity in the pairing ($s$) channel is regularized by a finite $\omega$.
One can choose different distribution functions, which amounts to choosing different RG prescriptions.
In Appendix \ref{app:nonpert_pairing}, we will take advantage of this freedom to simplify the flow equation of $\avn^{(r)}$.
$\left( P^{(\pm)} \right)^{j_2, j_2'}_{j_1, j_1'}$ projects the four-point function in the spin triplet ($+$) and singlet ($-$) channels, respectively, with multiplicities 
$\mathcal N^{(+)} =3$ and $\mathcal N^{(-)} =1$.
The factor of $\sqrt{{\bf K}_{F,\theta_1}{\bf K}_{F,\theta_2}}$ 
is introduced to account for the phase space of Cooper pairs available per angle in a manner that is symmetric between the incoming and outgoing Cooper pairs.
${\mathfrak V}^{(r)}$ is a matrix with continuous indices $\theta_1$ and $\theta_2$ composed of elements that are functions of $\omegaB$ and $\KFAVdim$. 
It represents the interaction vertex for the scattering of a Cooper pair of energy $\omegaB$ from angles  $(\theta_2, \theta_2+\pi)$ to $(\theta_1, \theta_1+\pi)$.
However, the effective coupling that appears in the RG flow equation is the interaction vertex multiplied by the density of states. 
Therefore, we multiply
the Cooper pair density of states,
\bqa
\omega \partial_\omega \int \frac{d \theta d\nu d\delta}{(2\pi)^3}
G\left(\omegaB+\nu,\delta,\theta\right) G\left(-\nu,\delta,\theta+\pi \right)  
\sim \omega^{\frac{2H_0}{z}}
\label{eq:Kprime}
\eqa
to $\avn^{(r)}$ to define an effective coupling function,
\bqa
\av^{(r)}(\omegaB;\KFAVdim) =
\omegaB^{\frac{2H_0}{z}}
\avn^{(r)}(\omegaB;\KFAVdim).
\eqa
Here, $H_0 =\frac{3(z-1)}{2} +2\eta_{\psi}$ is the scaling exponent associated with the disconnected four-point function.
In Fermi liquids, $H_0=0$,
and a non-zero $H_0$ captures the effect of single-fermion renormalization on superconducting fluctuations.
If fermions are incoherent
($z-1 + 2\eta_{\psiB,\theta} > 0$)
and the effective mass is divergent
($z>1$), $H_0$ is positive. 
This suppressed density of states is a pair-breaking effect\cite{
Balatsky01101993,
PhysRevLett.74.2575,
doi:10.1142/S0217979296000349,
PhysRevB.100.115132,
BORGES2023169221}. 

While it is difficult to compute the four-point function explicitly for strongly coupled theories, 
one can still constrain its scaling form at energies much lower than the Fermi energy.
This is possible because quantum corrections generated from the four-fermion couplings survive in the large $\KFAVdim$ limit only if each $V$ is accompanied by a factor $\KFAVdim$.
This occurs if and only if the four-fermion couplings successively scatter Cooper pairs with zero center of mass momentum, forming a ladder shown in  \fig{fig:1}.
In other words, the four-fermion coupling acts as a coupling with the tree-level scaling dimension $0$ in the Cooper channel, whereas it acts as a coupling with the dimension $-1$ otherwise.
In the ladder graphs, the four-fermion couplings factorize, and their contributions can be summed exactly
(see Appendix \ref{app:nonpert_pairing} for derivation).
All other graphs are suppressed by  $\omegaB^{1/z}/\KFAVdim$ at low energies 
(see Appendix \ref{app:nonladder}, for example).
A corollary of this is that the four-fermion coupling can be ignored for  the two-, three-, and bosonic four-point functions in the large $\KFAVdim$ limit.
This is because pairs of fermions scattered by the four-fermion coupling in those vertex functions generally have non-zero net momentum, and there is no enhancement by $\KFAVdim$
(see \fig{fig:largekFlimit}).

Our goal is to understand the evolution of the angle-dependent coupling function $\av_{\theta_1,\theta_2}(\omegaB;\KFAVdim)$  with decreasing $\omegaB$.
The manner in which $\av$ evolves as a function of $\omegaB$ depends on the infrared singularity of the Cooper pair self-energy defined by
\bqa
\Sigma=
\left( K + G^{(4)}_c \right)^{-1} - K^{-1}.
\label{eq:Sigma}
\eqa
Since the contribution of the four-fermion coupling is explicitly taken into account through the ladder graphs, the flow equation for $\av^{(r)}$ can be expressed solely in terms of $K$ and $\Sigma_0(\omega)$ that are evaluated at $V=0$. 
Being a diagonal matrix,  eigenvalues of $K$ scale as 
\eq{eq:Kprime}.
Suppose the eigenvalues of $\Sigma_0$ scale as $\omegaB^{-\alpha}$ in the low frequency limit.
The strength of the self-energy correction relative to the disconnected function is determined by the eigenvalues of $K \bullet \Sigma_0$ that scale as
$\omega^{-(\alpha - 2H_0/z )}$.
In general, there can be three possibilities:
\bqa \begin{cases} \mbox{Type I } & :   \alpha < 2H_0/z  \\ \mbox{Type II } & :  \alpha = 2H_0/z  \\ \mbox{Type III } & :  \alpha > 2H_0/z  \end{cases}. \eqa
In types I, II and III,
the pair self-energy is irrelevant, marginal, and relevant, 
respectively, relative to the disconnected four-point function.
We emphasize that the comparison is being made for the self-energy and the disconnected four-point functions that are already fully renormalized by the critical boson.
In theories that are strongly interacting at low energies, it is expected that the interactions renormalize the propagators and the fermion-boson vertex such that $K^{-1}$ and $\Sigma_0$ become comparable in the low-energy limit. 
This corresponds to type II non-Fermi liquids in which $K \bullet \Sigma_0$ has only logarithmic singularities, which are summed to an anomalous dimension for the four-fermion vertex in the low-energy limit.
All known examples of non-Fermi liquids belong to this type. 
In type I, the pair self-energy is strictly sub-leading relative to the disconnected function.
In type III, the self-energy is more singular. 
As is shown in  Appendix \ref{app:nonpert_pairing}, this leads to a scale-dependent anomalous dimension that runs algebraically in energy.
In the present paper, we do not consider such possibilities and focus on type II non-Fermi liquids.

In type II non-Fermi liquids, the functional renormalization group flow equation, which is exact in the large $\KFAVdim$ limit, is given by
\bqa
&& 
\hspace{-5mm}
\partial_l \av^{(r)} 
 =   -  
 \av^{(r)} \cdot
 \mK^{'(r)} 
 \cdot
 \av^{(r)}
  \nn 
&&  
-  \av^{(r)}\cdot 
 \mH^{'(r)^\dagger }
   -  
   \mH^{'(r)}
   \cdot \av^{(r)}
   + \mS^{'(r)},
\label{eq:general_fp_eq_for_av}
\eqa
where
$l =  \frac{1}{z} \log \frac{\Lambda}{\omega}$ is the logarithmic length scale.
$\av^{(r)}$,
$\mK^{'(r)}$,
$\mH^{'(r)}$
and
$\mS^{'(r)}$
are all matrices in angular space with inner product defined by
$(A\cdot B)_{\theta_1,\theta_2} = \int_{-\pi}^{\pi} \frac{d\theta}{2\pi}A_{\theta_1,\theta}B_{\theta,\theta_2}$. 
In Fermi liquids, $\mK^{'(r)}$ is independent of $\omega$, and $\mH'=\mS'=0$, 
which reproduces the well-known flow equation,
$\partial_l \av^{(r)} 
 =   - 
 \av^{(r)} \cdot 
 \mK^{'(r)}  \cdot
 \av^{(r)}$\cite{
SHANKAR,
PhysRevB.81.224505}.
The fate of Fermi liquids is determined by the sign of the bare coupling: an attractive (repulsive) pairing interaction grows (decays) logarithmically as the energy is lowered.
In non-Fermi liquids, the critical fluctuations modify the flow equation from that of Fermi liquids in the following ways.
\begin{itemize}
    \item 
$\mK^{'(r)}$:
The kernel of the quadratic term can be written as
$\mK^{'(r)}= 
\mK'_0
+ \mK^{'(r)}_1$.
$\mK'_0$ is the contribution of the disconnected Cooper pair propagator $K$.
Hence, it is proportional to 
$I_{\theta_1,\theta_2} 
= 2\pi \delta(\theta_1-\theta_2)$.
On the other hand, $\mK^{'(r)}_1$ results from the boson-generated connected four-point function. 
It describes the scatterings that virtual Cooper pairs undergo in the ladder between the rungs of the four-fermion vertices.

\item 
$\mH^{'(r)}$:
The kernel of the linear term can also be decomposed into the contributions from the disconnected propagator and the connected four-point function as
$
\mH^{'(r)}
=
\mH'_0
+
\mH^{'(r)}_1
$.
$\mH'_0 = H_0 I$ represents the anomalous dimension of the four-fermion coupling caused by the incoherence of fermions.
Being a single-particle property, it is diagonal in angular space and contributes equally to all angular momentum channels.
This has a pair-breaking effect.
$\mH^{'(r)}_1$ represents the mixing of Cooper pairs at different angles, which is mediated by the boson-generated connected four-point function.
It can be either pair-breaking or pair-forming, depending on the sign of the vertex correction.

\item $\mS^{'(r)}$:
Pairing interactions are generated by the critical fluctuations, even when the bare interaction is zero.
This is captured by
the source term that is independent of $\av^{(r)}$\cite{ PhysRevD.59.094019, PhysRevB.91.115111}.
The sign and range of the source term are determined by the type of the critical boson and the energy scale $\omegaB$.
This is entirely generated from the connected four-point function.
\end{itemize}
The full expressions of
$\mK^{'(r)}_{\theta_1,\theta_2}(\omega)$, $\mH^{'(r)}_{\theta_1,\theta_2}(\omega)$ and $\mS^{'(r)}_{\theta_1,\theta_2}(\omega)$ can be found 
in Appendix \ref{app:nonpert_pairing}.

\subsection{Diagonalization}
\label{sec:decoupling}

For a general classification of the solutions to the flow equation, we do not need the explicit form of the kernels.
In this section, we discuss the general properties of the kernels that we need.
The first key property is that  the kernels are entirely determined by the critical boson at $V=0$ because the effect of the four-fermion coupling is explicitly taken into account through the sum of the ladder graphs.
The second key property of the kernels is that the angular averages of $\mK^{('r)}_{\theta_1,\theta_2}(\omega)$, $\mH^{'(r)}_{\theta_1,\theta_2}(\omega)$ and $\mS^{'(r)}_{\theta_1,\theta_2}(\omega)$ are independent of $\omega$ in the small $\omega$ limit. 
This follows from the fact that in type II non-Fermi liquids, the irreducible pair self-energy gives logarithmic corrections that lead to a scale-independent flow of the coupling in the logarithmic energy scale in the s-wave limit (see Appendix \ref{app:nonpert_pairing} for details). 
However, the matrix elements are still functions of $\omega$ and $\KFAVdim$.
Since they describe scatterings of Cooper pairs from one angle to another, they are not diagonal in angular space.
The range of the off-diagonal elements is determined by the ratio between the momentum of the critical boson and $\KFAVdim$.

Since the boson carries typical momentum 
\bqa
q_\parallel(l) \sim  
\sqrt{\Lambda  \KFAVdim} e^{-l/2}
\label{eq:qparallel}
\eqa
at scale $l$,
the kernels are
sharply peaked at 
$|\theta_1-\theta_2| \sim 
q_\parallel(l)/\KFAVdim$.
We further assume that
the high-energy bosons do not renormalize low-energy Cooper pairs and that 
the kernels are
negligible
for $|\theta_1-\theta_2| \gg q_\parallel(l)/\KFAVdim$.
This implies that large-angle scattering is not important in the kernels,
and their eigenvalues are finite in the large $\KFAVdim$ limit.
The underlying intuition is simple.
In the momentum shell RG, integrating out the boson between momenta $q_\parallel(l)$ and $q_\parallel(l+dl)$ gives rise to quantum corrections whose support is bounded by $q_\parallel(l)$\cite{ PhysRevD.59.094019, PhysRevB.91.115111}.
Therefore, the low-energy boson at  scale $l$ cannot renormalize the off-diagonal scattering amplitude at
$|\theta_1-\theta_2| \gg q_{\parallel}(l)/\KFAVdim$,
even when the boson propagator itself decays slowly at large momentum due to a large anomalous dimension\cite{BORGES2023169221}.

\begin{figure}[ht]
\includegraphics[width=0.23\textwidth]{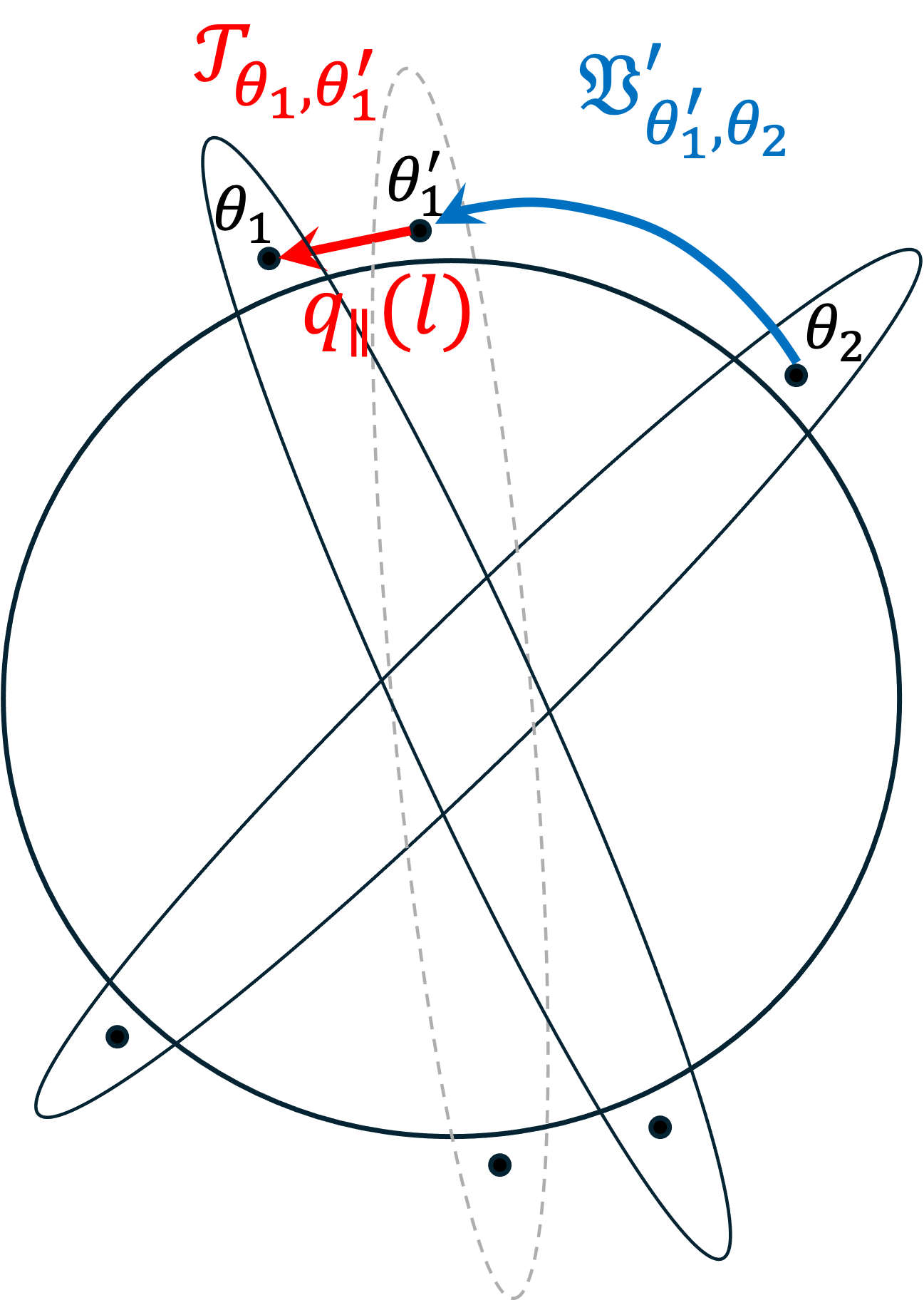}
\caption{ 
The fully renormalized pairing interaction 
$\av_{\theta_1,\theta_2}$ 
can mediate large-angle scatterings, in general.
However, its renormalization is governed by $\bT_{\theta_1,\theta_1'}(l)$ at scale $l$, which arises from 
small-angle scatterings with momentum transfer $q_\parallel(l)$.
}
\label{fig:VT}
\end{figure}

We emphasize that this does not necessarily imply that large-angle scatterings are suppressed in the vertex function.
In the low-energy limit, $\av_{\theta_1,\theta_2}$ can decay slowly in $\theta_1-\theta_2$, in which case large-angle scatterings are important\cite{BORGES2023169221,PhysRevLett.128.106402}.
What is guaranteed to decay fast is the kernels
in the flow equation,
which describe the contribution of the critical boson within the thin shell of 
scale between $l$ and $l + dl$.
On the contrary, $\av_{\theta_1,\theta_2}$ is the renormalized pairing interaction, which results from the quantum corrections that have accumulated from the short-distance cutoff scale all the way to scale $l$ as well as the short-ranged quartic four-fermion coupling that induces large-angle scatterings.
This is illustrated in \fig{fig:VT}.



Since  $\mK^{'(r)}_{\theta,\theta+\Delta \theta}$, $\mH^{'(r)}_{\theta,\theta+\Delta \theta}$ and $\mS^{'(r)}_{\theta,\theta+\Delta \theta}$ 
decay rapidly with increasing $\Delta \theta$,
they are only sensitive to the neighborhood of the Fermi surface at angle $\theta$. 
Hence, they can be computed within the patch theory that captures near forward scatterings around that angle.
Since every patch is identical to one another, up to a rescaling of momentum and fields, there is an emergent rotational symmetry that maps a patch to a nearby patch.
Consequently, there exists a rescaled angular coordinate $\bar \theta(\theta)$ in which the emergent rotational symmetry is realized as a shift: $\bar \theta \rightarrow \bar \theta + \bar \varphi$.
This is shown in Appendix \ref{app:diagonalization}. 
Therefore, the kernels of the flow equation can be diagonalized in the generalized angular momentum basis conjugate to $\bar \theta$.
In that angular momentum basis, 
\eq{eq:general_fp_eq_for_av} takes the form of the infinite dimensional matrix Riccati equation with diagonal kernels\cite{bittanti1991riccati},
\bqa
 \partial_l  
 \av_{nm} 
 =   -  
 \mK^{'}_{k}
 \av_{nk} 
 \av_{km} 
-  (\mH^{'}_{n}+\mH^{'}_{m})
   \av_{nm}
   + \mS^{'}_{m} \delta_{nm}. \nn
\label{eq:general_fp_eq_for_av2}
\eqa
Here, $k,n,m$ denote angular momentum indices.
$\av_{nm} = 
\left[U^{\dagger} \cdot \av^{(r)} \cdot U\right]_{nm}$ represents the effective coupling that describes the scattering of a Cooper pair from the angular momentum channel $m$ to $n$. 
Although the universal pairing interaction generated from the critical fluctuations is diagonal in this basis, the bare quartic coupling is generally not. 
Here, we suppress the superscript $(r)$ because the spin of a Cooper pair is fixed by its parity eigenvalue.
Furthermore,
$\av_{nm}=0$ if $n$ and $m$ have different parity.
$\mK^{'}_{n}$, $\mH^{'}_{n}$ and $\mS^{'}_{n}$ are the eigenvalues of the kernels,
$\mK^{'(r)}$,
$\mH^{'(r)}$
and $\mS^{'(r)}$, 
respectively.

\begin{figure}[ht]
\includegraphics[width=0.23\textwidth]{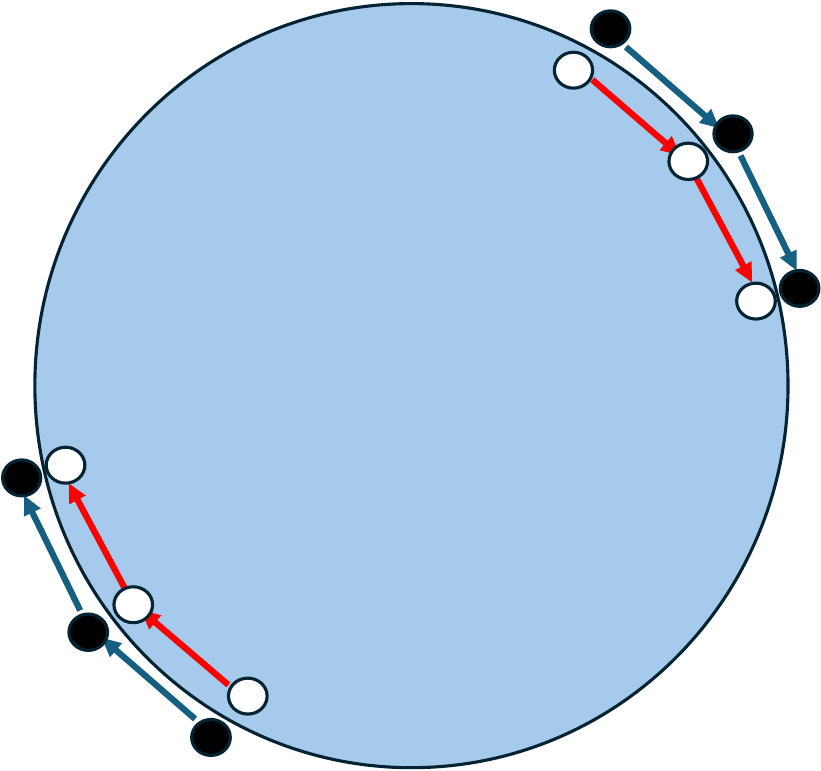}
\caption{
For every path of a Cooper pair of particles in the path integral,
there exists a time-reversal counter part of a Cooper pair of holes with the equal weight.
The phase spaces for the particles and holes are the same to the leading order in 
the large $\KFAVdim$ limit. 
}
\label{fig:PHSymmetry}
\end{figure}

In general, 
$\av_{nm}(\omegaB)$ 
is not a Hermitian matrix at non-zero $\omegaB$.
However, an emergent particle-hole symmetry forces it to be Hermitian to the leading order in $\Lambda/\KFAVdim$ because
for every particle state available near the Fermi surface,
there exists a corresponding hole state with the same energy,
as illustrated in \fig{fig:PHSymmetry}.
Although the phase spaces for particles and holes are not exactly the same, the difference is suppressed by $\Lambda/\KFAVdim$ in the large $\KFAVdim$ limit.
More precisely, 
the combination of a particle-hole transformation
($\hat \psi^{j}(\delta,\theta) \rightarrow \epsilon^{ji}  \hat \psi^{\dagger}_i(-\delta,\theta+\pi)$)
and time-reversal 
($\hat \psi^{j}(\delta,\theta) \rightarrow i (\sigma_y)^j_i \hat \psi^i(\delta,\theta+\pi)$)
leaves the theory invariant 
to the leading order in $\Lambda/\KFAVdim$ 
(see Appendix \ref{app:PHSymmetry}).
This gives a chain of equalities,
\bqa
& \av^{(r)}_{\theta_1,\theta_2}(\omegaB;\KFAVdim) 
= \av^{(r)PH}_{\theta_2,\theta_1}(-\omegaB;\KFAVdim) \nn
& = \av^{(r)T}_{\theta_2,\theta_1}(-\omegaB;\KFAVdim) 
= \av^{(r)}_{\theta_1,\theta_2}(-\omegaB;\KFAVdim)
\eqa
in the large $\KFAVdim$ limit,
where
$\av^{(r)PH}$
and
$\av^{(r)T}$
represent the pair-correlation function 
of the theories obtained after applying the particle-hole and time reversal transformations, respectively.
The first and last equalities hold generally, while the middle one is a consequence of the symmetry.
Consequently, $\av_{nm}(\omegaB)$ is even in $\omegaB$.
Combined with
$\av^*_{nm}(\omegaB)=  \av_{mn}(-\omegaB)$,
 we conclude that
$\av^*_{nm}(\omegaB)=  \av_{mn}(\omegaB)$.
Since 
$\mK^{'}_{n}$, $\mH^{'}_{n}$ and $\mS^{'}_{n}$ are the eigenvalues of the kernels made of $\av(\omegaB)$ at $V=0$,
they are real.

For $V=0$, $\av = \av_0$ is the solution of 
\eq{eq:general_fp_eq_for_av2}
by construction.
However, the fate of non-Fermi liquids in the low-energy limit generally depends on $V$, as is the case in Fermi liquids.
As will be shown below, whether or not there exists a finite $V$ for which a normal state remains stable down to zero temperature is an important criterion in classifying non-Fermi liquids.

\section{Superuniversality classes}
\label{sec:classification}

Being a first-order differential equation,  \eq{eq:general_fp_eq_for_av2} admits a unique solution for each bare coupling $\av(0)$.
Explicitly solving the flow equation is hard because it requires knowing the kernels 
$\mK^{'}_{n}$, $\mH^{'}_{n}$, and $\mS^{'}_{n}$ non-perturbatively.
Here, our goal is a more modest one.
We consider the bundle of particular solutions associated with all possible bare four-fermion couplings and classify the topology of the bundle.
As will be discussed, the topological classification can be done without knowing the full dynamical details.
Here, we assume that the fermion-boson coupling has already saturated to its asymptotic low energy value at scale $\Lambda$. 
In the large $\KFAVdim$ limit, the four-fermion coupling does not back-react to the two- and three-point functions.

\subsection{General solution of the flow equation} \label{subsec: general classification}

Since we are mainly interested in universal low energy physics, we focus on the flow of the coupling function below energy $\Lambda$.
The general solution of
\eq{eq:general_fp_eq_for_av2} can be written as\cite{bittanti1991riccati}
\bqa
\av_{nm}(l)
=
&&\left[\mathcal U_{21}(l;0)
+
\mathcal U_{22}(l;0)
\av(0)\right]_{nk}
\nn &&
\left[
\mathcal U_{11}(l;0)
+
\mathcal U_{12}(l;0)
\av(0)
\right]^{-1}_{km}.
\label{app:RiccatiSolution0}
\eqa
Here, $\av(0)$ is the coupling defined at scale $\Lambda$, which corresponds to $l=0$. 
$\mathcal U_{ab}$
for each $a,b=1,2$ is an infinite dimensional diagonal matrix in the angular momentum basis: 
$(\mathcal U_{ab}(l_2;l_1))_{nm} = u^{(n)}_{ab}(l_2;l_1)\delta_{nm}$.
For each $n$,
the $2 \times 2$ matrix 
$u^{(n)}(l_2;l_1) 
\equiv
\begin{pmatrix}
u_{11}^{(n)}(l_2;l_1) & u_{12}^{(n)}(l_2;l_1)\\
u_{21}^{(n)}(l_2;l_1) & u_{22}^{(n)}(l_2;l_1)
\end{pmatrix}
$
is given by
\bqa
u^{(n)}(l_2;l_1) 
= {\cal P}
\exp
\left[
\int_{l_1}^{l_2}
dl\,
\begin{pmatrix}
\mH_n'(l)  & \mK_n'(l)\\ \mS_n'(l)& -\mH_n'(l)
\end{pmatrix}
\right]. 
\label{eq:uabn}
\eqa
$\mathcal P$ is the RG path-ordering operator that places matrices with larger $l$ to the left of those with smaller $l$.

Superconducting instabilities are diagnosed by $\av(l)$ that flows to $-\infty$ at a finite length scale.
In Fermi liquids, 
$\mH_n'= \mS_n'=0$ and $\mK_n'=1$, which
results in
\bqa
\av_{nm}(l)
=
&& 
\av_{nk}(0)
\left[
1
+
l \av(0)
\right]^{-1}_{km}.
\label{app:RiccatiSolution0_FL}
\eqa
As expected, this shows that a superconducting instability arises if and only if $\av(0)$ has a negative eigenvalue\cite{SHANKAR}.
In non-Fermi liquids,  superconducting instabilities arise
if $X(l;0) + \av(0)$ has a zero eigenvalue,
where $X(l_2;l_1) \equiv 
\mathcal U_{12}(l_2;l_1)^{-1}
\mathcal U_{11}(l_2;l_1)$. 
With
$\mathcal U_{12}(0;0)=0$ and 
$\mathcal U_{11}(0;0) = I$, $X(0;0)=\infty$.
If $X_n(l;0) = -\infty$ at $l=l_s$ for some $n$, there exists a superconducting instability for any finite $\av(0)$:
since $X_n(l;0)$ changes from $\infty$ to $-\infty$ in $0<l<l_s$, some eigenvalues of $\av(0)+X(l;0)$ must cross zero in $0<l<l_s$. 
In this case, the metal becomes unstable against superconductivity, irrespective of the bare coupling.
On the other hand,  if $X_n(l;0)$ is bounded from below for all $l$ and $n$, the normal state remains stable down to zero temperature for a sufficiently repulsive bare coupling.



\eq{app:RiccatiSolution0} is a map from the couplings at one scale to those at another scale, as shown in
\eq{eq:category}.
Our goal is to classify the universal RG trajectories (projective fixed points) into topologically distinct classes.
In general, the couplings include off-diagonal elements unless the microscopic theory has exact rotational symmetry.
However, the restriction of  \eq{eq:category} on diagonal couplings contains enough information to fix $\mK'$, $\mH'$ and $\mS'$ because
the kernels are diagonal.
In Appendix
\ref{app:diagonal_completeness}, it is shown that 
$\mK'$, $\mH'$ and $\mS'$ can indeed be inferred from the solution restricted to the diagonal couplings.
Since the general flow equation is determined by $\mK'$, $\mH'$ and $\mS'$, the flow of the general off-diagonal coupling is entirely fixed by the flows in the subspace of diagonal couplings.
Therefore, we classify the general solutions using the diagonal couplings. 
Two non-Fermi liquids that exhibit topologically distinct behaviors for the diagonal solutions are surely in distinct superuniversality classes.
However, there can be, in principle, a finer classification once the off-diagonal solutions are included.

\subsection{Kinematic constraints}

Even in the subspace of diagonal couplings,
each angular momentum channel can, in principle, exhibit different behaviors. 
However, there are kinematic constraints that
limit what can possibly happen in different angular momentum channels.

Generally, $\mK^{'}_{n}$, $\mH^{'}_{n}$ and $\mS^{'}_{n}$ are determined by the following four momentum scales.
The first scale is the high-energy cutoff $\Lambda$.
The second is the Fermi momentum $\KFAVdim$. 
%
The third is the pitch of the Cooper pair wavefunction for $n \neq 0$: 
$2\pi \KFAVdim/n$.
%
At large $n$,
the pitch becomes arbitrarily small.
The final scale is the momentum carried by critical fluctuations.
At scale $l$,
the typical momentum that the boson carries is
$q_\parallel(l)$ in
\eq{eq:qparallel}.

Being dimensionless,
$\mK^{'}_{n}, \mH^{'}_{n}, \mS^{'}_{n}$ are functions of 
$q_\parallel/\KFAVdim$,
$n q_\parallel/{\bf k}_{F}$ and
$q_\parallel/\Lambda$.
Since 
$q_\parallel(l) \sim e^{-l/2}$ at large $l$,
we can take the $q_\parallel/\Lambda,
q_\parallel/\KFAVdim \rightarrow 0$ limit
at low energies. 
The flow equation is well defined in this limit because the high-energy modes do not affect the flow at low energies.
However, one cannot generally take
$n q_\parallel/{\bf k}_{F}$ to be small even at low energies because
$n$ is unbounded.
To denote the fact that the coefficients in the flow equation depend on $l$, $n$ and $\KFAVdim$ through only one parameter, 
we write the kernels of the flow equation as
\bqa
\left( \mK^{'}_{n}(l),
\mH^{'}_{n}(l),
\mS^{'}_{n}(l) \right)
= (\kappa_{y^{(n)}(l)}, h_{y^{(n)}(l)}, s_{y^{(n)}(l)}),
\label{eq:Cyty}
\eqa
where
\bqa 
y^{(n)}(l) \equiv \log\left(\frac{n q_\parallel(l)}{{\bf k}_{F}} \right)
\label{eq:yty}
\eqa
is the logarithmic scale of the boson momentum measured in units of the pitch of the Cooper pair wavefunction.
The logarithmic ratio between $q_\parallel(l)$ and the pitch decreases as $y^{(n)}(l)=y^{(n)}(0)-l/2$ with increasing $l$.
Although it is hard to compute the kernels $C_y  \equiv (\kappa_{y}, h_{y}, s_{y})$  explicitly for strongly coupled theories,  the following kinematic constraints of $C_y$ constrain the form of the general solutions.

\subsubsection*{Locality}
    
Since the underlying theory is local, quantum corrections generated at finite $l$ decay exponentially at large distances in real space.
In momentum space, this translates to the analytic property: 
$\Sigma_0$ 
at scale $l$ is smooth at angular scales less than 
$|\theta_1-\theta_2| \sim 
q_\parallel(l)/\KFAVdim$.
Consequently, its Fourier transformation becomes negligible if $n \gg \KFAVdim/q_\parallel(l)$.
In the large $n$ limit,
the contributions from
$\mK^{'(r)}_1$, 
$\mH^{'(r)}_1$ and $\mS^{'(r)}$ vanish.
The only non-Fermi liquid corrections that survive are 
$\mK'_0$ 
and $\mH'_0$,
which contribute to all angular momentum channels equally.
This gives the first kinematic constraint on $C_y$:
\bqa
\lim_{y \rightarrow \infty} C_{y}
=\Bigl( 
K_0,
H_0,
0 \Bigr),
\label{eq:constraint1}
\eqa
where $K_0$ is the 
eigenvalue of $\mK'_0$, 
which is independent of angular momentum and frequency.
It is noted that the crossover occurs when $q_\parallel(l)$ becomes comparable to ${\bf k}_{F}/n$ at
\bqa
l_n^* = \log \left( n^2\Lambda /\KFAVdim \right).
\label{eq:crossover}
\eqa
As $n$ increases, the crossover length scale increases without bound.

\subsubsection*{ Universal s-wave limit}

For $q_\parallel(l) \ll \KFAVdim/n$,
the critical boson scatters Cooper pairs essentially within one lobe of the Cooper pair wavefunction around the Fermi surface (see \fig{fig:pitch_vs_q}).
Consequently, 
for $l \gg l_n^*$,
the pairing interaction is renormalized in the same way as the pairing interaction in the s-wave channel.
This leads to the second kinematic constraint:
\bqa
\lim_{y \rightarrow -\infty}
C_{y} = 
\left(
\kappa_{-\infty}, h_{-\infty}, s_{-\infty}
\right).
\label{eq:constraint2}
\eqa
It is noted that the $y \rightarrow -\infty$ limit of $C_y$ exists because the large-angle scatterings do not contribute to the kernels
as discussed in Sec. \ref{sec:decoupling}.
For $l \gg l_n^*$, the flow equation no longer depends on $n$.
However, the crossover length scale $l_n^*$ is arbitrarily large for large $n$.
At any finite $l$, there always exist sufficiently large angular momentum channels for which $y^{(n)}(l) > 0$.

\subsection{Categorical classification}
\label{subsec: type I nfls}

\begin{figure}[ht]
\includegraphics[width=0.33\textwidth]{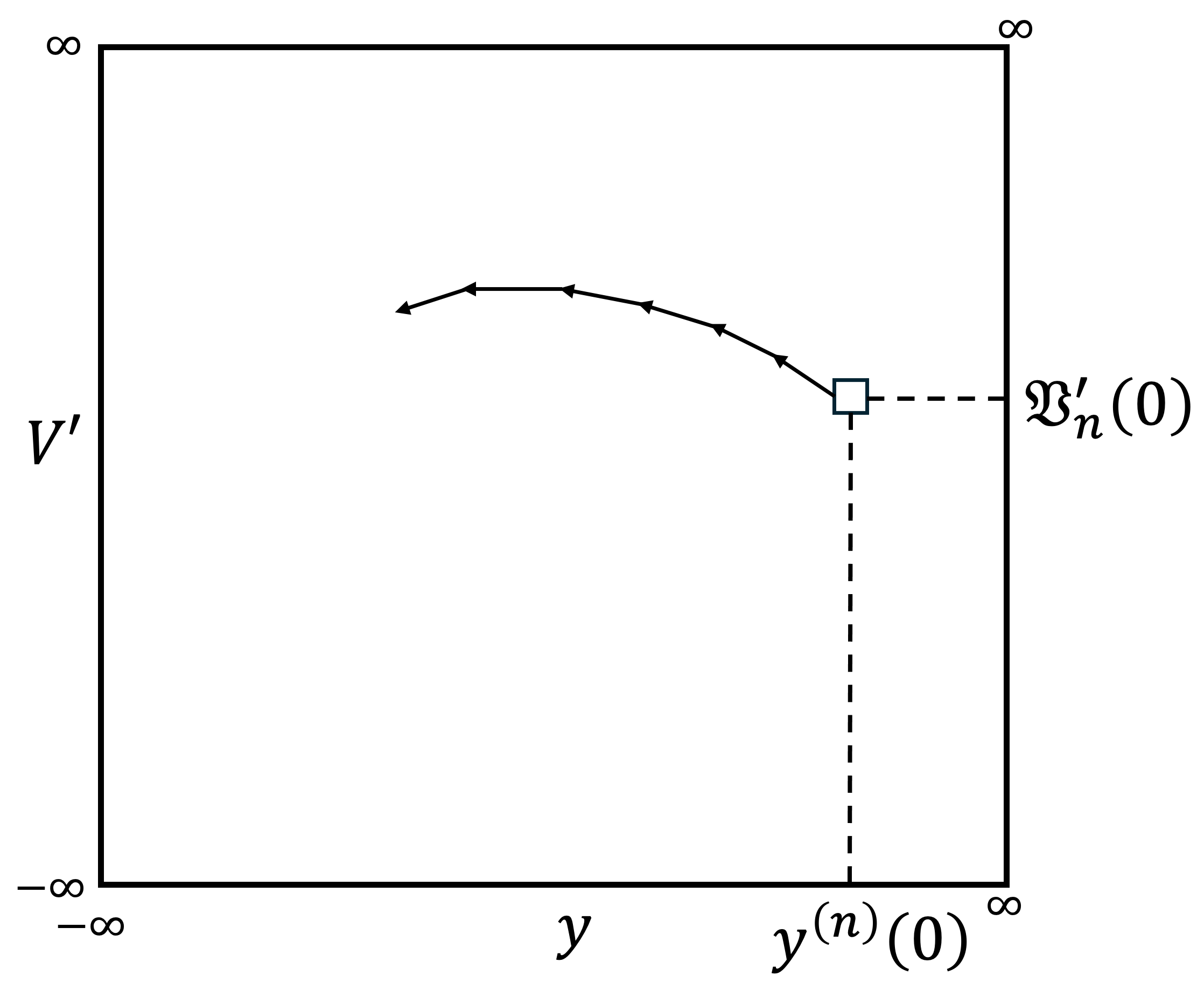}
\caption{
The coupling in channel $n$ 
at scale $l$ 
defines a point
$\left(y^{(n)}(l), \av_n(l) \right)$ 
in the $(y,\aV)$ plane. 
As $l$ increases to $l+dl$, the point shifts by 
$dl \Bigl(
 - 1/2,
 \beta_{
 y^{(n)}(l) }
 [\av_n(l)]
 \Bigr)$,
 where
 $
 \beta_y[\aV] =
- \kappa_{y} \aV^2 - 2 h_{y} \aV + s_{y}$.
The flow velocity depends on both coordinates of the plane 
because it depends on scale through $y$ as well as the coupling. 
Each trajectory corresponds to a particular solution of
\eq{eq:general_fp_eq_log}.
}
\label{fig:yVplane}
\end{figure}

The flow equation for the pairing interaction depends on $n$, $\KFAVdim$ and $l$ only through 
$y^{(n)}(l)$ at low energies. 
As a result, the flow velocity of the coupling in the angular momentum channel $n$ at scale $l$ is identical to that of 
the coupling in the angular momentum channel $n'$ at scale $l'$ 
as long as 
$y^{(n)}(l)= y^{(n')}(l')$.
This means that the flow of the coupling in all angular momentum channels is governed by one equation,
\begin{equation}
    \begin{aligned}
 -\frac{1}{2} \partial_y\aV_y =  
 - \kappa_y \aV^{{}^2{}}_y 
 - 2 h_y \aV^{{}}_y 
 + s_y,
\label{eq:general_fp_eq_log}
    \end{aligned}
\end{equation}
where $y$ is used as the scale along which the coupling runs based on $-1/2 \partial_y = \partial_l$.
\eq{eq:general_fp_eq_log}
completely captures how the universal pairing interaction evolves with increasing length scale.
Suppose that the bare coupling in the angular momentum channel $n$ is $\av_n(0)$ at $l=0$.
The coupling at larger $l$ is given by 
$\aV_{y^{(n)}(l)}$ with
$y^{(n)}(l)= y^{(n)}(0)  -l/2$, 
where 
$\aV_y$ is the solution of \eq{eq:general_fp_eq_log} that passes through $\left( y^{(n)}(0), \av_n(0) \right)$.
We call \eq{eq:general_fp_eq_log}  the \textit{projective fixed point} equation, and their solutions, {\it projective fixed points}.
Projective fixed points (PFPs) control the universal pairing interaction at low energies. 
Although PFPs play similar roles to conventional fixed points, one key difference is that they are functions of scales ($y$) and capture the universal crossovers.
The perturbative version of the PFP equation was first obtained in 
Ref. \cite{kukreja2026classificationnonfermiliquidsuniversal}.

In \eq{eq:general_fp_eq_log}, the flow speed depends not only on the coupling ($\aV$) but also on the energy scale through running $y$
(see \fig{fig:yVplane}).
The explicit scale-dependence of the flow equation prevents the couplings from settling into scale-invariant fixed point values.
Fixed point-like objects exist only in the large and small $y$ limits 
in which
$C_{y}$
approaches the $y$-independent asymptotic values
$C_{\infty}$
and
$C_{-\infty}$, respectively.
The fixed points of the flow equation in the $y=\pm \infty$ limit are referred to as $\pm \infty$ asymptotic fixed points.
In addition to the superconducting asymptotic fixed point $\aV^{SC}_{\pm \infty} = -\infty$, 
in each asymptotic limit,
there are two, one or zero
regular asymptotic fixed points 
if $\eta_{P,\pm \infty}$ is greater than, equal to or less than $0$, respectively, where
\bqa
\eta_{P,y} = 16\left(h_y^2 + \kappa_y s_y\right)
\label{eq:discriminant}
\eqa
is the discriminant.
\eq{eq:discriminant} is determined by the scale-dependent competition between the incoherence of fermions 
and the universal pairing interaction generated from critical fluctuations.
Highly incoherent fermions with large $h_y$ tend to make the discriminant positive, while a strong attractive interaction with negative $s_y$ tends to make it negative for $\kappa_y>0$\footnote{
In order for strong attractive interaction to flow to the superconducting fixed point, $\kappa_y$ should be positive.
}.
An asymptotic fixed point is stable, marginal, or unstable if the linearized flow is attractive, stationary, or repulsive, respectively. 
We denote the stable, marginal, and unstable asymptotic fixed points at $y=\pm \infty$
as 
$\aV^{\bullet}_{\pm \infty}$,
$\aV^{\halfominus}_{\pm \infty}$, and
$\aV^{\circ}_{\pm \infty}$, respectively.
In the large $y$ limit, there always exist a pair of stable and unstable asymptotic fixed points in addition to the superconducting fixed point because \eq{eq:constraint1} guarantees $\eta_{P,\infty}>0$\footnote{
If $h_\infty=0$, $\eta_{P,\infty}=0$.
This occurs in Fermi liquids in which $z=1$ and the anomalous dimension of the fermion is zero.
Here, we focus on non-Fermi liquids with $h_\infty > 0$.
}.
On the other hand, the number of $-\infty$-asymptotic fixed points varies, depending on the nature of the critical fluctuations.

In the angular momentum channels with $y^{(n)}(0) \ll 0$, the couplings `immediately' flow to one of the $-\infty$-asymptotic fixed points. 
Which asymptotic fixed points they flow into is determined by $\eta_{P,-\infty}$ and the bare coupling.
In the small $y$ asymptotic region, the flow is not different from the ordinary RG flow, as the flow equation becomes independent of $y$\cite{PhysRevD.59.094019,PhysRevB.91.115111}.
However, there always exist large angular momentum channels with $y^{(n)}(0) \gg 0$ and their flow cannot be understood without taking into account the scale-dependence of the flow equation.
Those couplings initially flow toward an $\infty$-asymptotic fixed point, say 
$\aV^\bullet_{\infty}$ if the bare coupling is above
$\aV^\circ_{\infty}$.
The couplings stay close to $\aV^\bullet_{\infty}$
for $l \ll l_n^*$.
Eventually, $y^{(n)}(l)$ flows out of the large $y$ asymptotic region and undergoes crossovers into the small $y$ asymptotic region for $l \gg l_n^*$.
The value of $\eta_{P,-\infty}$ and the coupling with which the RG trajectory enters the small $y$ asymptotic region determine the $-\infty$ asymptotic fixed points into which the coupling ultimately flows.
Since $l_n^*$ in  \eq{eq:crossover} increases indefinitely with increasing $n$,
the crossovers for increasingly large angular momentum channels continue down to the zero energy limit.

\begin{table}[]
    \centering
    \begin{tabular}{|c|c|c|}
    \hline
 ~~Class~~  & ~~$\eta_{P,-\infty} $ ~~ & Morphisms  \\
    \hline
    \hline
A &  $>0$ &  
~
$ \aV^\bullet_{-\infty} \leftarrow \aV^\bullet_{\infty}$,  ~$ \aV^\circ_{-\infty}  \leftarrow \aV^\circ_{\infty} $~ \\
    \hline
B &  $>0$ &  
~$ 
\aV^{SC}_{-\infty}
\leftarrow 
\aV^\bullet_{\infty} 
$,  ~$ 
\aV^{SC}_{-\infty} 
\leftarrow 
\aV^\circ_{\infty}  
$~ \\
    \hline
C &  $<0$ &  
~$ 
\aV^{SC}_{-\infty}
\leftarrow 
\aV^\bullet_{\infty} 
$,  ~$ 
\aV^{SC}_{-\infty} 
\leftarrow 
\aV^\circ_{\infty}
$~ \\
    \hline
AB &  $>0$ &  
~$ 
\aV^{\circ}_{-\infty}
\leftarrow 
\aV^\bullet_{\infty} 
$,  ~$ 
\aV^{SC}_{-\infty} 
\leftarrow 
\aV^\circ_{\infty}  
$~ \\
    \hline
AC &  $0$ &  
~$ 
\aV^{\halfominus}_{-\infty}
\leftarrow 
\aV^\bullet_{\infty}
$,  ~$ 
\aV^{\halfominus}_{-\infty} 
\leftarrow 
\aV^\circ_{\infty}  
$~ \\
    \hline
BC &  $0$ &  
~$ 
\aV^{SC}_{-\infty}
\leftarrow 
\aV^\bullet_{\infty} 
$,  ~$ 
\aV^{SC}_{-\infty} 
\leftarrow 
\aV^\circ_{\infty}  
$~ \\
    \hline
ABC &  $0$ &  
~$ 
\aV^{\halfominus}_{-\infty}
\leftarrow 
\aV^\bullet_{\infty} 
$,  ~$ 
\aV^{SC}_{-\infty} 
\leftarrow 
\aV^\circ_{\infty}  
$~ \\
     \hline
    \end{tabular}
\caption{
Categorical classification of the superuniversality classes.
The objects of the category is the set of asymptotic fixed points at $y=\pm \infty$, and
the morphisms are solutions of 
\eq{eq:general_fp_eq_log} that interpolate between 
the $\infty$ asymptotic fixed points and 
the $-\infty$ asymptotic fixed points.
}
\label{tab:classification}
\end{table}
The precise manner in which the couplings transition from $\infty$ to $-\infty$ asymptotic regions depends on the details of the theory.
However, the global pattern of crossovers has topological information that does not depend on the details.
The same topological data is shared by a group of non-Fermi liquids, called {\it superuniversality class}.
The data that specify the superuniversality class of non-Fermi liquids are as follows.
\begin{itemize}
\item 
Competition between the incoherence and the universal pairing interaction in the s-wave limit:

The number and stability of the $-\infty$ asymptotic fixed points are determined by the discriminant in the $-\infty$ asymptotic region, 
\bqa
\begin{array}{lc}
-~ \{ \aV^\bullet_{-\infty},
\aV^\circ_{-\infty},
\aV^{SC}_{-\infty} \}, & 
\eta_{P,- \infty}>0  \\ 
-~ \{ \aV^\halfominus_{-\infty}, \aV^{SC}_{-\infty} \}, & 
\eta_{P,- \infty}=0 \\
-~ \{ \aV^{SC}_{-\infty}\}, &
\eta_{P,- \infty}<0 
\end{array}.
\eqa

\item

PFPs between asymptotic fixed points:

Since $y$ decreases with increasing length scale, RG trajectories that emanate from $\infty$-asymptotic fixed points flow to $-\infty$-asymptotic fixed points.
This defines asymptotic PFPs that interpolate between $\infty$-asymptotic fixed points and $-\infty$-asymptotic fixed points.
For example, if there is a PFP that connects 
$\aV^\bullet_\infty$ to $\aV^\bullet_{-\infty}$,
we denote it by 
$
\aV^\bullet_{-\infty} 
\leftarrow
\aV^\bullet_{\infty} 
$.
We call PFPs that interpolate asymptotic fixed points as morphisms. 
\end{itemize}
The asymptotic fixed points and the morphisms between them completely determine the topology of the universal RG trajectories.
Mathematically, this can be viewed as a category in which the asymptotic fixed points are objects and the PFPs that connect them are morphisms (arrows)\footnote{
A category consists of a collection of objects, $\{ A, B, C,...\}$ and a collection of morphisms, $\{ f, g, h, ... \}$. A morphism `relates' a source object and a target object.  If $f$ relates $A$ and $B$, we denote it as  $f : A \rightarrow B$. Out of two morphisms, $f : A \rightarrow B$ and $g : B \rightarrow C$, one can construct a composite  $g \circ f : A \rightarrow C$. The composition obeys associativity:  $(h \circ g) \circ f = h \circ (g \circ f)$. Each object is equipped with the identity morphism such that $f \circ 1_A = 1_B \circ f = f$.
}.
%
One of the morphisms that plays an important role in the classification is the separatrix.
Each $-\infty$ asymptotic fixed point defines a basin of attraction:
the set of $(y,\aV)$ that flow to that $-\infty$ asymptotic fixed point in the low-energy limit.
In all classes with $\eta_{P,-\infty} \geq 0$, there exists a stable or marginal asymptotic fixed point at $y=-\infty$ in addition to $\aV^{SC}_{-\infty}$.
The boundary between the basin of attraction of 
$\aV^\bullet_{-\infty}$ (or $\aV^\halfominus_{-\infty}$)
and that of $\aV^{SC}_{-\infty}$ is a PFP itself. 
It is referred to as the separatrix PFP and denoted as $\aV^S_y$.

\begin{figure}[ht]
    \centering

    \begin{subfigure}{0.15\textwidth}
        \centering
        \includegraphics[width=\linewidth]{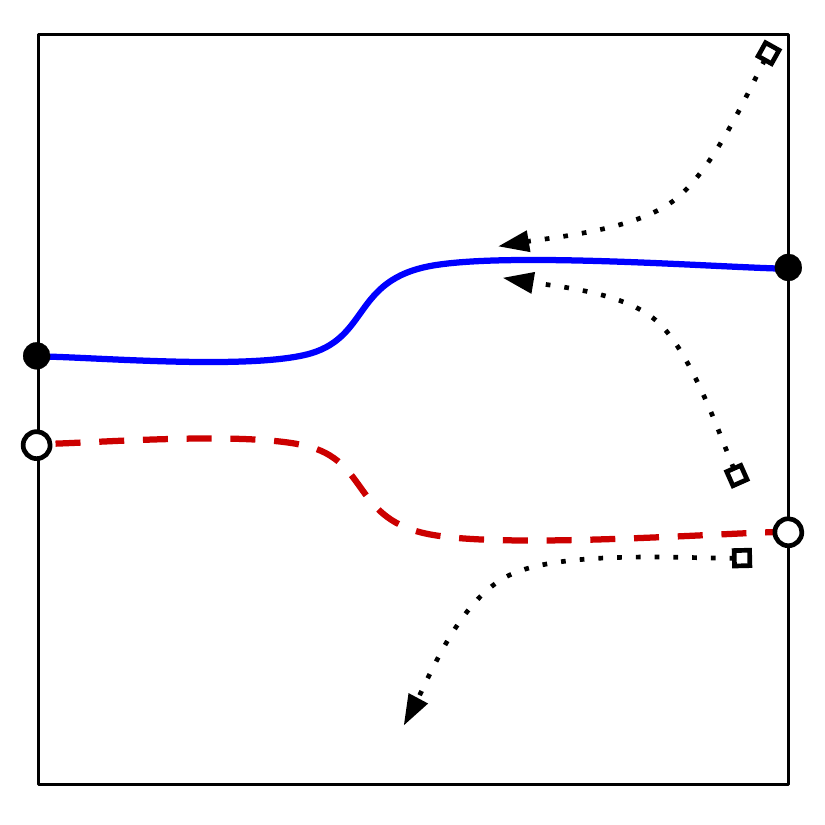}
        \caption{Class A}
    \end{subfigure}
    \begin{subfigure}{0.15\textwidth}
        \centering
        \includegraphics[width=\linewidth]{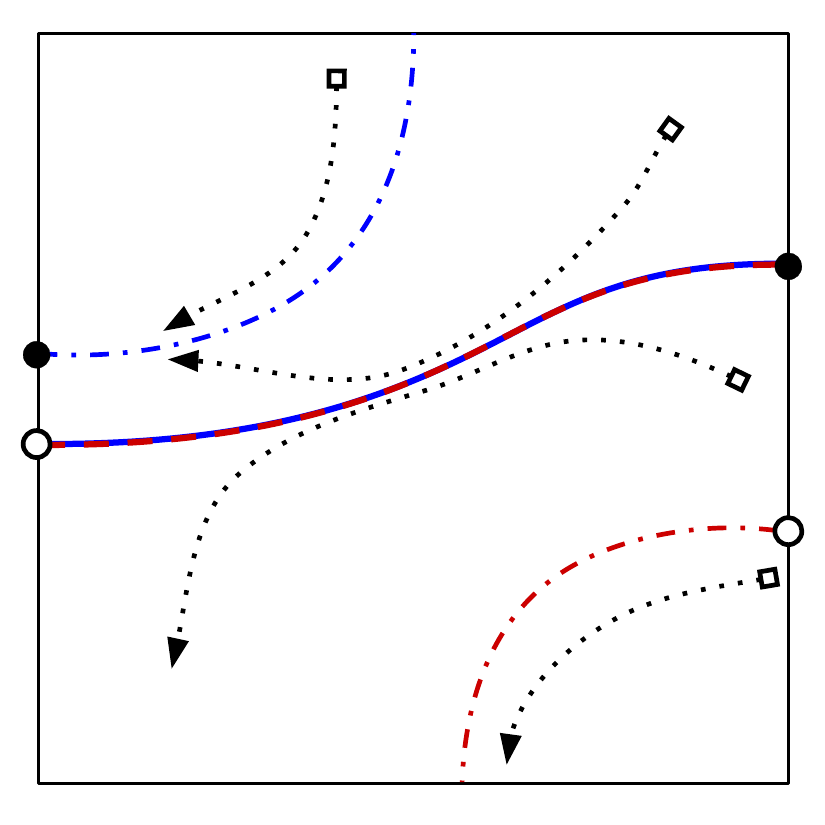}
        \caption{Class AB}
    \end{subfigure}
    \begin{subfigure}{0.15\textwidth}
        \centering
        \includegraphics[width=\linewidth]{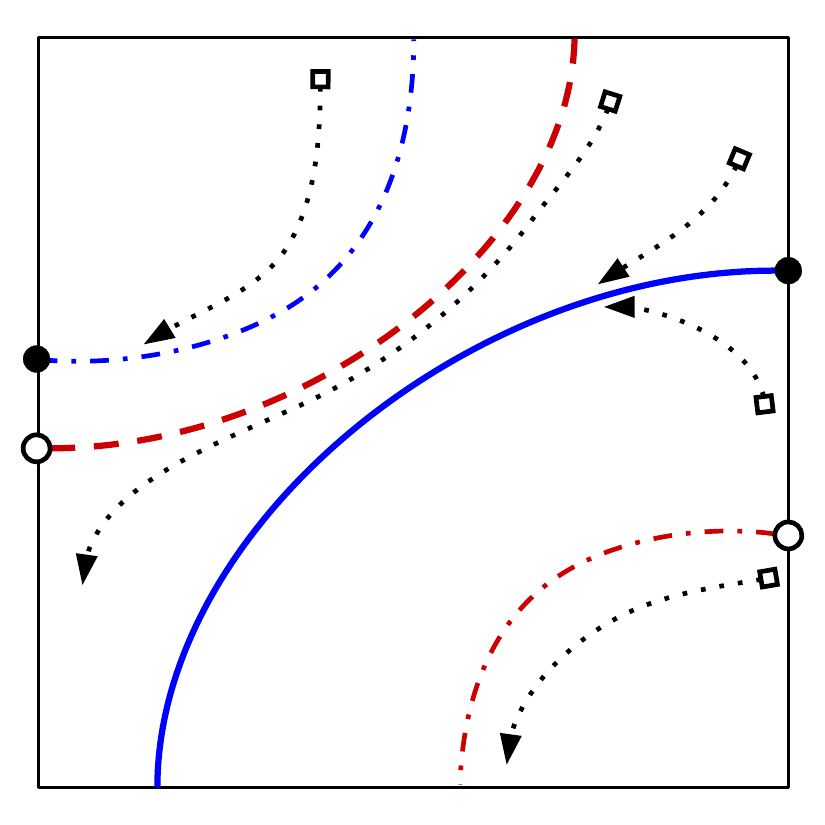}
        \caption{Class B}
    \end{subfigure}

    \begin{subfigure}{0.15\textwidth}
        \centering
        \includegraphics[width=\linewidth]{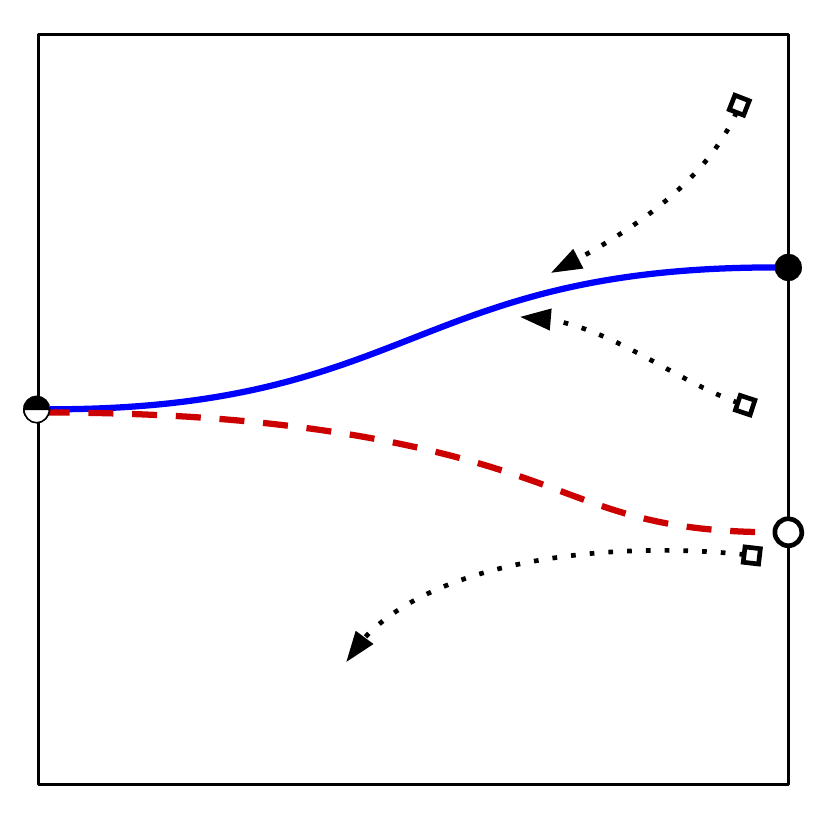}
        \caption{Class AC}
    \end{subfigure}
    \begin{subfigure}{0.15\textwidth}
        \centering
        \includegraphics[width=\linewidth]{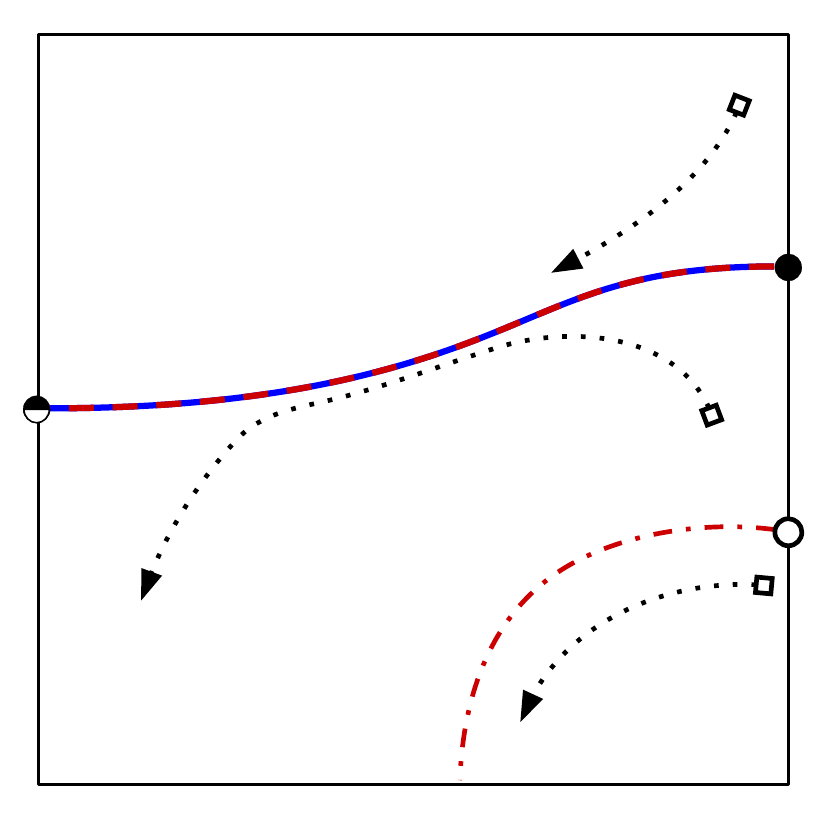}
        \caption{Class ABC}
    \end{subfigure}
    \begin{subfigure}{0.15\textwidth}
        \centering
        \includegraphics[width=\linewidth]{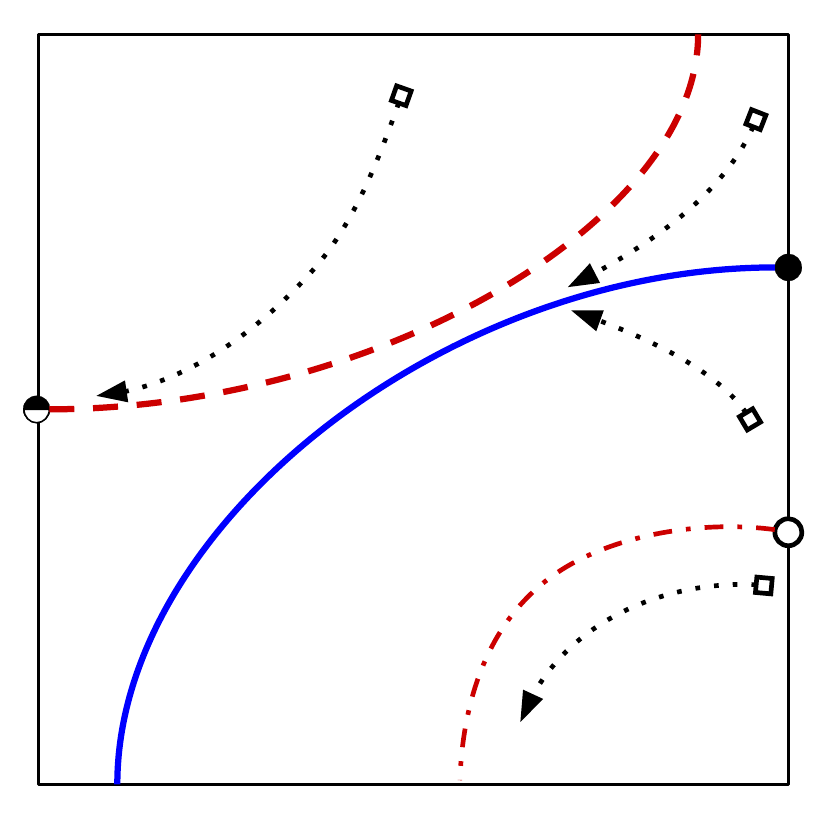}
        \caption{Class BC}
    \end{subfigure}

    \begin{subfigure}{0.15\textwidth}
        \centering
        \includegraphics[width=\linewidth]{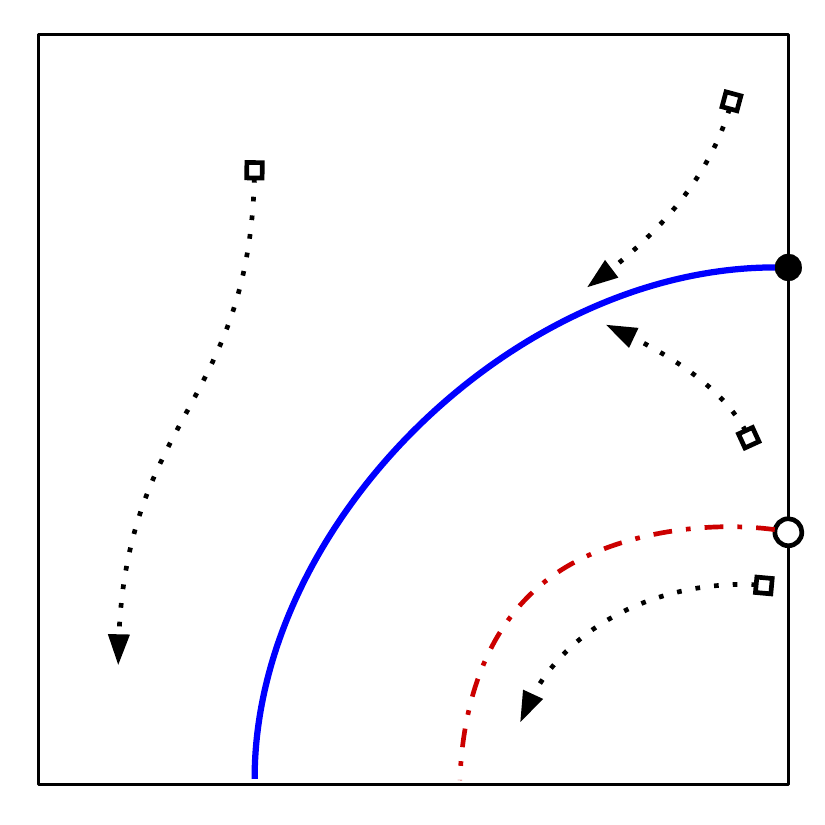}
        \caption{Class C}
    \end{subfigure}
\caption{
Seven topologically distinct phase portraits of PFPs in the $(y,\aV)$ plane.
In all panels, the horizontal and vertical axes are $y$ and $\aV$, respectively.
The infinite interval of $(-\infty, \infty)$ has been compactified in both axes.
At $y=\infty$, there always exist a pair of stable ($\bullet$)
and unstable ($\circ$) asymptotic fixed point
beside the superconducting fixed point.
At $y=-\infty$, there can be a pair of stable ($\bullet$) and unstable ($\circ$) asymptotic fixed point, 
a marginal 
(\protect\halfominus)
fixed point,
or no fixed point besides the superconducting fixed point.
The number and type of the asymptotic fixed points and the morphisms between them completely fix how others PFPs behave.
Here, the PFPs that emanate from $\aV^\bullet_\infty$ are shown as blue solid lines, while the sepratrix PFPs are shown as red dashed lines.
In classes AB and ABC, the two coincide.
}
    \label{fig:main}
\end{figure}

There are seven topologically distinct classes, as summarized in Table \ref{tab:classification}.
The corresponding portraits of PFPs are shown in \fig{fig:main}.
The present non-perturbative classification coincides with the one obtained from the perturbative flow equation\cite{kukreja2026classificationnonfermiliquidsuniversal}.
This is because the flow equation remains quadratic in the large $\KFAVdim$ limit, and the topological classification does not depend on the dynamical details.
Therefore, the characterization of each superuniversality class 
in Ref. \cite{kukreja2026classificationnonfermiliquidsuniversal} follows through in the present case.
Here, we summarize the main features of each superuniversality class.

In class A, there are two regular morphisms between asymptotic fixed points: 
$
\aV^\bullet_{-\infty} \leftarrow  \aV^\bullet_{\infty} 
$ and 
$\aV^\circ_{\infty} \leftarrow \aV^\circ_{-\infty}$.
The former is called the metallic projective fixed point (PFP), and the latter is the separatrix PFP.
Since two different RG trajectories cannot cross in the $(y,\aV)$ plane, all points above the separatrix should remain above it under the RG flow.
Couplings that start above the separatrix are attracted towards the metallic PFP at low energies because
$\aV^\bullet_{-\infty}$ and  $\aV^\bullet_{\infty}$ are stable asymptotic fixed points.
This means that, in the low energy limit, the system flows to a stable metallic state whose universal pairing interaction is controlled by the metallic PFP as long as the bare coupling in every angular momentum channel is above the separatrix.
Nonetheless, the stable non-Fermi liquid state does {\it not} exhibit scale invariance at low energies because the couplings keep running within the metallic PFP even after they have flowed into it.

In class C,  
$\eta_{P,-\infty}$ is negative, and 
there is a one-way flow to $\aV=-\infty$ in the small $y$ region.
Therefore, the superconducting instability is inevitable in the s-wave channel, irrespective of the bare coupling.
The upper bound for the superconducting length scale is 
$l_{SC,max} = 4\pi/\sqrt{|\discinf|}$,
which translates to the minimum $T_c$, $ T_{c,min} = \Lambda  e^{-z l_{SC,max}}$ for s-wave superconductivity\cite{
PhysRevB.91.115111,
kukreja2026classificationnonfermiliquidsuniversal}.
In fact, no angular momentum channel is safe from superconducting instability because 
the PFP that emanates from 
$\left( y^{(n)}(0), \av_n(0) \right)$ for any $n$ and $\av_n(0)$ reaches the small $y$ region in $l > l_n^*$ and eventually diverges to $-\infty$.
In practice, however, superconducting instabilities will not occur in high angular momentum channels unless their bare couplings lie below the separatrix.
A coupling that lies above the separatrix at large $y^{(n)}(0)$ is first attracted toward $\aV^\bullet_{\infty}$ and stays there up to $l \sim l_n^*$ before 
it begins to grow attractive only for $l>l_n^*$.
If $l_n^* > l_{SC,max}$, however, the s-wave instability occurs before the instability in channel $n$ sets in.
The actual pairing symmetry is determined by whichever channel becomes unstable first, 
which depends on the bare coupling.

In class B,  $\eta_{P,-\infty}$ is positive, and there is no  superconducting instability in the s-wave channel for bare coupling above $\aV^\circ_{-\infty}$.
However,
there exist superconducting instabilities at high angular momentum channels 
because  $\aV^\bullet_{\infty}$ flows to
$\aV^{SC}_{-\infty}$ under the RG flow.
This behavior is caused by an intermediate region of $y$ in which $\eta_{P;y}$ is sufficiently negative.
Namely, there is a range of energy scales in which the pairing interaction generated by critical fluctuations at non-zero angular momentum channels is strong enough to overcome the pair breaking effect of incoherence.
Couplings at large angular momentum channels first flow close to $\aV^\bullet_{\infty}$ at an intermediate energy scale and diverge to $-\infty$ before they enter the small $y$ asymptotic region.
Since there is no such instability for small enough $n$,
there exists a critical angular momentum $n_S$ above which superconducting instability is unavoidable, irrespective of the bare coupling. 
Unlike class C, superconducting instabilities in class B are driven  by scale-dependent but universal low-energy fluctuations.

These three types (classes A,B,C) of non-Fermi liquids can be realized without fine-tuning (except for the tuning that is potentially required to keep the boson gapless).
As one tunes some marginal parameters or kinematic parameters, such as the contents of the critical bosons, one can drive transitions from one class to another.
There are three critical classes (AB, AC and BC) that require one such fine tuning.
Class AC is obtained by making the universal pairing interaction $s_y$ more attractive relative to $h_y$ starting within class A until $\eta_{P,-\infty}$ becomes zero.
This causes 
$\aV^\bullet_{-\infty}$
and
$\aV^\circ_{-\infty}$
to merge into
$\aV^\halfominus_{-\infty}$ while
the morphism from
$\aV^\bullet_{\infty}$ remains within the basin of attraction of $\aV^\halfominus_{-\infty}$.
In this critical superuniversality class, non-Fermi liquids support the strongest universal s-wave pairing interaction that a normal state can bear.
In class BC, such a merger of $-\infty$-asymptotic fixed points occurs within class B.
For repulsive bare couplings, the universal s-wave pairing interaction becomes critical, while the superconducting instability occurs at a high angular momentum channel. 
The critical class AB is somewhat different from AC and BC.
As one drives the transition from class A to B, the morphism that emanates from 
$\aV^\bullet_{\infty}$ 
switches its target object from 
$\aV^\bullet_{-\infty}$ 
to
$\aV^{SC}_{-\infty}$.
At the critical point (class AB), the morphism from
$\aV^\bullet_{\infty}$ flows to the unstable asymptotic fixed point
$\aV^\circ_{-\infty}$.

Finally, there exists a double critical class (ABC) that requires tuning two parameters.
Class ABC can be realized by tuning the s-wave pairing interaction to a critical value such that $\discinf=0$ in the critical class AB.
It differs from class AC in that the morphism from  $\aV^\bullet_{\infty}$ is on the boundary of (not within) the basin of attraction of 
$\aV^\halfominus_{-\infty}$. 
Consequently, a small deformation of the bare coupling in a large angular momentum channel below $\aV^\bullet_{\infty}$ immediately gives rise to a superconducting instability, unlike in class AC (see \fig{fig:main}).

\begin{figure}[ht]
\includegraphics[width=0.45\textwidth]{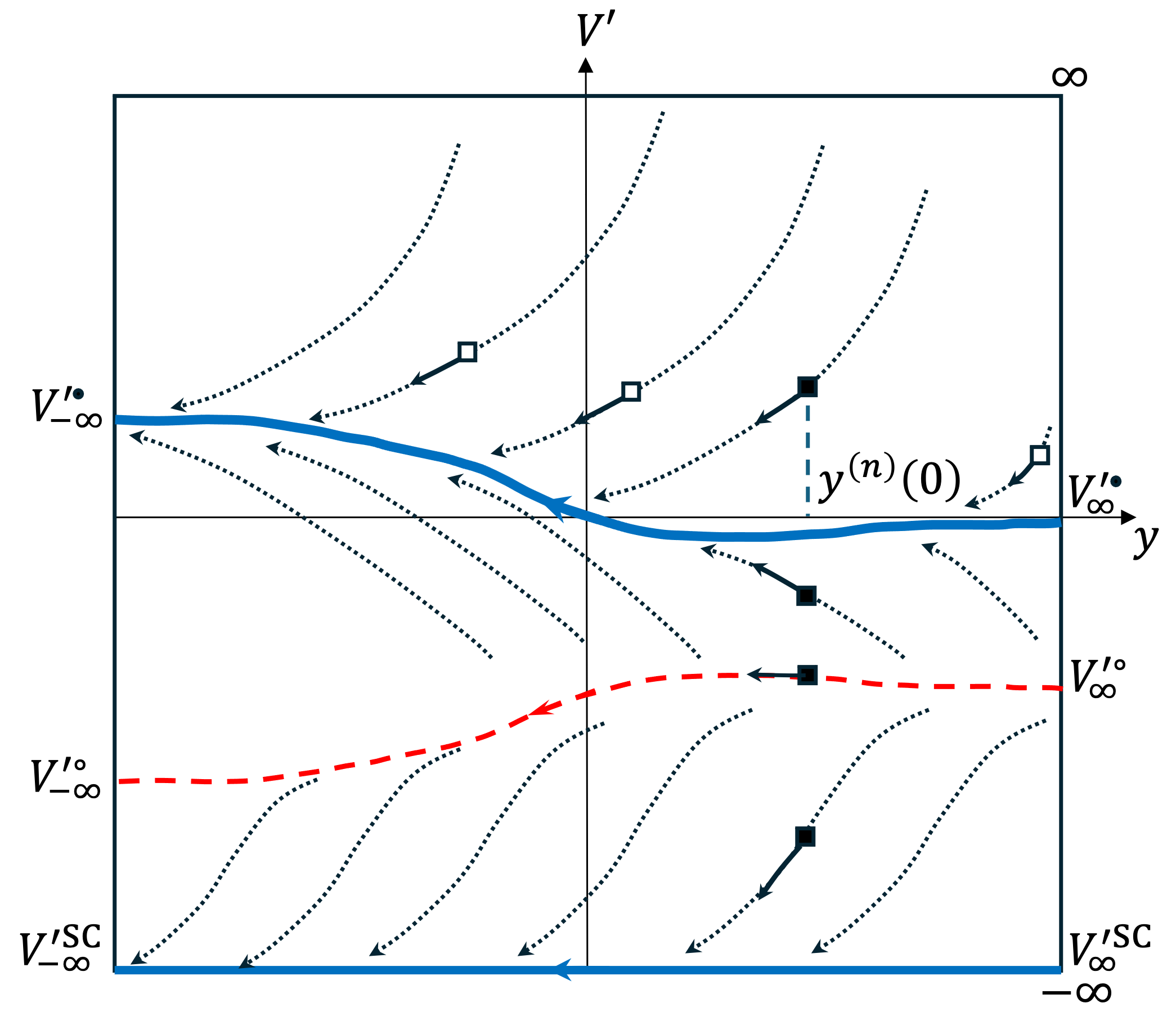}
\caption{
In class A, 
$\eta_{P,-\infty}>0$,
and
$\aV^\bullet_\infty$
and 
$\aV^\circ_\infty$
are connected to
$\aV^\bullet_{-\infty}$ and $\aV^\circ_{-\infty}$ 
through the metallic PFP (blue solid curve)
and the separatrix PFP (red dashed curve), respectively.
The metallic PFP is an attractor towards which the bare couplings above the separatrix PFP flow at low energies.
On the other hand, bare couplings below (on) the separatrix flow to $-\infty$ (stays on the separatrix).
}
\label{fig:portrait}
\end{figure}

Each superuniversality class generally includes multiple universality classes of non-Fermi liquids that can be reached by tuning the bare coupling.
In superuniversality class A, for example, a stable non-Fermi liquid is realized if the bare couplings are strictly above the separatrix PFP,
$ 
\aV^\circ_{-\infty}
\leftarrow 
\aV^\circ_{\infty}
$.
If the bare coupling 
in one or more angular momentum channels
is tuned to the separatrix, a critical non-Fermi liquid is realized, which exhibits a universal pairing interaction different from the stable non-Fermi liquid.
If the bare coupling is below the separatrix in any angular momentum channel, the metal becomes superconducting at sufficiently low energies.
This is illustrated in \fig{fig:portrait}.
In general, the separatrix can lie above the $\aV=0$ line for some $y$.
In that case, the theory with bare coupling $\aV=0$ becomes unstable against superconductivity,
although the instability is suppressed for a sufficiently repulsive bare coupling above the separatrix.

Predicting the superuniversality class to which a specific non-Fermi liquid belongs requires further dynamical input.
Near the upper critical dimension, in which the perturbative calculation is controlled,
the U(1) gauge theory and the Ising-nematic critical metal are predicted to be in class A\cite{kukreja2026classificationnonfermiliquidsuniversal}.
As the space dimension is lowered towards $d=2$, the perturbative calculation suggests that the former theory remains in class A, while the latter undergoes a transition to class C through AC.
However, non-perturbative corrections may change those designations in two space dimensions.

\section{Universal phase diagram}
\label{sec:predictions}

How do distinct superuniversality classes manifest themselves in experiments?
Before we answer this question, we first note that the superuniversality class does not classify individual universality classes.
Instead, it classifies the collection of universality classes
that can be accessed by tuning the bare four-fermion coupling.
It is a higher-level classification of metallic universality classes that groups individual universality classes based on the topology of the bundles of universal RG trajectories associated with a collection of bare four-fermion couplings.

Two non-Fermi liquids that realize the same ground state, in the sense that they are smoothly connected, may belong to different superuniversality classes.
For example, a non-Fermi liquid in class A and a non-Fermi liquid in class B may become superconductors with the same pairing symmetry for particular choices of bare coupling. 
What differentiates them is their responses to tuning the bare coupling.
In class A, $T_c$ vanishes as the bare coupling approaches a critical coupling, obeying a universal scaling form\cite{kukreja2026classificationnonfermiliquidsuniversal}.
In class B, superconducting instabilities cannot be suppressed, and $T_c$ remains non-zero, irrespective of the bare coupling.
The pairing symmetry may change as a function of the bare coupling, but a superconducting instability is unavoidable in angular momentum channels above a non-zero critical angular momentum $n_s$.
This difference arises because the bundle of PFPs exhibits distinct topologies in different classes.
Class C is similar to class B in that $T_c$ is always non-zero. 
In class C, however, there is no critical angular momentum because all channels, including the s-wave channel, are subject to superconducting instabilities, irrespective of the bare coupling.

One `super order parameter' that differentiates the superuniversality class is the phase diagram in the space of temperature and bare coupling. 
Although the full phase diagram is sensitive to the microscopic details,
the low-temperature part of the phase diagram, far below the UV cutoff energy scale, captures the universal crossovers.
In the following, we discuss the universal phase diagram expected for each superuniversality class.
We use the critical frequency at which the four-point function exhibits a singularity at zero temperature as a proxy for the superconducting transition temperature $T_c$.
The categorical data completely fixes the topology of the RG flow portrait in the plane of $(y, \aV)$, as shown in \fig{fig:main}.
The topology of the phase portrait, in turn, determines how 
$T_c$ varies as a function of the bare coupling.

To begin our discussion, consider a theory with a set of bare couplings 
$\{\av_n(0) | n \in \mathbb{Z} \}$. 
This defines a set of initial points 
$\left\{ \left(y^{(n)}(0), \av_n(0) \right) | n \in \mathbb{Z} \right\} $ in the $(y,\aV)$ plane.
Under the RG flow, each point flows, tracking a  PFP that obeys \eq{eq:general_fp_eq_log}.
In effect, one has a collection of PFPs that describe the flow of couplings in various angular momentum channels, embedded in one of the phase portraits in \fig{fig:main}. 
If the trajectory for the angular momentum channel $n$ diverges to $-\infty$ at a finite $y$, denoted as $y_{n,-\infty}^*$,
we define $T_{c,n} \equiv \Lambda e^{-2z(y^{(n)}(0) - y_{n,-\infty}^*)}$. 
Physically, $T_{c,n}$ represents the superconducting temperature that would be realized in angular momentum channel $n$ if other channels do not become superconducting above that temperature. 
Although one can only access the actual  superconducting transition temperature $T_c = \max_{n} T_{c,n}$ in experiments,
it is still useful to first understand what information $T_{c,n}$ contains regarding the superuniversality class.
While $T_{c,n}$ generally depends on the individual universality class, the categorical data in Table \ref{tab:classification} leave the following distinctive fingerprints that  depend only on the superuniversality class.
\begin{itemize}
\item $T_{c,n}$ in the highly repulsive bare coupling limit:

One important criterion for classifying non-Fermi liquids is whether superconducting instabilities can be suppressed by tuning the bare coupling sufficiently repulsive.
This information is contained in $T_{c,n}$ in the obvious way.
If the theory has any channel for which $T_{c,n} >0$ in the large repulsive bare coupling limit, it is bound to become a superconductor at low energies, irrespective of the bare coupling.
This follows from the fact that an RG trajectory associated with a finite bare coupling must stay below that of the infinite bare coupling at all scales.

\item Competition between incoherence and pairing glue in the s-wave limit ($\eta_{P,-\infty}$):

One categorical data that differentiates superuniversality classes is the discriminant in the small $y$ asymptotic region.
This information is encoded in how $T_{c,n}$ depends on the bare coupling in the small $y^{(n)}(0) \ll 0$ limit.
In this limit, $T_{c,n}$ is fixed by $C_{-\infty}=(\kappa_{-\infty}, h_{-\infty}, s_{-\infty})$ as
\begin{widetext}
\bqa
\frac{T_{c,n}}{\Lambda} \approx \begin{cases}
\Theta\Big(\av_{c,n}^\star - \av_n(0) \Big)\Bigg[\frac{2\kappa_{-\infty}\big(\av_{c,n}^\star - \av_n(0)\big)}{\sqrt{\eta_{P,\infty}}-2\kappa_{-\infty}\big(\av_{c,n}^\star - \av_n(0)\big)} \Bigg]^{\frac{2z}{\sqrt{\eta_{P,-\infty}}}} \quad &\text{for } \eta_{P,-\infty} > 0, \\[0.5ex]
\Theta\Big(\av_{c,n}^\star - \av_n(0) \Big)\exp\Big\{\frac{-z}{\kappa_{-\infty}\big(\av_{c,n}^\star- \av_n(0)\big)}\Big\} &\text{for } \eta_{P,-\infty} = 0,\\[0.5ex]
\exp\Big(\frac{-4z}{\sqrt{-\eta_{P,-\infty} }} \Big\{\frac{\pi}{2} + \arctan\Big[\frac{4\kappa_{-\infty}}{\sqrt{-\eta_{P,-\infty}}}\Big(\av_n(0) + \frac{h_{-\infty}}{\kappa_{-\infty}}\Big)  \Big]  \Big\}\Big) &\text{for } \eta_{P,-\infty} < 0.
\end{cases}
\label{eq:main_Tcn_smalln}
\eqa
\end{widetext}
Here, $\av_{c,n}^\star \equiv  \aV^S_{y^{(n)}(0)}$ is the critical coupling below which $T_{c,n}$ turns on for $\eta_{P,-\infty} \geq 0$.
\eq{eq:main_Tcn_smalln} exhibits qualitatively different behaviors depending on  
$\eta_{P,-\infty}$.
For $\eta_{P,-\infty}<0$,
where the universal pairing glue dominates over incoherence,
$T_{c,n}$ is non-zero, irrespective of the bare coupling.
For $\eta_{P,-\infty}\geq0$, 
where the pairing glue is subdominant ($>$) to or balanced with ($=$) incoherence,
$T_{c,n}$ becomes non-zero only below the critical bare coupling. 
With $y^{(n)}(0) \ll 0$,
the critical coupling is close to
$\aV^\circ_{-\infty}$
and
$\aV^\halfominus_{-\infty}$
for
$\eta_{P,-\infty}>0$ 
and
$\eta_{P,-\infty}=0$, respectively.
$T_{c,n}$ is algebraically (exponentially) small for small deformations below the critical coupling for 
$\eta_{P,-\infty}>0$ ($=0$);
a small attractive deformation added to 
$\aV^\circ_{-\infty}$
($\aV^\halfominus_{-\infty}$)
grows algebraically (logarithmically) in energy scale.

\begin{figure}[t]
    \centering
    \begin{subfigure}{0.325\textwidth}
        \centering
        \includegraphics[width=\linewidth]{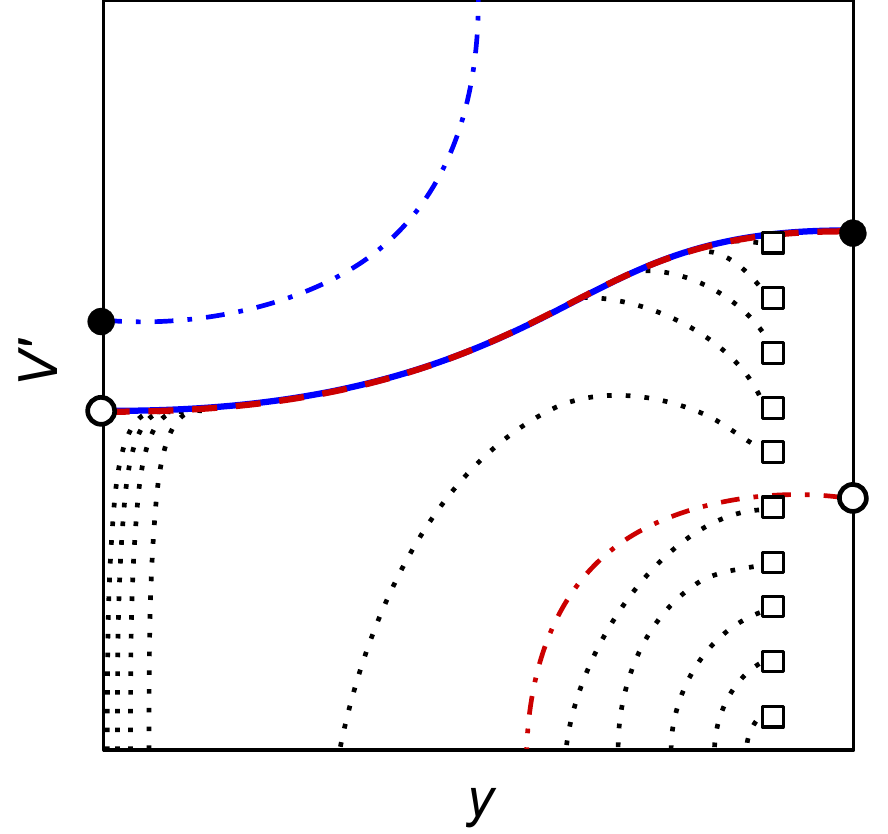}
        \caption{}
        \label{fig:TcnAB}
    \end{subfigure}
    \begin{subfigure}{0.325\textwidth}
        \centering
        \includegraphics[width=\linewidth]{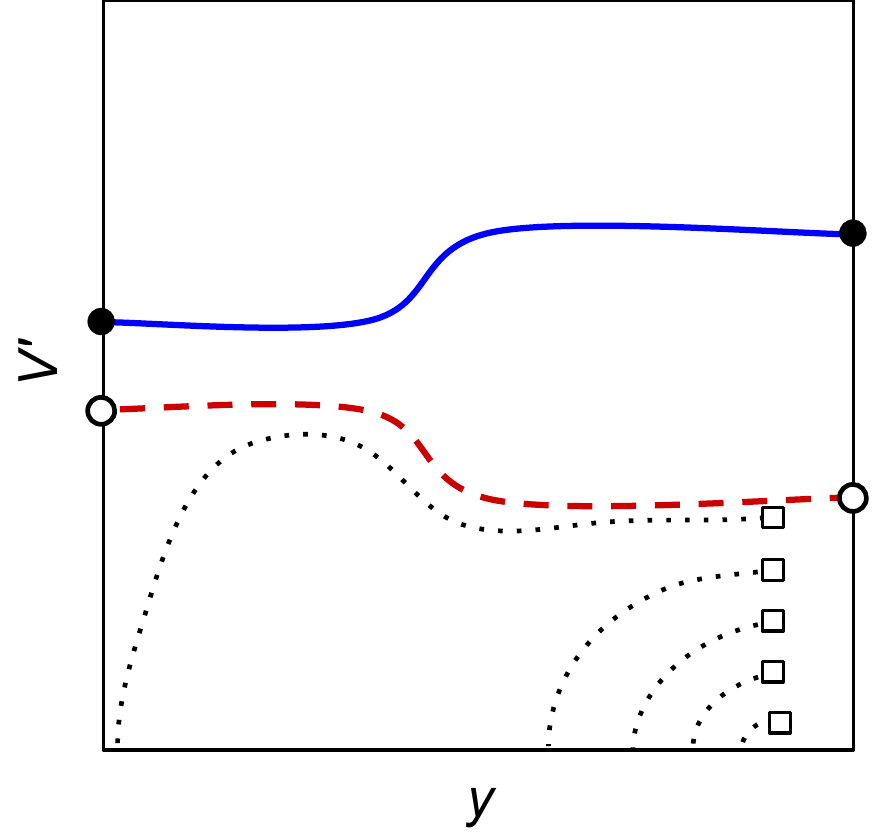}
        \caption{}
        \label{fig:porTcnA}
    \end{subfigure}
    \caption{
Illustration of how the phase portrait topology influences $T_{c,n}$ at large $n$ in classes (a) AB and  (b) A. 
The squares represent bare couplings at a large but fixed $n$. 
In class AB, the attractive nature of the separatrix at large $y$ ensures that bare couplings placed between $\aV^\bullet_\infty$ and $\aV^\circ_\infty$
first attract towards the stable asymptotic fixed point before diverging only at small $y$, leading to a suppression of $T_{c,n}$.
There is no such suppression for bare coupling smaller than $\aV^\circ_\infty$.
This gives rise to
a two-step structure in $T_{c,n}$ as a function of the bare coupling.
In class A, no two-step structure exists since the separatrix saturates to the unstable fixed point.
}
    \label{fig:AvsAB_exp}
\end{figure}

\item
$-\infty$-asymptotic fixed point `under'
$\aV^\bullet_\infty$:

Another feature that is crucial for differentiating the superunivesality classes is the $-\infty$ asymptotic fixed point to which $\aV^\bullet_\infty$ flows under the RG flow.
This information is naturally captured by $T_{c,n}$ in the large $n$ limit because a bare coupling chosen above $\aV^\circ_{\infty}$ first flows toward $\aV^\bullet_{\infty}$ at large $y^{(n)}(0)$ and its fate in the low-energy limit is determined by the PFP that emanates from $\aV^\bullet_{\infty}$.
Especially, $T_{c,n}$ at large $n$ develops a distinct behavior as a function of the bare coupling when $\aV^\bullet_\infty$ is connected to the separatrix PFP, as in
classes AB and ABC.
%
%
In those classes, there is a finite gap between the separatrix and $\aV^\circ_{\infty}$ in the large $y$ region. 
A coupling placed within the gap at large $y^{(n)}(0)$ first flows upward towards the separatrix because the separatrix is attractive in the large $y$ region.
The coupling keeps approaching the separatrix up to scale $\sim l_n^*$ before beginning to grow negative and eventually diverging to $-\infty$ at a length scale larger than $l_n^*$. 
This leads to the suppression of $T_{c,n}$  in the large $n$ limit as
\begin{widetext}
\bqa
\frac{T_{c,n}}{\Lambda} \sim \Theta\Big(\av_{c,n}^\star - \av_n(0) \Big) n^{-2z} \begin{cases}
    \bigg[1 + \mathcal{C} n^{\sqrt{\eta_{P,\infty}}}\bigg(\frac{\av_n(0)-\aV^\circ_{\infty}}{\av_{c,n}^\star-\av_n(0)}\bigg)\bigg]^{-\frac{2z}{\sqrt{\eta_{P,-\infty}}}}\quad &\text{for Class AB},\\[0.5ex]
   \exp{-\mathcal{D} n^{\sqrt{\eta_{P,\infty}}}\Big( \frac{\av_n(0)-\aV^\circ_{\infty}}{\av_{c,n}^\star-\av_n(0)}\Big)} &\text{for Class ABC},
\end{cases}
\label{eq:ABABC}
\eqa
\end{widetext}
where $\mathcal{C}$ and $\mathcal{D}$ are constants
(see Appendix \ref{sec:app_exp} for details).
Again, $\av_{c,n}^\star$ is the critical coupling set by the separatrix at $y^{(n)}(0)$, which is
close to $\aV^\bullet_{\infty}$
for $y^{(n)}(0) \gg 0$.
In class AB, $T_{c,n}$ is suppressed algebrically in $n$
because $l_n^*$ is linear in $y^{(n)}(0)$.
In class ABC, it is further suppressed by an exponential factor.
This stronger suppression originates from the fact that the coupling at a large angular momentum channel first flows close to $\aV^\halfominus_{-\infty}$ at length scale $\sim l_n^*$ before it beings to grow more attractive.
Since the separatrix is connected to the marginal asymptotic fixed point $\aV^\halfominus_{-\infty}$ in class ABC, the coupling grows only logarithmically in energy.
As a result, $T_{c,n}$ is exponentially suppressed in the renormalized coupling, which is already suppressed at large $n$.
Consequently, $T_{c,n}$ 
for the bare coupling in 
$\aV^\circ_{\infty} < \av_n(0) < \aV^\bullet_{\infty}$
is highly suppressed as $n$ increases for a fixed bare coupling chosen between the gap.
This suppression of $T_{c,n}$ disappears if the bare coupling is tuned below $\aV^\circ_{\infty}$ because the coupling placed below the unstable asymptotic fixed point immediately begins to grow more attractive with increasing length scale. 
This distinction is shown schematically in Figure \ref{fig:AvsAB_exp} for classes A and AB, but a similar illustration applies to classes AC and ABC. 
This causes a two-step structure in the profile of $T_{c,n}$ as a function of the bare coupling for large $n$, as is shown 
in \fig{fig:15b}.
In classes A and AC, on the other hand, the separatrix connects to $\aV_{\infty}^\circ$, and 
$T_{c,n}$ for the bare coupling below the critical coupling is not suppressed in the large $n$ limit.
Hence, $T_{c,n}$ exhibits a simple on-set behavior below the critical coupling as in  \fig{fig:exp_figure_A_AC}.

\end{itemize}


Using this information, one can diagnose the superuniversality class through $T_c$ measured as a function of the bare couplings.
%
%
First, consider a theory in which the bare couplings are strongly repulsive
$\av_n(0) = 
\av^{UV} > 0$ in all angular momentum channels.
If $T_c \neq 0$ in the large $\av^{UV}$ limit,
it is in class B, BC, or C. 
If $T_c = 0$ or approaches zero in the large $\av^{UV}$ limit, it is in class A, AB, AC, or ABC. 
We refer to the former (latter) as the unstable (stable) classes.
From here, a further differentiation between the classes can be made by tuning the coupling in one angular momentum channel to be more attractive. 

\begin{figure}[th]
    \centering
    \begin{subfigure}{0.325\textwidth}
        \centering
        \includegraphics[width=\linewidth]{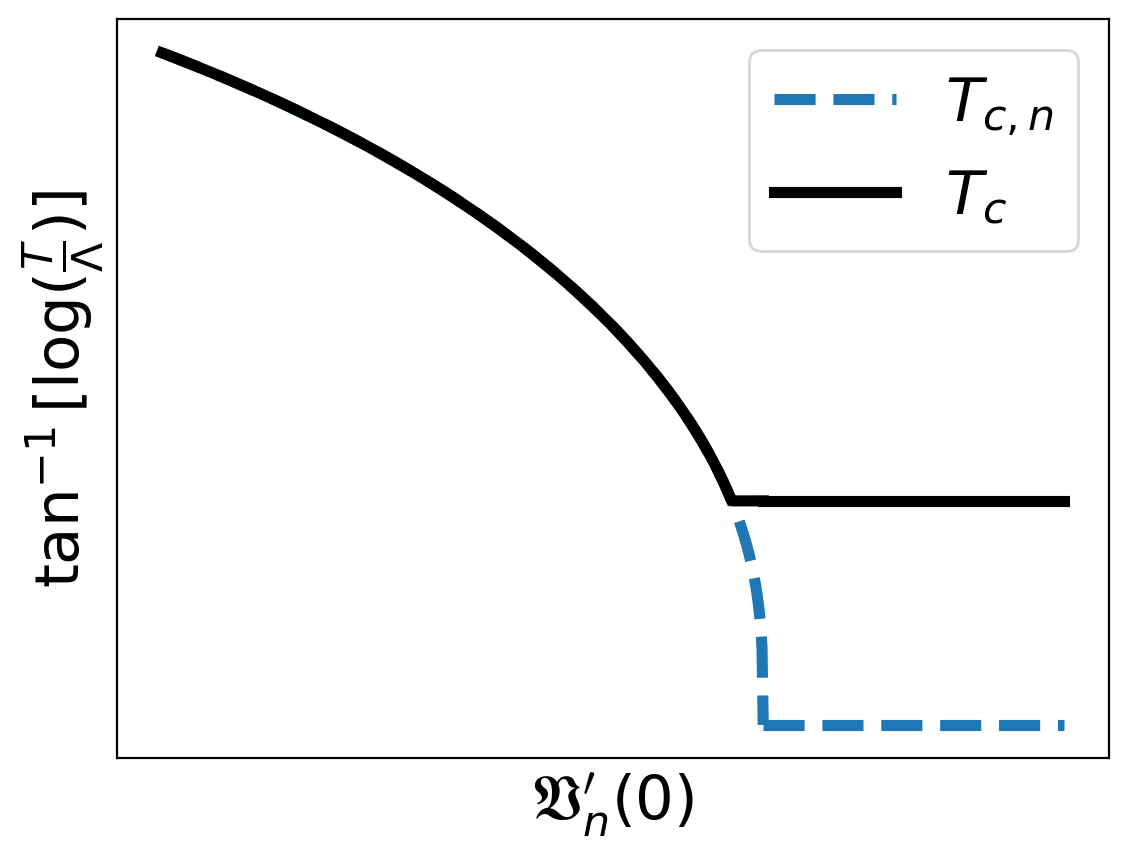}
        \caption{}
    \end{subfigure}
    
    \begin{subfigure}{0.325\textwidth}
        \centering
        \includegraphics[width=\linewidth]{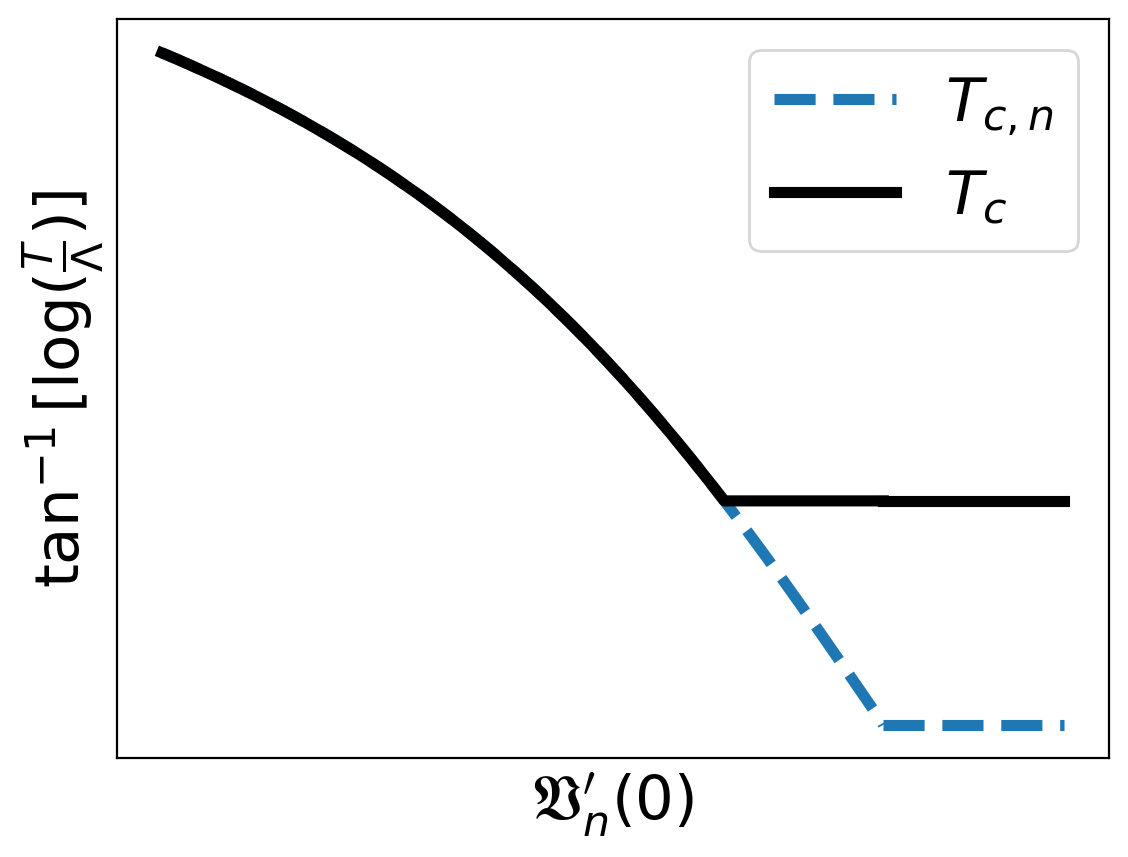}
        \caption{}
    \end{subfigure}
    
    \begin{subfigure}{0.325\textwidth}
        \centering
        \includegraphics[width=\linewidth]{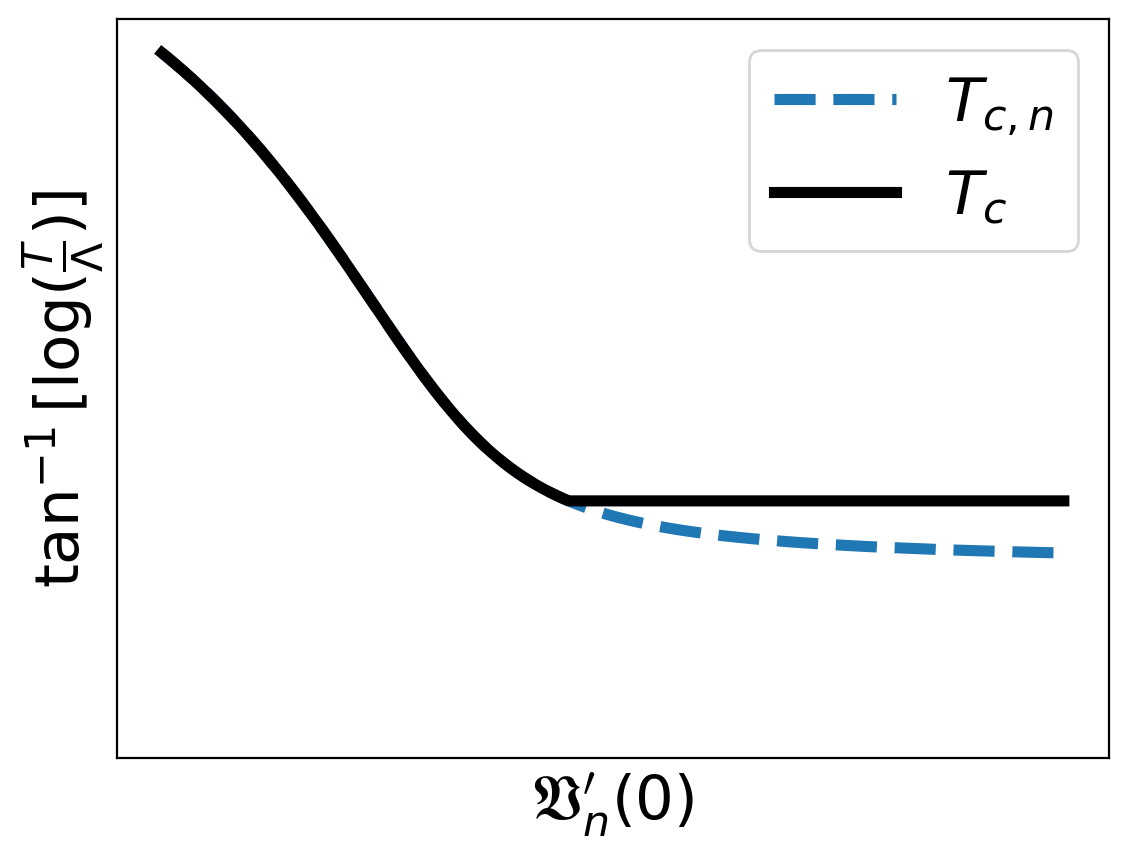}
        \caption{}
    \end{subfigure}
\caption{
$T_{c,n}$ and $T_c$ vs $\av_n(0)$ at small $y^{(n)}(0) \ll 0$ for classes (a) B, (b) BC, and (c) C. 
When the bare couplings are highly repulsive in all angular momentum channels, superconductivity occurs in angular momentum channel $n_O$ at $T_c$ denoted by the solid plateau.
As $\av_n(0)$ with $n \neq n_O$ and $y^{(n)}(0) \ll 0$ is lowered,
the pairing channel jumps from $n_O$ to $n$ across the critical coupling.
The way $T_c$ rises beyond the critical coupling depends on the asymptotic discriminant as shown in  \eq{eq:main_Tcn_smalln}. 
}
\label{fig:exp_sc_classes}
\end{figure}

We first discuss the unstable classes.
Even in the limit that the bare couplings are highly repulsive in all angular momentum channels, non-Fermi liquids in these classes become superconductors at low temperatures.
In this limit, those in classes B and BC must be a non-s-wave superconductor because $\eta_{P.-\infty}\geq 0$.
On the other hand, class C does not prevent the s-wave superconductivity.\footnote{Actually, it is most likely to be the s-wave superconductor in class C\cite{kukreja2026classificationnonfermiliquidsuniversal}}
To further differentiate class B and BC, 
suppose that the metal becomes superconducting in channel $n_O$ in the large $\av^{UV}$ limit.
If one tunes the bare coupling in a new channel $n$ to be attractive enough that 
$T_{c,n} > T_{c,n_O}$, 
the pairing wavefunction changes from $n_O$ to $n$ across the critical coupling, $\av^\star_{SC,n}$.
Since classes B, BC, and C are distinguished uniquely by $\eta_{P,-\infty}$,
we can select $n$ with $y^{(n)}(0) \ll 0$ to differentiate them. 
Below the critical coupling, the dependence of $T_c$ on $\av_n(0)$ is given by
\eq{eq:main_Tcn_smalln}.
Each class can be distinguished by how $T_c$ depends on the bare coupling below the critical coupling. 
The difference becomes more sharply visible when the baseline $T_{c,n_O}$ is small, revealing the profile of $T_{c,n}$ down to the lower temperature.
This is illustrated in Fig. \ref{fig:exp_sc_classes}.

\begin{figure}[th]
    \centering
    \begin{subfigure}{0.325\textwidth}
        \centering
        \includegraphics[width=\linewidth]{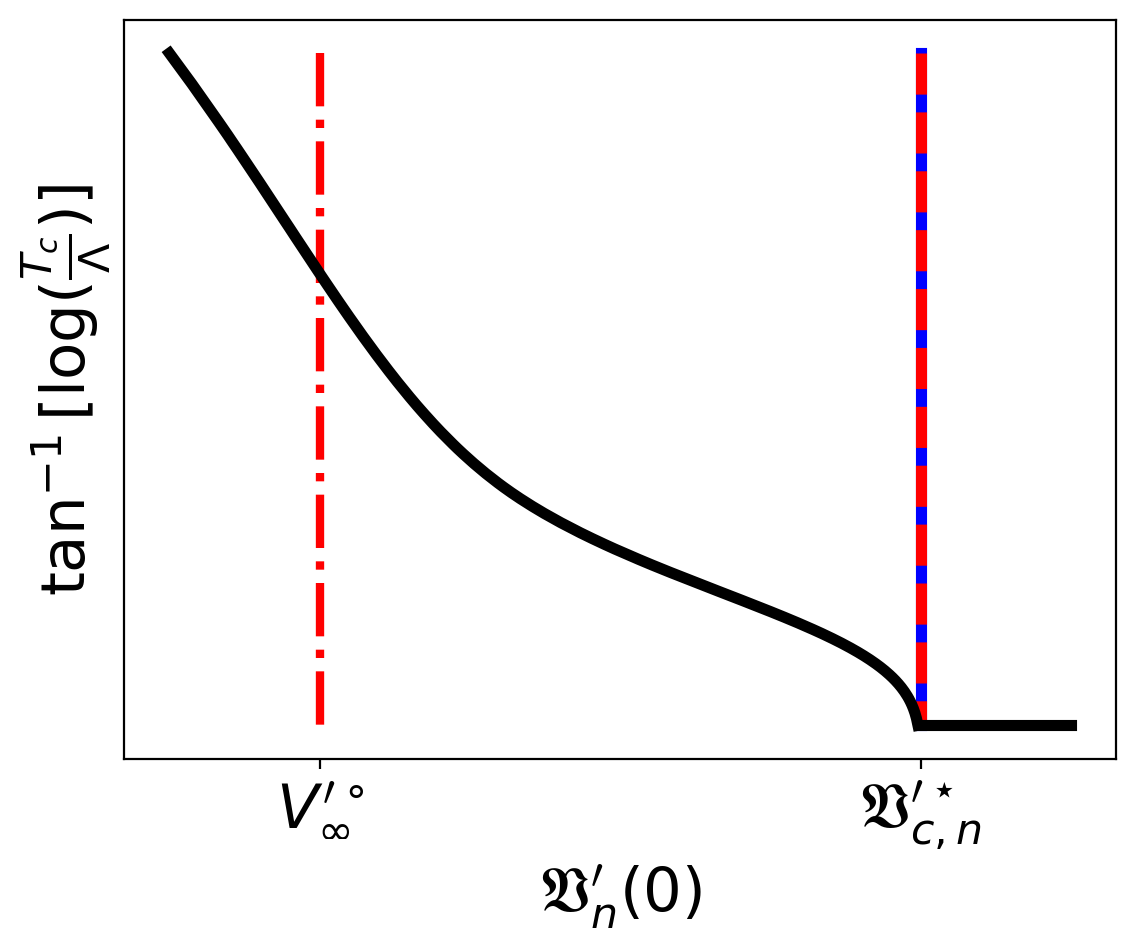}
        \caption{}
        \label{fig:15a}
    \end{subfigure}
    \begin{subfigure}{0.325\textwidth}
        \centering
        \includegraphics[width=\linewidth]{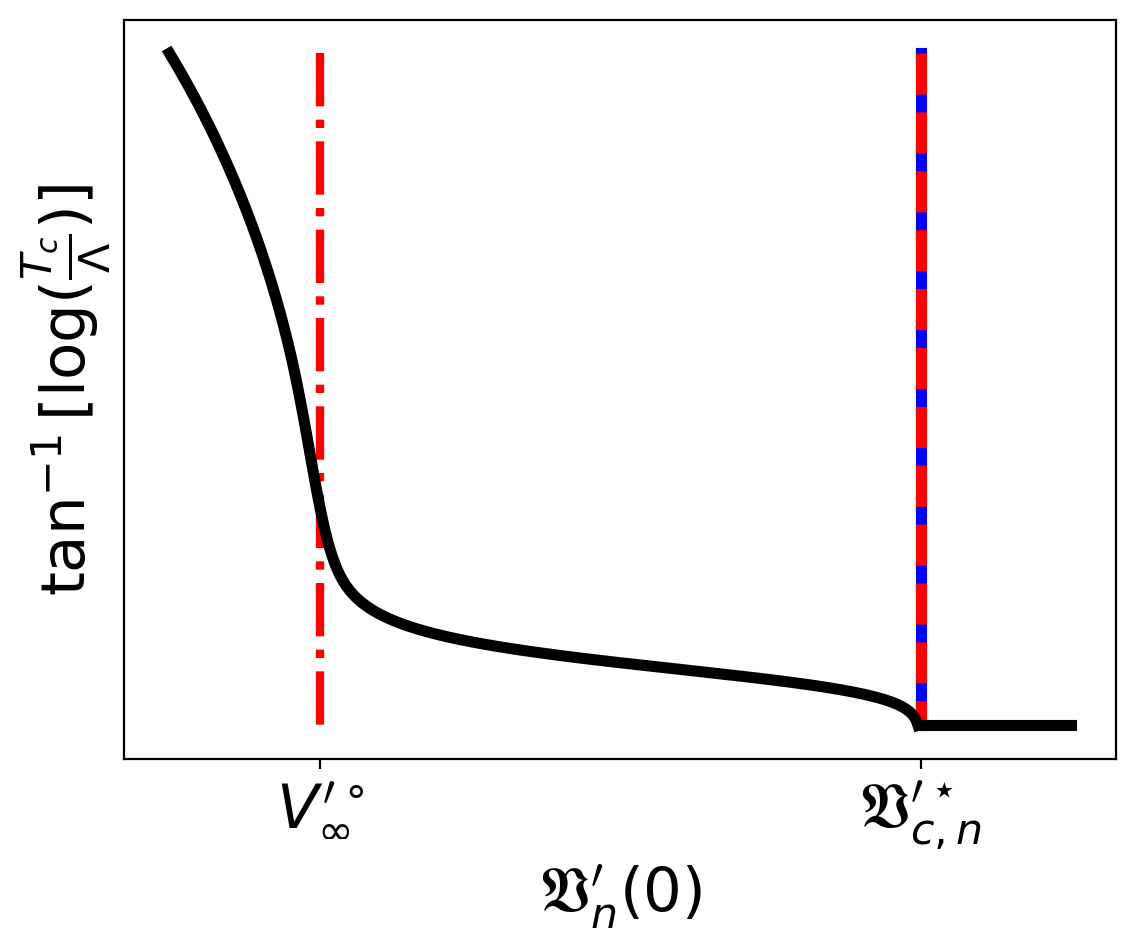}
        \caption{}
        \label{fig:15b}
    \end{subfigure}
\caption{
$T_c$ versus bare coupling $\av_n(0)$ at large angular momenta with $y^{(n)}(0)\gg 0$
in class AB, with an otherwise repulsive UV theory. 
$T_c$ turns on below  
the critical coupling
$\av_{c,n}^* = 
\aV^S_{y^{(n)}(0)}$ but remains small until the bare coupling is lowered below  
$\aV^\circ_\infty$
due to the attractive nature of the separatrix,
as explained in \fig{fig:TcnAB}.
As $n$ increases from (a) to (b), the two-step structure becomes more pronounced as the first step extends towards $\aV^\circ_\infty$ and goes down.
}
\label{fig:exp_figure_AB} 
\end{figure}

\begin{figure}[th]
    \centering
    \begin{subfigure}{0.325\textwidth}
        \centering
        \includegraphics[width=\linewidth]{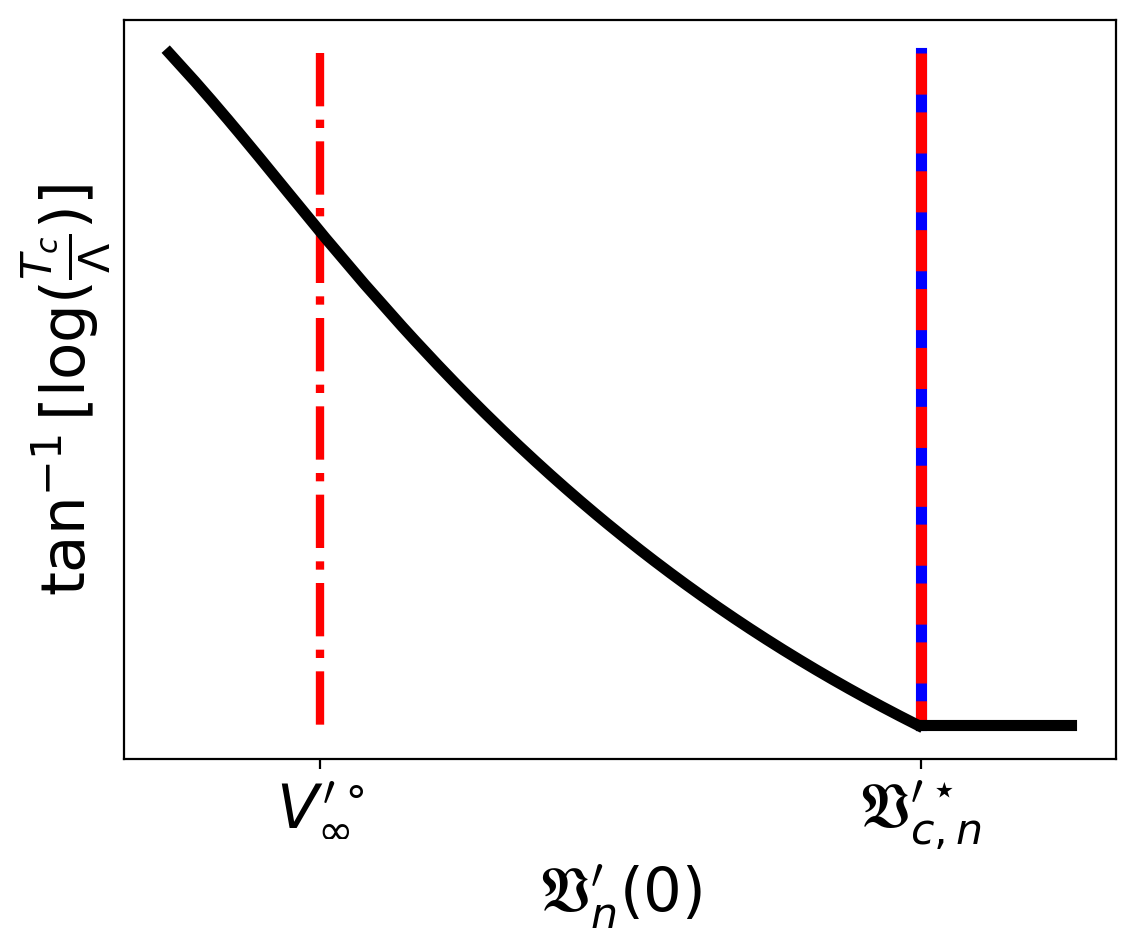}
        \caption{}
    \end{subfigure}
    \begin{subfigure}{0.325\textwidth}
        \centering
        \includegraphics[width=\linewidth]{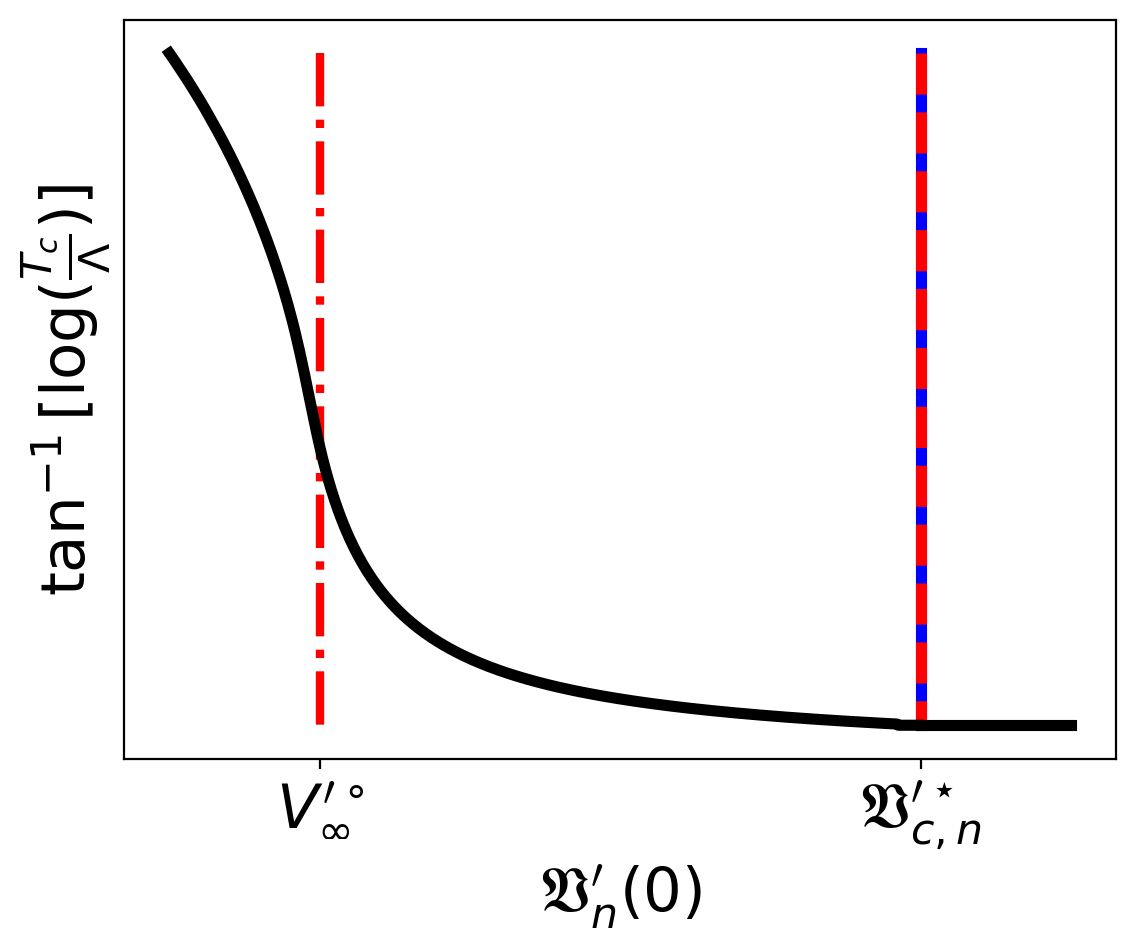}
        \caption{}
    \end{subfigure}
\caption{$T_c$ versus $\av_n(0)$ at large angular momenta with $y^{(n)}(0)\gg 0$
in class ABC, with an otherwise repulsive UV theory. 
The two-step structure, which becomes clearer as $n$ increases from (a) to (b), is the same as class AB.
The difference from class AB is the slower onset of $T_c$ below the critical coupling,
which is caused by
the marginal nature of the $-\infty$ asymptotic fixed point.
}
\label{fig:exp_figure_ABC}
\end{figure}

To differentiate the stable classes, we again start with a highly repulsive bare coupling and
tune the bare coupling in a large angular momentum channel with $y^{(n)}(0) \gg 0$.
As the bare coupling is tuned below the critical value set by the separatrix PFP, $T_c$ turns on.
As discussed above, the behavior of $T_c$ as a function of the bare coupling is controlled by the $-\infty$ asymptotic fixed point to which the PFP from $\aV^\bullet_\infty$ flows.
In classes AB and ABC, the PFP flows to the unstable and marginal asymptotic fixed points, respectively.
Consequently, the two-step structure appears in the profile of $T_c$ as a function of the bare coupling.
\footnote{
This structure persists at all $n$ that is large enough.}
This is shown in Figures \ref{fig:exp_figure_AB} and \ref{fig:exp_figure_ABC}.
Classes AB and ABC are further differentiated from each other by the strength of the large $n$ suppression of $T_c$ within the first step,
as shown in
\eq{eq:ABABC}.
In class AB (ABC)
$T_c$ is algebraicaally (exponentially) suppressed in $n$ within the first step.

\begin{figure}[ht]
    \centering
    \begin{subfigure}{0.325\textwidth}
        \centering
        \includegraphics[width=\linewidth]{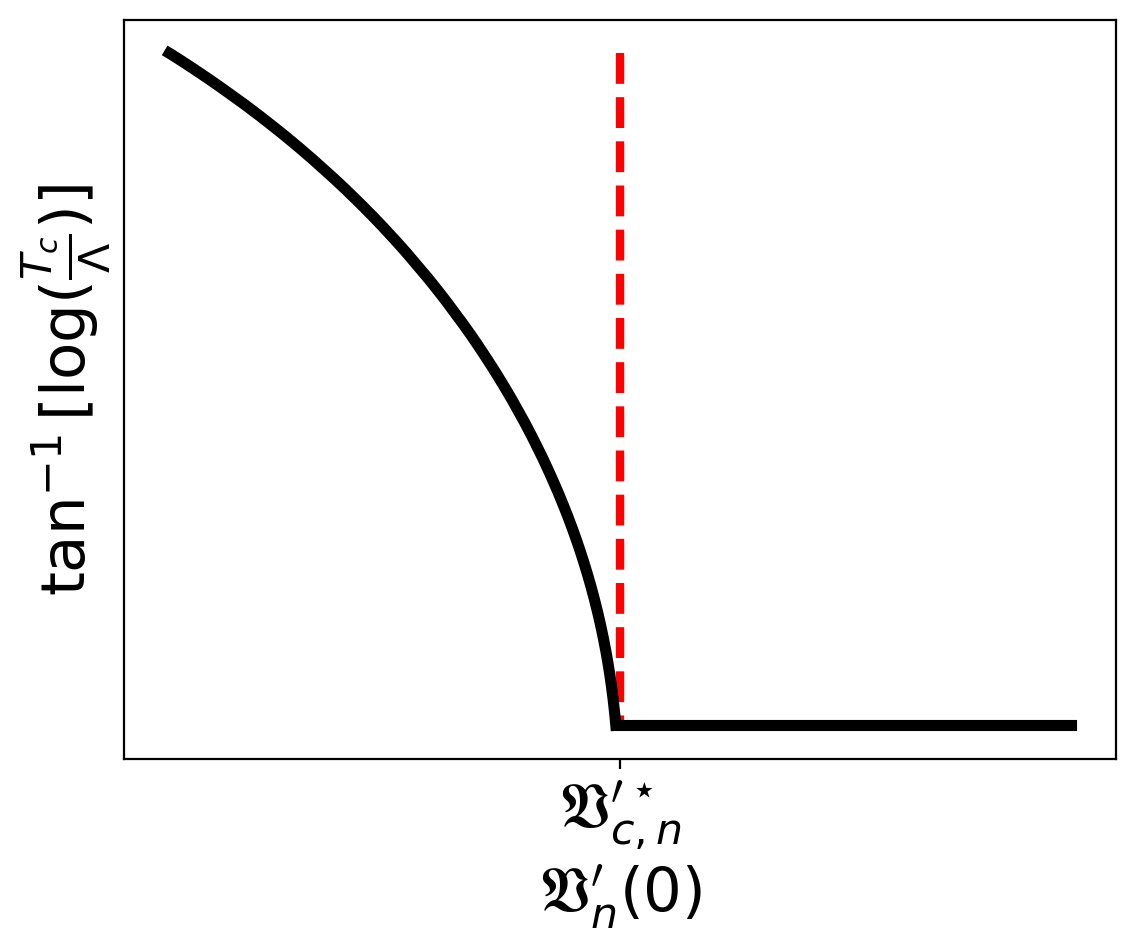}
        \caption{}
        \label{fig:exp_figure_A_AC_a}
    \end{subfigure}
    \begin{subfigure}{0.325\textwidth}
        \centering
        \includegraphics[width=\linewidth]{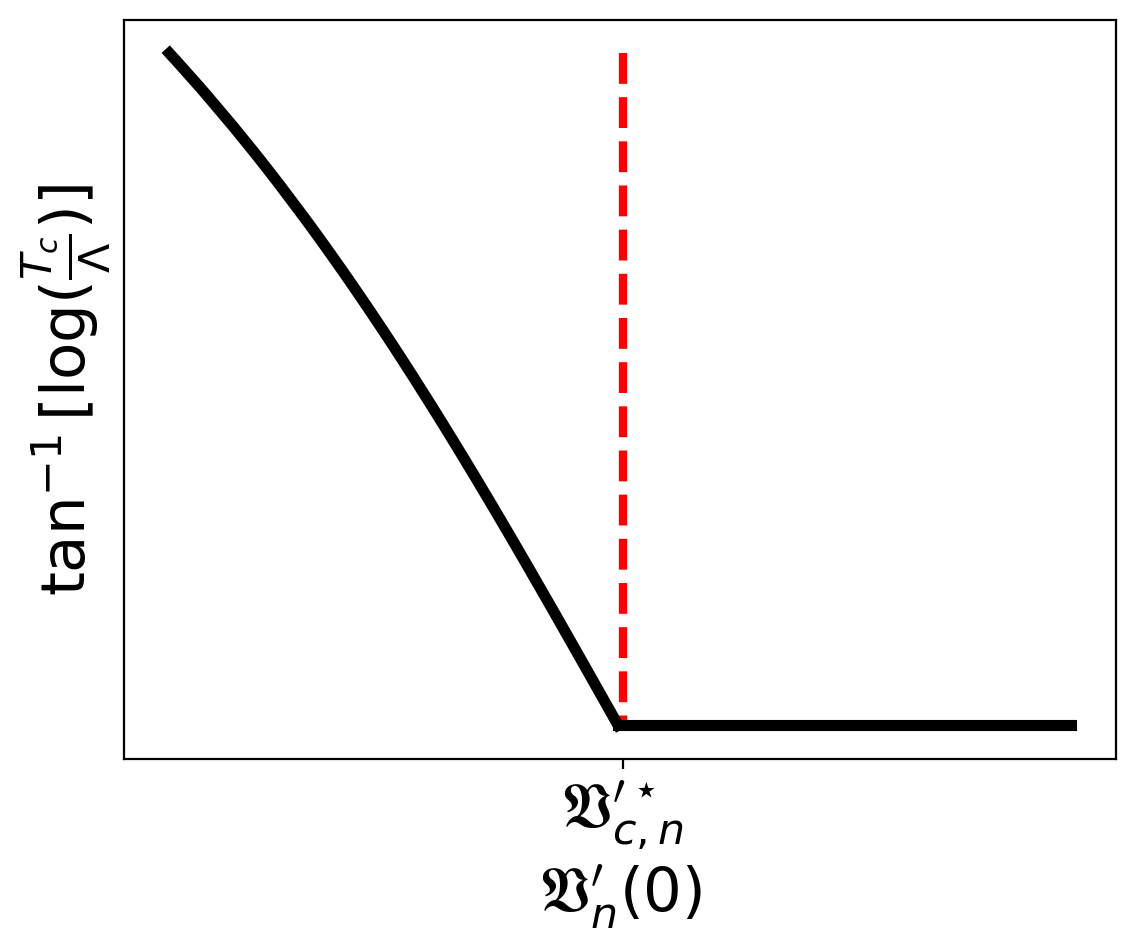}
        \caption{}
        \label{fig:exp_figure_A_AC_b}
    \end{subfigure}
\caption{
$T_c$ versus bare coupling for small $n$ with $y^{(n)}(0) \ll 0$ in classes (a) A and (b) AC. 
The slower onset of $T_c$ in class AC is due to the marginality of the $-\infty$ asymptotic fixed point.
}
\label{fig:exp_figure_A_AC}
\end{figure}

In classes A and AC, the 
two-step structure is absent at large $y^{(n)}(0)$, 
as illustrated in
\fig{fig:exp_figure_A_AC}.\footnote{
For large $n$, the profile is insensitive to $\eta_{P,-\infty}$ unless $\av_n(0)$ is very close to the separatrix within $\sim e^{-\sqrt{\eta_{P,\infty}}y^{(n)}(0)}$ as discussed in Appendix \ref{sec:app_exp}.} 
Classes A and AC can be further differentiated from each other by the onset behavior of $T_c$ when the bare coupling is tuned below the critical coupling in an angular momentum channel with $y^{(n)}(0) \ll 0$, as shown in \eq{eq:main_Tcn_smalln}
(See Fig. \ref{fig:exp_figure_A_AC}).
The universal features of the phase diagram in each superuniversality class are summarized in
Tables \ref{tab:stableclasses}
and \ref{tab:unstableclasses}.

Besides the low-temperature phase diagram, there are physical observables that are unique to each superuniversality class.
They include the angular profile of the universal pairing interaction that emerges in the normal state and the universal relation between $T_c$ and $\KFAVdim$ for some superconducting non-Fermi liquids \cite{kukreja2026classificationnonfermiliquidsuniversal}.
We don't repeat them here because the same predictions hold even for strongly interacting non-Fermi liquids.
Those predictions 
only depend on the topological aspects of the phase portraits of PFPs although they were first discovered in the context of perturbative non-Fermi liquids.

\begin{table*}[t]
    \centering
    \begin{tabular}{|c|c|c|c|}\hline 
       ~Class~  & 
       ~$T_c$ in the highly ~&~ 
       Onset of $T_c$ below critical coupling ~&~ 
       Suppression of $T_c$ below critical coupling \\
       ~&~ repulsive limit ~&~
       in the s-wave limit
       ~&~ in high angular momentum channels \\\hline 
       A & \multirow{4}{*}{$0$}  &  \multirow{2}{*}{
       algebraic 
       }& none \\ 
       \cline{1-1}
       \cline{4-4}
       AB &  &  & algebraic \\\cline{1-1}\cline{3-4}
       AC &   & \multirow{2}{*}{
       essential singularity 
       } & none \\ \cline{1-1}\cline{4-4}
       ABC &  &  & exponential \\ \hline
    \end{tabular}
  \caption{
Universal features of the phase diagrams in the stable superuniversality classes.
}
    \label{tab:stableclasses}
\end{table*}
\begin{table*}[t]
    \centering
    \begin{tabular}{|c|c|c|}\hline 
       ~Class~  & 
       ~ Unstable channel in the highly ~&~ 
       Growth of $T_c$  below critical coupling ~ 
        \\
       ~&~ repulsive limit ~&~
       in the s-wave limit ~
        \\\hline 
       B & 
       \multirow{2}{*}{
       must be non-s-wave
       }
       &
        algebraic  \\ 
       \cline{1-1}
       \cline{3-3}
       BC & 
       & 
       essential singularity
        \\ \hline
       C &  can be s-wave & 
       smooth 
        \\\hline
    \end{tabular}
  \caption{
Universal features of the phase diagrams in the unstable superuniversality classes. 
}
    \label{tab:unstableclasses}
\end{table*}


\section{Summary and discussion}
\label{sec:summary}

Metals do not exhibit scale invariance due to the Fermi momentum scale.
Therefore, one needs to generalize the notion of RG fixed points to universal RG trajectories (called projective fixed points) to classify metallic universality classes.
In this paper, $2+1$-dimensional non-Fermi liquids are classified into seven superuniversality classes based on the topology of the bundle of projective fixed points associated with the tunable four-fermion couplings in the pairing channel. 
The earlier classification obtained from the perturbative RG flow equation remains valid even for strongly coupled non-Fermi liquids in $2+1$ dimensions in the large $\KFAVdim$ limit with fixed dimensionless couplings, thanks to the non-perturbative kinematic constraints.
Each superuniversality class exhibits a unique low-temperature phase diagram in the space of temperature and the bare four-fermion coupling. 
The predicted features of the phase diagram are superuniversal in the sense that they are shared by all individual non-Fermi liquids within a superuniversality class.
We close with a few open questions.\\

\begin{itemize}

\item Tunability of the four-fermion coupling:

In this paper, we provide a way to differentiate distinct superuniversality classes, assuming that the bare pairing interaction in each angular momentum channel is independently tunable.
While the bare couplings defined at a non-zero energy scale are tunable to a certain extent, in real experiments, 
one does not have infinitely many knobs to tune all couplings independently.
For example, one may tune only one or two parameters by changing the density and velocity of electrons in a layer that is used to screen the Coulomb repulsion
\cite{Arora2020,Park2021,Hao2021,Zhou2021,Yankowitz2019,Lu2019,Saito2020,Stepanov2020,Ye2012,Costanzo2016}.
It will be of interest to understand the exact differentiability of superuniversality classes achievable with such a limited tunability of bare couplings.

\item 
Emergent symmetry and further
experimental probes:

Each non-Fermi liquid universality class exhibits a distinct emergent symmetry at low energies.
As discussed above, the stable classes (A, AB, AC, ABC) host non-Fermi liquids that are stable down to the zero temperature limit, while non-Fermi liquids in the unstable classes (B, C, BC) become superconductors at low temperatures.
The symmetry that emerges at low temperatures, but above $T_c$ if $T_c$ is non-zero, depends not only on the superuniversality class but also on the individual universality class.
Some non-Fermi liquids exhibit the LU(1) group as the emergent symmetry\cite{PhysRevX.11.021005} while some exhibit its subgroup, such as the OLU(1).
Under the full LU(1), the fermion can be transformed as $\psi(\theta) \rightarrow e^{i \zeta(\theta)} \psi(\theta)$ for general smooth functions 
$\zeta(\theta)$.
The OLU(1) is a subgroup in which the angle-dependent phase is restricted to 
$\zeta(\theta+\pi)+
\zeta(\theta)=0$.
The LU(1) is broken down to OLU(1) if the universal pairing interaction supports large-angle scatterings,
which prevents the particle number from being conserved independently at each patch but keeps the odd angular momentum components. 
We refer the readers to Ref. \cite{kukreja2026classificationnonfermiliquidsuniversal} for the details.
It is desirable to have experimental probes that are sensitive to the emergent symmetry. 
One potential observable is the angle-resolved dynamical pairing susceptibility, which measures the response of Cooper pairs at one angle to a pairing field applied at another angle.

\item Ultraviolet/Infrared mixings:

In metals, there can be, in principle, two distinct types of ultraviolet/infrared (UV/IR) mixing\cite{BORGES2023169221}.
The first is the mixing between the large-momentum processes and the low-energy physics.
The second is the mixing between the high-energy processes and the low-energy physics.
We call the former horizontal UV/IR mixing and the latter vertical UV/IR mixing, regarding the horizontal and vertical directions as the momentum and energy axes, respectively.
It is important to distinguish the two because, in metals, large momenta do not necessarily mean high energies.
It is perfectly natural to have horizontal UV/IR mixing because low-energy fermions carry momenta comparable to $\KFAVdim$\cite{IPSITA} and can exchange large momenta through pairing interactions\cite{SHANKAR, PhysRevD.59.094019, PhysRevB.91.115111, BORGES2023169221}. 
However, the vertical UV/IR mixing is a fundamental departure from the framework of low-energy effective theories.
In this paper, we focused on non-Fermi liquids without vertical UV/IR mixing, but allowing horizontal UV/IR mixing. 
Although most non-Fermi liquids seem to be free from vertical UV/IR mixing as expected, it is still unclear whether and how the marginal Fermi liquids avoid it\cite{
VARMALI,
HALPERINHALF}.
In the marginal Fermi liquids, there is an indication that large-angle scatterings mediated by high-energy bosons may contribute to the anomalous dimension of the critical boson at `strong' bare coupling\cite{PhysRevLett.128.106402}.
It remains an open problem to understand whether non-perturbative effects remove or positively confirm the vertical UV/IR mixing.

\item 
Physical realizations:

Needless to say, it is of great interest to find physical examples that realize each superuniversality class.
It seems likely that the $C_4$ symmetric Ising-nematic quantum critical metal is in class C, while the U(1) gauge theory at finite density is in class A in $2+1$ dimensions\cite{kukreja2026classificationnonfermiliquidsuniversal}.
However, there is no candidate for the superuniversality class B.
It will be interesting to explore a wider class of non-Fermi liquids, including those that host critical fluctuations beyond conventional order parameters\cite{bp6l-46z7,Li:2021aa,PhysRevLett.102.186401,2026arXiv260827453D}.

\item
Type III non-Fermi liquids:

The classification and the physical predictions discussed in this paper are applicable to type II non-Fermi liquids
in which the fully renormalized two-particle irreducible self-energy is comparable with the fully renormalized disconnected four-point function.
While there is no known example of non-Fermi liquids of type I or III, there appears to be no general argument that excludes them at the moment.
In type I non-Fermi liquids, the pair self-energy is irrelevant. 
Since the pair-breaking effect of the incoherence dominates the pairing glue contained in the self-energy, type I non-Fermi liquids exhibit the similar behavior as superuniversality class A in type II.
Type III non-Fermi liquids are significantly more different from type II. 
They obey a similar flow equation, but with respect to the renormalization group scale $l$ that grows algebraically in energy, 
$l = 
\frac{1}{\alpha z - 2H_0} \left[ \left( \frac{\Lambda}{\omega} \right)^{\alpha - 2H_0/z}-1 \right]$
(see Appendix \ref{app:nonpert_pairing}).
We defer the search for such exotic examples and  their classification as future work.

\end{itemize}


\appendix
\onecolumngrid

\newcommand{\dvi}{\Delta\aV^*_{\infty}}
\newcommand{\mvi}{\aV^*_{\text{mid},\infty}}

\newcommand{\dviii}{\Delta\aV^*_{-\infty}}
\newcommand{\mviii}{\aV^*_{\text{mid},-\infty}}

\newpage
\appendix
\section{
Derivation of the flow equation
}
\label{app:nonpert_pairing}

In this appendix, we derive Eq. (\ref{eq:general_fp_eq_for_av}).
In the large $\KFAVdim$ limit, the connected four-point function 
for Cooper pairs with net three momentum ${\bf q}=(\omegaB,\vec q=0)$
is given by the sum of diagrams shown in \fig{fig:1},
\begin{equation}
\begin{aligned}
    \Gc{G}
      &=
    \Gce{G}
    -
    (K+ \Gce{G})\bullet
    \left[
        \tilde{V}
        -\tilde{V}\bullet(K+\Gce{G})\bullet\tilde{V}
        \right.\\&\left.
        + \tilde{V}\bullet(K+\Gce{G})\bullet\tilde{V}\bullet(K+\Gce{G})\bullet\tilde{V}
        + \cdots
   \right]\bullet
    (K+ \Gce{G}),
\label{app:net_two_body_nperturb_sum_1}
\end{aligned}
\end{equation}
where 
\begin{equation}
    \begin{aligned}
        \left(\Gc{G}\right)^{(j_1,j_1');(\nu_1,\delta_1,\theta_1)}_{(j_2,j_2');(\nu_2,\delta_2,\theta_2)}  &=
        \frac{1}{(2\pi)^3\delta^3(0)}\Bigg[\Big\langle   
        \psiB^{j_1}\left(\mathbf k_1\right)
        \psiB^{j_1'}\left(\mathbf k_1^\prime\right)
        \psiB_{j_2'}^{\dagger}\left(\mathbf k_2^\prime\right)
        \psiB_{j_2}^{\dagger}\left(\mathbf k_2\right)
        \Big\rangle
        -   
        \Big\langle
        \psiB^{j_1'}\left(\mathbf k_1^\prime\right)
        \psiB_{j_2'}^{\dagger}\left(\mathbf k_2^\prime\right)
        \Big\rangle
         \Big\langle
        \psiB^{j_1}\left(\mathbf k_1\right)
        \psiB_{j_2}^{\dagger}\left(\mathbf k_2\right)
        \Big\rangle
         \\&
        + 
        \Big\langle
        \psiB^{j_1'}\left(\mathbf k_1^\prime\right)
        \psiB_{j_2}^{\dagger}\left(\mathbf k_2\right)
        \Big\rangle
         \Big\langle
        \psiB^{j_1}\left(\mathbf k_1\right)
        \psiB_{j_2'}^{\dagger}\left(\mathbf k_2^\prime\right)
        \Big\rangle
        \Bigg]
\label{app:Gc4_definition}
\end{aligned}
\end{equation} 
denotes the connected four-point function for 
incoming fermions with momenta 
$\mathbf{k}_2=(\omegaB/2+\nu_2,\delta_2,\theta_2)$, $\mathbf{k}_2'=(\omegaB/2-\nu_2,\delta_2,\theta_2+\pi)$ and spins $j_2, j_2'$,
and
outgoing fermions with
$\mathbf{k}_1=(\omegaB/2+\nu_1,\delta_1,\theta_1)$, $\mathbf{k}_1'=(\omegaB/2-\nu_1,\delta_1,\theta_1+\pi)$ and $j_1, j_1'$.
All objects, including $\Gc{G}$, are viewed as matrices whose indices run over the spins and the relative momentum of a Cooper pair: $\{(j,j');(\nu,\delta,\theta)\}$.
The matrix product is defined as 
$(A\bullet B)^{(j_1,j_1');(\nu_1,\delta_1,\theta_1)}_{(j_2,j_2');(\nu_2,\delta_2,\theta_2)}=\sum_{i,j}\int \frac{d\nu\,d\delta}{(2\pi )^2}\int_{-\pi}^{\pi}\frac{
\KFthetadim\,
d\theta}{2\pi}A^{(j_1,j_1');(\nu_1,\delta_1,\theta_1)}_{(i,j);(\nu,\delta,\theta)}B^{(i,j);(\nu,\delta,\theta)}_{(j_2,j_2');(\nu_2,\delta_2,\theta_2)}$.
$K$ is the disconnected four-point function for $V=0$:
$K^{(j_1,j_1');(\nu_1,\delta_1,\theta_1)}_{(j_2,j_2');(\nu_2,\delta_2,\theta_2)}
=  2
\frac{
(2\pi)^3
}{\mathbf{K}_{F,\theta_1}}
\delta^{j_1}_{j_2}\delta^{j_1'}_{j_2'}\tdelta(\nu_1-\nu_2)\tdelta(\delta_1-\delta_2)\tdelta(\theta_1-\theta_2)G(\omegaB/2+\nu_1,\delta_1,\theta_1)
G(\omegaB/2-\nu_1,\delta_1,\theta_1+\pi)$,
where  $G(\omegaB,\delta,\theta)$
is the exact Green's function for $V=0$.
The factor of two arises from the contributions of the direct and exchange diagrams. 
\(\Gce{G}\) is the connected four-point function for \(V=0\), which includes the fermion-boson coupling to all orders. 
$\tilde{V}^{(i',j');(\nu',\delta',\theta')}_{(i,j);(\nu,\delta,\theta)}
=
V^{i'j'}_{ij} (\theta',\theta;\vec q=0)
$,
which is independent of $\nu$, $\nu'$, $\delta$ and $\delta'$.
Fermionic statistics implies
$
\tilde{V}^{(i',j');(\nu',\delta',\theta')}_{(i,j);(\nu,\delta,\theta)}
=
-
\tilde{V}^{(j',i');(-\nu',\delta',\theta'+\pi)}_{(i,j);(\nu,\delta,\theta)}
=-
\tilde{V}^{(i',j');(\nu',\delta',\theta')}_{(j,i);(-\nu,\delta,\theta+\pi)}$.

The terms in the square bracket in 
\eq{app:net_two_body_nperturb_sum_1}
form a geometric series and can be compactly written after projecting onto the spin channel as 
\begin{equation}
    \Gcr
    =
        \Gcre
        -
     (K'+ \Gcre)
     \circ
    \left[
        \left(V^{(r)}\right)^{-1}+ (K'+ \Gcre)
    \right]^{-1}
    \circ
     (K'+ \Gcre).
   \label{app:VB_resummed}
\end{equation}
Here,
$V^{(r)}_{\theta',\theta} 
=
\frac{1}{\mathcal N^{(r)}}\left(P^{(r)}\right)^{i',j'}_{i,j}
V^{ij}_{i'j'} (\theta',\theta;\vec q=0)$
is the four-fermion coupling projected to spin channel $r$ 
and
\bqa
K ' =2 \frac{(2\pi)^3}{\mathbf K_{F,\theta_1}} \tdelta(\nu_1-\nu_2)\tdelta(\delta_1-\delta_2)\tdelta(\theta_1-\theta_2)G(\omegaB/2+\nu_1,\delta_1,\theta_1)G(\omegaB/2-\nu_1,\delta_1,\theta_1+\pi).
\eqa
$\left(\Gcr\right)^{(\nu_1,\delta_1,\theta_1)}_{(\nu_2,\delta_2,\theta_2)} = \frac{1}{\mathcal N^{(r)} }\left(P^{(r)}\right)^{j_2,j_2'}_{j_1,j_1'}\left(\Gc{ G}\right)^{(j_1,j_1');(\nu_1,\delta_1,\theta_1)}_{(j_2,j_2');(\nu_2,\delta_2,\theta_2)}$ and $\Gcre = \Gcr|_{V=0} $ with $\mathcal N^+ = 3$ and $\mathcal N^- = 1$.
$(A\circ B)^{(\nu_1,\delta_1,\theta_1)}_{(\nu_2,\delta_2,\theta_2)}=\int \frac{d\nu\,d\delta}{(2\pi )^2}\int_{-\pi}^{\pi}\frac{
\KFthetadim\,
d\theta}{2\pi}A^{(\nu_1,\delta_1,\theta_1)}_{(\nu,\delta,\theta)}B^{(\nu,\delta,\theta)}_{(\nu_2,\delta_2,\theta_2)}$.
We can solve Eq. (\ref{app:VB_resummed}) for $V^{(r)}$ as
\begin{equation}
    \begin{aligned}
\left(V^{(r)}\right)^{-1}
     =
    -\left(K'+
    \Gcre
\right)\circ
\left(
\Gcr -
    \Gcre
  \right)^{-1}\circ  
  \left(K'+
    \Gcre
    \right)-\left(K'+
    \Gcre
    \right).
\label{app:VB_isolated}
\end{aligned}
\end{equation}
By applying $d/dl = 
-z\omegaB\partial_{\omegaB}
$ on both sides of Eq. (\ref{app:VB_isolated}), 
and using the fact that $V$ is independent of $\omegaB$,
we obtain 
\begin{equation}
    \begin{aligned}
       \frac{d\Gcr}{dl} = 
         \Gcr  \circ 
        \mathsf K^{(r)}
        \circ
         \Gcr
        -
         \Gcr \circ 
     \mathsf H^{(r) \dagger}
        -
        \mathsf H^{(r)}
        \circ  \Gcr
        - \mathsf S^{(r)},
        \label{eq:general_fp_eq_for_av_1}
    \end{aligned}
\end{equation}
where the kernels are determined by
\begin{equation}
    \begin{aligned}
        \mathsf K^{(r)} =
- \frac{d(G_{0}^{(4,r)})^{-1}}{dl},
~~~~~        
\mathsf H^{(r)} = 
       K' \circ 
 \frac{d (G_{0}^{(4,r)})^{-1}}{dl},
~~~~~
\mathsf S^{(r)} &= 
        K' \circ 
 \frac{d (G_{0}^{(4,r)})^{-1}}{dl}
        \circ K'
         +
        \frac{d K'}{dl},
\label{app:prime_kernels}
    \end{aligned}
\end{equation}
and $G_{0}^{(4,r)}=K'+\Gcre$ is the full four-point function for $V=0$.
To isolate the four-point vertex function, we remove that external leg by defining
$\Gammar=
-K'^{-1}
\circ 
\Gcr
\circ 
K'^{-1}$.
The flow equation for 
$\Gamma^{(r)}$ becomes
\bqa
\partial_l \Gammar 
 =   -  
 \Gammar 
 \circ
 k^{(r)}  
 \circ
 \Gammar
-  \Gammar \circ
   h^{(r)\dagger} 
   -  
   h^{(r)}
   \circ \Gammar
   + s^{(r)}, 
\label{eq:general_fp_eq_for_av_app8}
\eqa
where
\begin{equation}
    \begin{aligned}
 k^{(r)}  
       =
-K' \circ 
\frac{d(G_{0}^{(4,r)})^{-1}}{dl}
\circ K',
~~~~~        
   h^{(r)} 
= 
\frac{d(G_{0}^{(4,r)})^{-1}}{dl}
\circ K' + K'^{-1} \circ 
\frac{dK'}{dl},
~~~~~
s^{(r)} 
 &= 
\frac{d(G_{0}^{(4,r)})^{-1}}{dl}
 +
  K'^{-1} \circ \frac{d K'}{dl} \circ K'^{-1}.
\label{app:prime_kernels}
    \end{aligned}
\end{equation}

The four-point 
vertex function for Cooper pairs
with zero center of mass momentum and total energy $\omegaB$,
$\Gamma^{(r); (\nu_1,\delta_1,\theta_1)}_{~~~~(\nu_2,\delta_2,\theta_2)}(\omegaB)$ can be viewed as a matrix whose rows and columns are labeled by the relative three-momenta $(\nu,\delta,\theta)$.
However, it is not necessary to keep track of the dependence of the vertex function on all variables 
because $\nu_i \sim \omegaB$ and $\delta_i \sim \omegaB^{1/z}$ at low energies.
Hence, it is sufficient to keep track of the dependence of the vertex function on the angles and $\omegaB$ to characterize the universal superconducting fluctuations.
For this, we define the vertex function evaluated at 
$\nu_i \sim \omega$ 
and
$\delta_i \sim \omega^{1/z}$ 
as
\bqa
\bar \Gamma^{(r)}_{\theta_1,\theta_2}(\omegaB)
=
\int d \nu_1 d\delta_1 d \nu_2 d\delta_2 ~
W^*_{\nu_1,\delta_1,\theta_1}(\omega)
\Gamma^{
(r);
(\nu_1,\delta_1,\theta_1)
}_{
~~~~(\nu_2,\delta_2,\theta_2)
}(\omegaB)
W_{\nu_2, \delta_2, \theta_2}(\omega). 
\eqa
Here, $W_{\nu,\delta,\theta}(\omega)$ is a smooth distribution function of $\nu$ and $\delta$ centered around $\nu=\delta=0$ with widths $\omega$ and $\omegaB^{1/z}$, respectively.
In order to derive the flow equation for
$\bar \Gamma^{(r)}_{\theta_1,\theta_2}(\omegaB)$,
we decompose the full vertex function as
$\Gamma^{(r); (\nu_1,\delta_1,\theta_1)}_{~~~~(\nu_2,\delta_2,\theta_2)}(\omegaB)
=
\Gamma^{(r); (\nu_1,\delta_1,\theta_1)}_{0~~~(\nu_2,\delta_2,\theta_2)}(\omegaB)
+
\Gamma^{(r); (\nu_1,\delta_1,\theta_1)}_{1~~~(\nu_2,\delta_2,\theta_2)}(\omegaB)
$,
where the former is the vertex function for $V=0$ and the latter contains at least one four-fermion coupling.
In $\Gamma^{(r); (\nu_1,\delta_1,\theta_1)}_{1~~~(\nu_2,\delta_2,\theta_2)}(\omegaB)$,
the incoming and outgoing Cooper pairs are separated by $V$ which is independent of
$(\nu, \delta)$.
This causes its dependence
on 
$(\nu_1,\delta_1)$ and $(\nu_2,\delta_2)$ to factorize for $V$ that varies smoothly in angles.
To see this, we write 
$\Gamma^{(r); (\nu_1,\delta_1,\theta_1)}_{1~~~(\nu_2,\delta_2,\theta_2)}(\omegaB)$ as
\begin{equation}
    \begin{aligned}
\Gamma^{(r); (\nu_1,\delta_1,\theta_1)}_{1~~~(\nu_2,\delta_2,\theta_2)}(\omegaB)
=        \left[(\mathbb I+  K'^{-1}\circ\Gcre)
     \circ
    \left[
        \left(V^{(r)}\right)^{-1}+ (K'+ \Gcre)
    \right]^{-1}
    \circ
     (\mathbb I+ \Gcre\circ K'^{-1})
     \right]^{(\nu_1,\delta_1,\theta_1)}_{(\nu_2,\delta_2,\theta_2)},
    \end{aligned}
\label{eq:fW0}
\end{equation}
where $\mathbb I^{(\nu_1,\delta_1,\theta_1)}_{(\nu_2,\delta_2,\theta_2)} = \frac{(2\pi)^3}{\mathbf K_{F,\theta_1}}\tdelta(\nu_1-\nu_2)\tdelta(\delta_1-\delta_2)\tdelta(\theta_1-\theta_2)$. 
We note that
$\Bigg[\left(
        \left(V^{(r)}\right)^{-1}+ (K'+ \Gcre)
    \right)^{-1}\Bigg]^{(\nu_1,\delta_1,\theta_1)}_{(\nu_2,\delta_2,\theta_2)} =
    \Bigg[\left(
        \left(V^{(r)}\right)^{-1}+ (K'+ \Gcre)
    \right)^{-1}\Bigg]^{\theta_1}_{\theta_2}$
while
$(\mathbb I+ \Gcre\circ K'^{-1})\Big]^{(\nu_1,\delta_1,\theta_1)}_{(\nu_2,\delta_2,\theta_2)}$ is sharply peaked at $\theta_1 =\theta_2$ in the large $\KFAVdim$ limit.\footnote{
Since 
the off-diagonal elements 
of $\left(\mathbb I+ \Gcre\circ K'^{-1} \right)^{(\nu_1,\delta_1,\theta_1)}_{(\nu_2,\delta_2,\theta_2)}$ are generated by the boson,
its angular profile 
is controlled by the boson propagator.
At energy $\omega$, the boson propagator scales as 
$D(q,\omega) \sim$ 
$1/\omega^{(3-z-2\eta_\phiB)/2z}$,
$1/q^{ 3-z-2\eta_\phiB }$
and
$1/q^2$
for
$q< \omega^{1/2z}$, 
$\omega^{1/2z}< q <  \sqrt{\Lambda \KFAVdim}$, 
and
$\sqrt{\Lambda \KFAVdim}<q$, 
respectively.
This follows from the fact that the boson is strongly dressed by particle-hole excitations only below momentum $\sqrt{\Lambda \KFAVdim}$
and its infrared divergence is cut off at $q \sim \omega^{1/2z}$.
For $3-z-2\eta_\phiB>1$, the dominant contribution arises from $q \sim \omega^{1/2z}$.
For $3-z-2\eta_\phiB \leq 1$, 
the bosons mediate large angle-scatterings, which are cut off at $\sqrt{\Lambda \KFAVdim}$.
In any case, the main support of the boson-mediated connected four-point function is within the angular width of $\sqrt{\Lambda /\KFAVdim}$.
}
For smooth four-fermion couplings, which do not vary appreciably within $\Delta \theta \sim 
\sqrt{\Lambda/ \KFAVdim}$,
we can perform the internal three-momentum integrations in
\eq{eq:fW0} with
$\left[ \left(V^{(r)}\right)^{-1}+ (K'+ \Gcre)
\right]^{-1}$ evaluated at $\theta_1$ and $\theta_2$
to the leading order in
$\sqrt{\frac{\Lambda}{\KFAVdim}} \frac{\partial_{\theta_1} V_{\theta_1,\theta_2}}{V_{\theta_1,\theta_2}}$
and
$\sqrt{\frac{\Lambda}{\KFAVdim}} \frac{\partial_{\theta_2} V_{\theta_1,\theta_2}}{V_{\theta_1,\theta_2}}$,
\begin{equation}
    \begin{aligned}
\Gamma^{(r); (\nu_1,\delta_1,\theta_1)}_{1~~~(\nu_2,\delta_2,\theta_2)}(\omegaB)
=        
    \mathcal A_{\nu_1,\delta_1,\theta_1}^*
    \left( \left[
        \left(V^{(r)}\right)^{-1}+ (K'+ \Gcre)
    \right]^{-1} \right)^{\theta_1}_{\theta_2}
    \mathcal A_{\nu_2,\delta_2,\theta_2},
    \end{aligned}
\label{eq:fW1}
\end{equation}
where $\mathcal A_{\nu,\delta,\theta} = 
\int \frac{d\nu'd\delta' d \theta'}{(2\pi)^3}
{\bf K}_{F,\theta'}
(\mathbb I+ \Gcre\circ K'^{-1})\Big]^{(\nu',\delta',\theta')}_{(\nu,\delta,\theta)}$.
This factorization allows us to write 
$\Gamma^{(r); (\nu_1,\delta_1,\theta_1)}_{1~~~(\nu_2,\delta_2,\theta_2)}(\omegaB)
=
f^*_{\nu_1,\delta_1,\theta_1}(\omegaB)
\bar \Gamma^{(r)}_{1,\theta_1,\theta_2}(\omegaB) 
f_{\nu_2,\delta_2,\theta_2}(\omegaB)$, where
$\bar \Gamma^{(r)}_{a,\theta_1,\theta_2}(\omegaB)=
\int d \nu_1 d\delta_1 d \nu_2 d\delta_2 ~
W^*_{\nu_1,\delta_1,\theta_1}(\omega)
\Gamma^{
(r);
(\nu_1,\delta_1,\theta_1)
}_{
a~~~(\nu_2,\delta_2,\theta_2)
}(\omegaB)
W_{\nu_2, \delta_2, \theta_2}(\omega)$ 
with $a=1,2$,
and
$f_{\nu,\delta,\theta} = \frac{\mathcal A_{\nu,\delta,\theta}}{\int d\nu' d\delta' 
\mathcal A_{\nu',\delta',\theta}
W_{\nu',\delta',\theta}
   }$.
   Therefore, we can write the vertex function as
\bqa
\Gamma^{(r); (\nu_1,\delta_1,\theta_1)}_{~~~~(\nu_2,\delta_2,\theta_2)}(\omegaB)
=
f^*_{\nu_1,\delta_1,\theta_1}(\omegaB)
\bar \Gamma^{(r)}_{\theta_1,\theta_2}(\omegaB)
f_{\nu_2,\delta_2,\theta_2}(\omegaB)
+
\Gamma^{'(r); (\nu_1,\delta_1,\theta_1)}_{~~~~(\nu_2,\delta_2,\theta_2)}(\omegaB),
\label{eq:factorization}
\eqa
where
$
\Gamma^{'(r); (\nu_1,\delta_1,\theta_1)}_{~~~~(\nu_2,\delta_2,\theta_2)}(\omegaB)
=
\Gamma^{(r); (\nu_1,\delta_1,\theta_1)}_{0~~~(\nu_2,\delta_2,\theta_2)}(\omegaB)
-
f^*_{\nu_1,\delta_1,\theta_1}(\omegaB)
\bar \Gamma^{(r)}_{0, \theta_1,\theta_2}(\omegaB)
f_{\nu_2,\delta_2,\theta_2}(\omegaB)
$
is a contribution that is independent of $V$.

Now, we insert \eq{eq:factorization}
to 
\eq{eq:general_fp_eq_for_av_app8}, and evaluate the matrix equation at 
$\left( \hat \nu_{\theta_1}(\omega),\hat \delta_{\theta_1}(\omega),\theta_1 \right)$
and
$\left( \hat \nu_{\theta_2}(\omega),\hat \delta_{\theta_2}(\omega),\theta_2 \right)$,
where 
$\hat \nu_{\theta}(\omega)$
and $\hat \delta_{\theta}(\omega)$ are chosen to satisfy
$\partial_\omega f_{\nu,\delta,\theta}(\omegaB)\Big|_{\nu = \hat \nu_\theta(\omega),\delta = \hat \delta_\theta(\omega)} = 0$. 
Then, the flow equation for
\bqa \avn^{(r)}_{\theta_1,\theta_2}(\omegaB)
=
\sqrt{\mathbf K_{F,\theta_1}\mathbf K_{F,\theta_2}}
\bar \Gamma^{(r)}_{\theta_1,\theta_2}(\omegaB)
\eqa
becomes
\bqa
&& 
\partial_l \avn^{(r)} 
 =   -  
 \avn^{(r)} \cdot
 \mK^{(r)}  
 \cdot
 \avn^{(r)}
-  \avn^{(r)}\cdot 
 \mH^{(r)\dagger} 
   -  
   \mH^{(r)}
   \cdot \avn^{(r)}
   + \mS^{(r)}. 
\label{eq:general_fp_eq_for_av_app}
\eqa
Here,
$(A\cdot B)_{\theta_1,\theta_2} = \int_{-\pi}^{\pi} \frac{d\theta}{2\pi}A_{\theta_1,\theta}B_{\theta,\theta_2}$.
The kernels, which are matrices in the angular space, are
\bqa
\mK^{(r)}_{\theta_1,\theta_2}
 &=&
F^*(\theta_1) \circ
( \dot K' - K' \circ \dot \Sigma_0^{(r)} \circ K' ) \circ F(\theta_2), \nn
\mH^{(r)}_{\theta_1,\theta_2}
 &=&
{\cal I}^*(\theta_1) \circ\Big[
\dot  \Sigma^{(r)}_0
\circ
K' + {{ \Gamma^{'(r)}}} \circ  ( \dot K' - K' \circ \dot \Sigma_0^{(r)} \circ K' )  \Big]\circ
F(\theta_2), \nn
\mS^{(r)}_{\theta_1,\theta_2}
&=&
{\cal I}^*(\theta_1) \circ \Big[
\dot  \Sigma^{(r)}_0
- \dot  \Gamma^{'(r)}
-  \dot  \Sigma^{(r)}_0
\circ
K'\circ {{ \Gamma^{'(r)}}} - {{ \Gamma^{'(r)}}}\circ K' \circ \dot \Sigma_0^{(r)} - {{\Gamma^{'(r)}}} \circ  ( \dot K' - K' \circ \dot \Sigma_0^{(r)} \circ K' ) \circ {{ \Gamma^{'(r)}}} \Big]\circ
{\cal I}(\theta_2).
\nn
\eqa
$\dot A \equiv \frac{dA}{dl}$
represents the derivative with respect to the logarithmic length scale $l=\frac{1}{z} \log \frac{\Lambda}{\omega}$.
\bqa
\Sigma_0^{(r)} = 
(G_{0}^{(4,r)})^{-1} - K'^{-1}
\eqa
is the pair self-energy.
$F(\theta_1)$ 
and
${\cal I}(\theta_1)$ 
are vectors whose elements are given by
$[F(\theta_1)]_{\nu,\delta,\theta}= \frac{2\pi}{\sqrt{\mathbf K_{F,\theta_1}}} \tdelta (\theta-\theta_1)
f_{\nu,\delta,\theta}$
and
$[{\cal I}(\theta_1)]_{\nu,\delta,\theta}= \frac{(2\pi)^3}{ \sqrt{\mathbf K_{F,\theta_1}}}
\tdelta (\theta-\theta_1)
\tdelta (\nu-\hat \nu_\theta(\omegaB))
\tdelta(\delta-\hat \delta_\theta(\omegaB))
f^{-1}_{\nu,\delta,\theta}$.
It is understood that all matrices depend on $\omegaB$, although not shown to avoid clutter in notation.

The flow of the coupling function is determined by the eigenvalues of the matrices
$\mK^{(r)}$,
$\mH^{(r)}$
and
$\mS^{(r)}$.
In particular, the eigenvalue in the $s$-wave limit plays the important role in the low-energy limit.
To estimate the leading frequency dependence of the eigenvalues in the $s$-wave limit, we consider the angular average of the kernels, 
\bqa
\bar \mK^{(r)}_{a}
 =  \frac{1}{(2\pi)^2} \int d\theta_1 d \theta_2 \mK^{(r)}_{a;\theta_1,\theta_2}, ~~~
\bar \mH^{(r)} 
=  \frac{1}{(2\pi)^2} \int d\theta_1 d \theta_2
\mH^{(r)}_{\theta_1,\theta_2}, ~~~
\bar \mS^{(r)}
=  \frac{1}{(2\pi)^2} \int d\theta_1 d \theta_2
\mS^{(r)}_{\theta_1,\theta_2},
\eqa
where
$\mK_{0;\theta_1,\theta_2}
 =
F^*(\theta_1) \circ  \dot K'  \circ F(\theta_2)$
and
$ \mK^{(r)}_{1;\theta_1,\theta_2}
=- F^*(\theta_1) \circ 
 K' \circ \dot \Sigma_0^{(r)} \circ K'  \circ F(\theta_2)$.
All kernels are made of 
$K'$, 
$\dot \Sigma_0^{(r)}$
and
$\Gamma^{'(r)}$
and 
the frequency dependence of the kernels is fixed by the exponents for 
$\bar \mK_0$
and
$\bar \mS^{(r)}$.
The frequency dependence of 
$\bar \mK_0$ is fixed by the anomalous dimension of the single-fermion excitation:
$\bar \mK_0 \sim \omega^{\frac{2H_0}{z}}$.
If we write $\bar \mS^{(r)} \sim \omega^{-\alpha}$ at low frequencies, the scaling forms of the remaining terms are fixed to be
\bqa
\bar \mK^{(r)}_1 \sim \omega^{\frac{4H_0}{z}-\alpha},
~~~
\bar \mH_1^{(r)} \sim \omega^{\frac{2H_0}{z}-\alpha}.
\eqa
Here, we have used the fact that 
$\Gamma^{'(r)}$
scales as $\Sigma_0^{(r)}$ ($K'^{-1}$) when 
$K'\circ \Sigma_0^{(r)} \ll 1$ 
($K'\circ \Sigma_0^{(r)} \gg 1$).
The factor of $\omegaB^{\frac{2H_0}{z}}$ in 
$\bar \mK_0$ can be readily absorbed into the coupling as $\av^{(r)} = \omegaB^{\frac{2H_0}{z}}   {\avn}^{(r)}$.
The flow equation for $\av^{(r)}$ takes the following forms, depending on the relative magnitude of $\frac{2H_0}{z}$ and $\alpha$.
\begin{itemize}
\item 
$\frac{2H_0}{z} > \alpha$:
The pair self-energy is sub-dominant compared to the disconnected function.
Hence, $\mK^{(r)}_1$,
$\mH^{(r)}$ and
$\mS^{(r)}$,
which arise from the self-energy,
are negligible compared to 
$\mK_0$.
To the leading order in $\omegaB$, the flow equation becomes
\bqa
\partial_l 
\av^{(r)} = 
- \av^{(r)} \cdot 
\mK^{'(r)} 
\cdot \av^{(r)} 
- \av^{(r)}\cdot \mH^{\prime}_0 
- \mH'_0\cdot \av^{(r)},
\eqa
where
$\mK^{'(r)} 
= \mK'_0$ with $\mK'_0 = \omega^{-\frac{2H_0}{z}}
\mK_0$
and $\mH'_0 =  H_0I$
 with $I_{\theta_1,\theta_2} = 2\pi \delta(\theta_1-\theta_2)$.
\item $\frac{2H_0}{z}=\alpha$:
The self-energy and the disconnected function are comparable.
All terms are equally important, and the flow equation reads
\bqa
\partial_{l} 
\av^{(r)} = 
- \av^{(r)} \cdot 
\mK^{'(r)} 
\cdot \av^{(r)} 
- \av^{(r)} \cdot \mH^{'(r)\dagger} - \mH^{'(r)}\cdot \av^{(r)} 
+ \mS^{'(r)},
\eqa
where
$\mK^{'(r)}  = 
\mK^{\prime}_0 + \mK_1^{\prime(r)}$
with
$\mK_a^{\prime(r)}=
\omega^{-\frac{2H_0}{z}}
\mK^{(r)}_a$,
$\mH^{'(r)}  = \mH^{\prime}_0+\mH^{\prime(r)}_1$ with $\mH^{\prime(r)}_1 = \mH^{(r)}$,
and
$\mS^{'(r)}  = \omega^{\frac{2H_0}{z}} \mS^{(r)}$
.
\item 
$\frac{2H_0}{z} < \alpha$:
The self-energy is dominant,
and
$\mK_0$
is negligible compared to 
$\mK^{(r)}_1$.
Due to the power-law singularity of the self-energy, 
it is more convenient to describe the flow in terms of the algebraic (not logarithmic) length scale. 
With respect to $l' = \frac{1}{\alpha z - 2H_0} \left[ \left( \frac{\Lambda}{\omega} \right)^{\alpha - 2H_0/z}-1 \right]$,
the flow equation becomes
\bqa
\partial_{l'} 
\av^{(r)} = 
- \av^{(r)} \cdot 
\mK^{'(r)} 
\cdot \av^{(r)} 
- \av^{(r)} \cdot \mH^{'(r)\dagger} - \mH^{'(r)}\cdot \av^{(r)} 
+ \mS^{'(r)},
\eqa
where
$\mK^{'(r)}  =  \Lambda^{\frac{2H_0}{z}-\alpha}\omega^{-\frac{4H_0}{z}+\alpha} \mK^{(r)}_1$,
$\mH^{'(r)}  = \Lambda^{\frac{2H_0}{z}-\alpha}\omega^{-\frac{2H_0}{z} +\alpha} \mH^{(r)}$,
$\mS^{'(r)}  = \Lambda^{\frac{2H_0}{z}-\alpha}\omega^{\alpha} \mS^{(r)}$.
\end{itemize}
In all cases, the flow equation takes the form of 
Eq. (\ref{eq:general_fp_eq_for_av}),
where the angular average of the kernels becomes independent of frequency in the low-energy limit.

\section{Non-ladder quantum correction}
\label{app:nonladder}

\begin{figure}[H]
\centering
\begin{subfigure}{0.25 \textwidth}
\includegraphics[width=\textwidth]{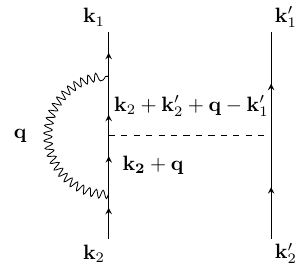}
\caption{}
\label{fig:vertex1_2}
    \end{subfigure}
\begin{subfigure}{0.35\textwidth}
\includegraphics[width=\textwidth]{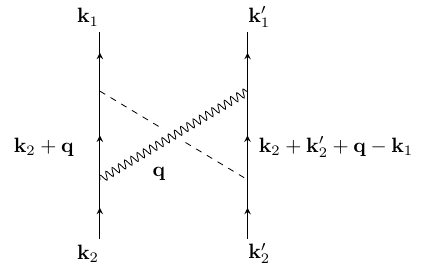}
        \caption{}
        \label{fig:vertex1_1}
\end{subfigure}
\caption{Examples of non-ladder quantum corrections involving at least one crossed bosonic exchange across a short-range four-fermion interaction.
}
\label{fig:nonladder_quantum_corrections}
\end{figure}

The goal of this appendix is to illustrate why non-ladder quantum corrections 
do not contribute to the eigenvalues of the kernels in Eq. (\ref{eq:general_fp_eq_for_av}) to the leading order in 
the small $\omega/\KFAVdim$ limit. 
As examples, some one-loop non-ladder diagrams are shown in
Fig.~\ref{fig:nonladder_quantum_corrections}.
In these diagrams, internal fermions are in different patches of the Fermi surface for generic choices of external momenta. 
Because they cannot simultaneously stay close to the Fermi surface while exchanging non-zero momentum, the infrared singularity is weakened.
The infrared singularity is enhanced only for specific choices of external momenta, for example,  $\vec k_1 = \pm \vec k_2$ in Fig.~\ref{fig:nonladder_quantum_corrections}, which allows internal fermions to be in one patch or antipodal patches. 
Since the phase space for such special external momenta becomes vanishingly small at low energies,
their contribution to the eigenvalues of the kernels 
is suppressed by positive powers of $\omega/\KFAVdim$ at low energies.
%
%

To see the suppression of the infrared singularity in non-ladder diagrams explicitly, let us consider the following factor in
\fig{fig:vertex1_2}
(similar factors are present in other non-ladder graphs, such as in \fig{fig:vertex1_1}),
\begin{equation}
    \begin{aligned}
I(\vec k_1,\vec k_2) = 
\int d\mathbf q 
D_1(\mathbf q)G(\mathbf k_2+\mathbf q)G(\mathbf k_2+\mathbf k_2^\prime-\mathbf k_1^\prime+\mathbf q),
        \label{app:NL_1}
    \end{aligned}
\end{equation}
Here, we ignore the smooth momentum dependence of the coupling functions, which do not affect the singularity. 
The external momenta are   
$ \mathbf k_1 = (3\omegaB,\vec k_1)$,
$\mathbf k_1^\prime = (-\omegaB,-\vec k_1)$,
$\mathbf k_2 = (2\omegaB,\vec k_2)$,
$\mathbf k_2^\prime = (0,-\vec k_2)$ with
$\vec k_1 = (\delta_1,\theta_1)$ and $\vec k_2 = (\delta_2,\theta_2)$.
We first evaluate the quantum correction for $\theta_1-\theta_2$ away from $0$ and $\pi$. 
At such generic angular separations, the patches centered at $\theta_1$ and $\theta_2$ are not parallel or anti-parallel.
To check that there is no infrared singularity in such cases, we set $\omega=0$,
\begin{equation}
 I(\vec k_1,\vec k_2)=\int _{0}^{2\pi}\frac{d\varphi}{2\pi}\int \frac{dq_0}{(2\pi)}\int \frac{qdq}{2\pi}
\frac{
D_1(\mathbf q)
}
{\left[i\gamma|q_0|^{2/3}\mathrm{sgn}(q_0)+
v_{F}\cos(\varphi-\theta_2)q
\right]
\left[i\gamma|q_0|^{2/3}\mathrm{sgn}(q_0)+
v_{F}\cos(\varphi-\theta_1)q
\right]}
.
\label{eq:kernel}
\end{equation}
Here, we have used the one-loop dressed boson and fermion propagators, where
$\gamma$ is a constant,
and
$D_1(\mathbf q) =  \frac{1}{q^2+f\frac{|q_0|}{q}}$ with $f  = \frac{1}{2}\frac{\KFAVdim e^2}{v_{F}^2}$\cite{kukreja2026classificationnonfermiliquidsuniversal}.
For simplicity, we consider a circular Fermi surface, but the same principle applies to general Fermi surfaces. 
Two non-parallel patches of the Fermi surfaces provide the natural oblique coordinate system $(q_1,q_2) =  (q \cos(\varphi-\theta_1),q \cos(\varphi-\theta_2))$,
in which
\eq{eq:kernel} reads
\begin{equation}
 I(\vec k_1,\vec k_2)=\frac{1}{|\sin(\theta)|}\int\frac{dq_0}{(2\pi)}\int \frac{d q_1}{2\pi} \int \frac{dq_2}{2\pi}
\frac{\tilde D_1(q_1,q_2)}
{\left[i\gamma|q_0|^{2/3}\mathrm{sgn}(q_0)+
v_{F}q_1
\right]
\left[i\gamma|q_0|^{2/3}\mathrm{sgn}(q_0)+
v_{F}q_2
\right]},
\label{eq:kernel_oblique}
\end{equation}
where 
$\tilde D_1(q_1,q_2) = D\left(
\frac{\sqrt{q_1^2+q_2^2-2q_1q_2\cos(\theta)}}{|\sin(\theta)|} \right)$. 
To estimate the infrared singularity, we can assign the scaling dimension to each component of momentum as $[q_1]=[q_2]=1$ and $[q_0]=3$.
Under this scaling, the self-energy of the fermion is unimportant at low energies, except for the sign structure.
\eq{eq:kernel_oblique} has the net scaling dimension $1$.
Therefore, there is no infrared singularity.
The absence of IR singularity can be checked explicitly.

For 
$\theta=0, \pi$, the patches become parallel or anti-parallel, which can give rise to an IR singularity. 
To understand the strength of the IR singularity and how the infrared singularity disappears away from those special angles, we introduce a non-zero external frequency $\omega$ as an infrared cutoff and 
set \(\theta_2=0\), \(\theta_1=\pi-\delta\theta\). 
In the local Cartesian coordinates, in which $q_x$ ($q_y$) is perpendicular (parallel) to the patch at $\theta_2$, 
\eq{eq:kernel} becomes
\begin{equation}
\begin{aligned}
    \tilde I(\delta\theta,\omegaB) &= 
    \int \frac{dq_0 dqx dq_y}{(2\pi)^3}\frac{1}{i\gamma|q_0+2\omegaB|^{2/3}\mathrm{sgn}(q_0+2\omegaB)+v_F\left(q_x+\frac{q_y^2}{2\KFAVdim}\right)}
    \\
    &\times
    \frac{1}{i\gamma|q_0+3\omegaB|^{2/3}\mathrm{sgn}(q_0+3\omegaB)+v_F\left(-q_x+\frac{q_y^2+\KFAVdim^2(\delta\theta)^2+2\KFAVdim q_y\delta\theta }{2\KFAVdim}\right)}
    \frac{|q_y|}{|q_y|^3+f|q_0|},
    \end{aligned}
\end{equation}
where we set the external frequencies
to be on the order of $\omegaB$.
Here, we have ignored $q_x$-dependence in the boson propagator that is subleading. 
Integrating over $q_x$, we obtain
\begin{equation}
\begin{aligned}
    \tilde I(\delta\theta,\omegaB) =& \frac{i}{v_F}\int \frac{dq_0 dq_y}{(2\pi)^2}\Bigg[-\frac{\Theta(q_0+2\omegaB)}{i\gamma(|q_0+2\omegaB|^{2/3}+|q_0+3\omegaB|^{2/3})+\frac{v_F}{\KFAVdim}\left[\left(q_y+\frac{\KFAVdim \delta\theta}{2}\right)^2+\frac{\KFAVdim^2(\delta\theta)^2}{4}\right]}
    \\
    +&\frac{\Theta(-q_0-3\omegaB)}{-i\gamma(|q_0+2\omegaB|^{2/3}+|q_0+3\omegaB|^{2/3})+\frac{v_F}{\KFAVdim}\left[\left(q_y+\frac{\KFAVdim \delta\theta}{2}\right)^2+\frac{\KFAVdim^2(\delta\theta)^2}{4}\right]}\Bigg]
    \frac{|q_y|}{|q_y|^3+f|q_0|}.
\label{app:theta_pi_q0_integrated}
    \end{aligned}
\end{equation}
\eq{app:theta_pi_q0_integrated}
is logarithmically divergent.
There are two infrared scales
$q^{*1}_y\sim f^{1/3}\omegaB^{1/3}$
and $q^{*2}_y\sim \KFAVdim \delta\theta$.
The larger of the two cuts off the infrared singularity.
Therefore, $\tilde I(\delta\theta,\omegaB)$ exhibits a logarithmic singularity in $\omegaB$ within the patch of width $\delta\theta^*\sim 
\Lambda^{1/3}\omegaB^{1/3}/\KFAVdim^{2/3}$
around $\theta_1=\pi$.
The logarithmic singularity in $\omegaB$ is cut off outside the patch. 
Because the width of the patch decays in powers of $\omega$, the contribution of \fig{fig:nonladder_quantum_corrections} to the eigenvalue of $\bT$ vanishes at low energies.

\section{Diagonalization of the kernels}
\label{app:diagonalization}

\begin{figure}[ht]
\includegraphics[width=0.3\textwidth]{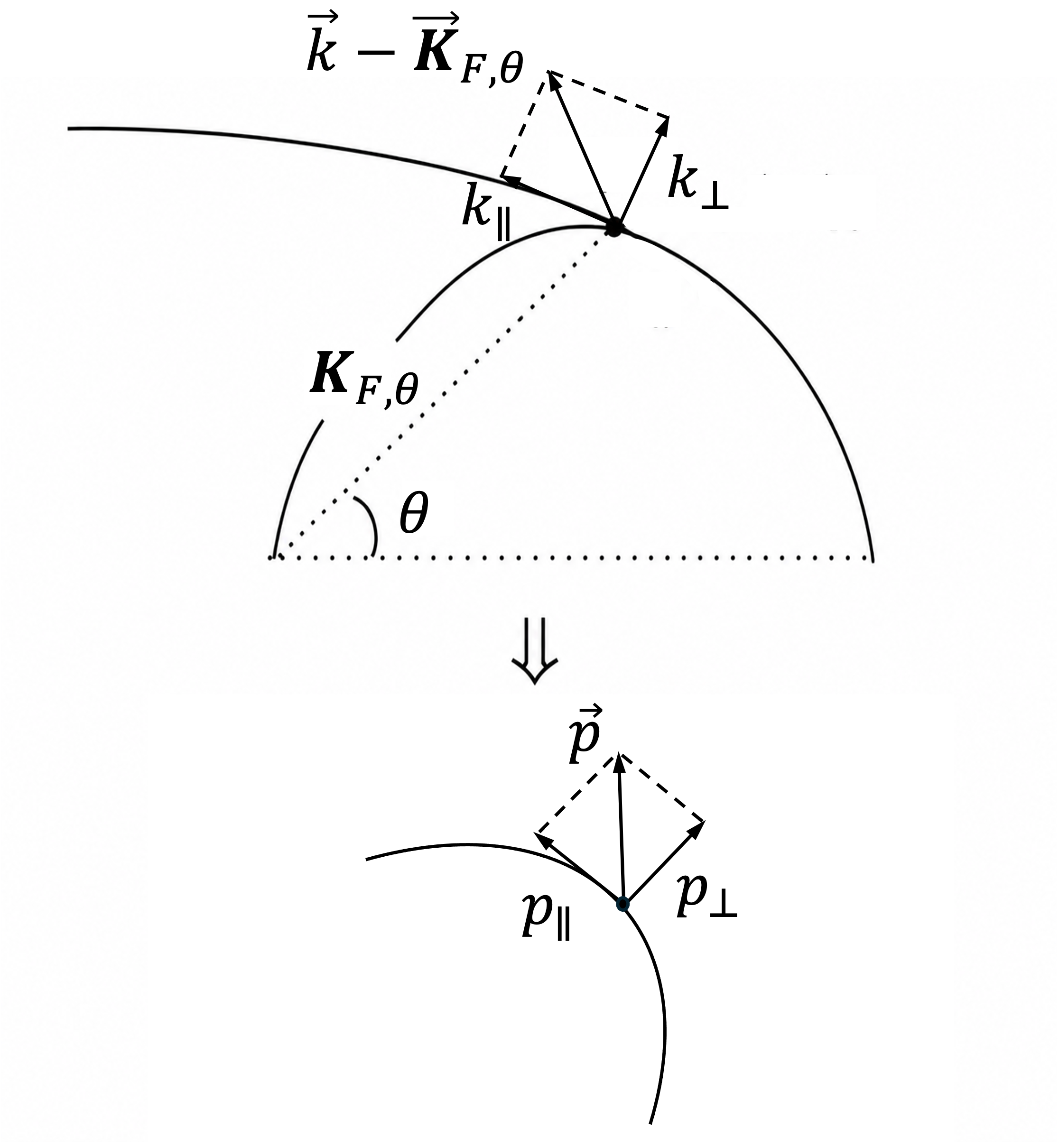}
\caption{
Through an angle-dependent rescaling of momentum, any patch of a convex Fermi surface can be locally mapped into the canonical parabola with dispersion $E_{\vec p} = p_\perp + p_\parallel^2$.
}
\label{fig:patch}
\end{figure}

In this appendix, we find the basis in which the kernels of the flow equation are diagonalized.
This is done by taking advantage of the fact that the kernels decay quickly in angular space and can be computed from the patch theory.
To transition from \eq{action_low_energy}
to patch theory, we 
choose a Fermi momentum
$\KFthetadim$ to be the center of a patch
and
introduce a local Cartesian coordinate to denote the deviation of the fermion momentum from $\KFthetadim$, as 
$\vec k - \vec {\mathbf{K}}_{F,\theta} = k_\perp \hat k_{\perp,\theta} + k_\parallel \hat k_{\parallel, \theta}$,
where $\hat k_{\perp,\theta}$ 
and $\hat k_{\parallel,\theta}$ are the unit vectors perpendicular and parallel to the Fermi surface at angle $\theta$, respectively (see \fig{fig:patch}).
One can then scale each component of momentum and fields to remove $v_{F,\theta}$ and $\KFthetadim$ in the patch action\cite{POLCHINSKI2}.
The action for the anti-podal patch can be obtained in the same way.
The theory that describe small-angle scatterings around $\theta$ and $\theta+\pi$ reads
\begin{align}
S &=
\sum_{s=\pm}
\int d \mathbf p \,
\psiB^\dagger_{s,j}(\mathbf p
)
\left[
i p_0
+
s p_\perp + 
p_\parallel^2
\right]
\psiB_s^j(\mathbf p)
\nonumber \\ &
+
\frac12
\int d \mathbf p \,
p_\parallel^2
|\phi_a(\mathbf p)|^2
\nonumber \\ &
+
\frac{1}{\sqrt 2}
\sum_{s=\pm}
e_s
\int d \mathbf p d \mathbf p' \,
~
\phiB_a( \mathbf p')
\psiB^\dagger_{s,i}( \mathbf p 
+ \mathbf p')
T^{a,i}_{j}
\psiB_s^j(\mathbf p) \nn
& + \frac{u}{4} \int  d \mathbf k  d \mathbf k' d \mathbf q \, 
\phiB_a( - \mathbf q-\mathbf k')
\phiB_a(\mathbf k')
\phiB_b(\mathbf k)
\phiB_b(\mathbf q-\mathbf k). 
\label{eq:patch_theory}
\end{align}
Here, $s=+$ and $-$ represent the 
patch centered at angle $\theta$ and $\theta+\pi$, respectively.
$e_+=e_\theta( \varphi_\theta)$
and
$e_-=e_{\theta+\pi}( \varphi_\theta)$, where
$ \varphi_\theta$ is the angle of a vector tangent to the Fermi surface at $\theta$, measured with respect to the positive $x$-axis and pointing in the direction of increasing $\theta$.
We set $V=0$ because the kernels 
depend on the four-point function solely generated from boson-fermion coupling.
\eq{eq:patch_theory} is independent of angle because every patch of a convex Fermi surface is locally identical upon rescaling.
The information on $v_{F,\theta}$ and $\KFthetadim$ is contained in the relation between
the rescaled momentum $(p_\perp,p_\parallel)$
and
the original momentum
$(k_\perp, k_\parallel)$: 
\bqa
p_\perp &=& 
v_{F,\theta}~ 
        \mathrm{cosec}( \varphi_\theta -\theta )k_\perp,     
                \nn
p_\parallel^2  &=& 
\frac{ v_{F,\theta}  \mathscr{G}_{ \varphi_\theta,\theta}}{ \KFthetadim} k_\parallel^2,
\label{eq:pxyk}
\eqa
where 
$\mathscr{G}_{\varphi,\theta} =  
\frac{1}{2}\left( 
\sin^2(\varphi -\theta )\left[1-\frac{{\bf K}^{\prime\prime}_{F,\theta}}{\KFthetadim}\right]+\sin\left(2\left(\varphi-\theta\right)\right)\frac{{\bf K}^{\prime}_{F,\theta}}{\KFthetadim}\right)$
and
$\KFthetadim'$ is the derivative of $\KFthetadim$ with respect to $\theta$.
It is noted that $p_\parallel \sim k_\parallel /\sqrt{\KFthetadim}$ has dimension $1/2$.

We can compute the kernels using this patch theory.
Let $\bT_{\theta_1,\theta_2}(\omega)$ 
be any of the kernels
$\left\{ \mK^{('r)}_{\theta_1,\theta_2}(\omega), 
\mH^{'(r)}_{\theta_1,\theta_2}(\omega), \mS^{'(r)}_{\theta_1,\theta_2}(\omega) \right\}$.
According to the patch theory, 
$\bT_{\theta,\theta + \Delta \theta}$ should depend on $\Delta \theta$ through 
$p_\parallel/\omegaB^{1/2z}$,
where $p_\parallel$ is the rescaled momentum associated with $k_\parallel$ that connects 
${\bf K}_{F,\theta}$ and ${\bf K}_{F,\theta + \Delta \theta}$.
Using $k_\parallel = 
\KFthetadim
\Delta \theta 
/
\sin( \varphi_\theta-\theta)$
and
\eq{eq:pxyk},
we can relate $\Delta \theta$ to $p_\parallel$ as
$p_\parallel = 
\sqrt{\frac{ v_{F,\theta}  \mathscr{G}_{ \varphi_\theta,\theta}\KFthetadim}{
\sin^2( \varphi_\theta-\theta)
}}  
\Delta \theta
 $.
From this, we conclude that $\bT$ must take the form of
\bqa
\bT_{\theta + \Delta \theta, \theta}
=
\frac{1}{\Delta_\theta(\omegaB)}
T \left(
\frac{\Delta \theta}{
\Delta_\theta(\omegaB)} 
\right),
\label{eq:T_angular_form}
\eqa
where
$
\Delta_\theta(\omegaB) 
= \sqrt{\frac{ 
\sin^2( \varphi_\theta-\theta)}{ 
v_{F,\theta}  \mathscr{G}_{ \varphi_\theta,\theta}\KFthetadim }}
\omegaB^{1/2z} 
$ represents the angular width of a patch at the energy scale $\omegaB$ and $T(x)$ is a universal function.
In the small $\omegaB$ limit, $\Delta_\theta(\omegaB)$ becomes small, allowing one to ignore the variation of $v_{F,\theta}$ and $\KFthetadim$ within the peak of $T(x)$.
The overall factor of $1/\Delta_\theta(\omegaB)$ is present because the weight $\int d \Delta \theta \bT_{\theta+\Delta \theta,\theta}$ is independent of $\omegaB$ and $\theta$.
The independence of the weight with respect to $\omegaB$ is a property of type II non-Fermi liquids.
The independence with respect to $\theta$ comes from the fact that every patch is locally identical to each other.
One can confirm this form explicitly through perturbative calculations\cite{kukreja2026classificationnonfermiliquidsuniversal}.
The decoupling of high-energy bosons at low energies guaranties that $T(x)$ decays fast at large $x$, so that $\int_{-\infty}^\infty dx T(x)$ is finite.
The eigenvalues of the kernels have no infrared singularity either, because the small but non-zero $\omegaB$ cuts off infrared divergence.

The equivalence of patches
promotes the sliding symmetry of each patch theory\cite{MAX0} to  full rotational invariance. 
This emergent rotational invariance acts linearly 
to a coordinate $\bar \theta$ related to the original angle through a non-linear coordinate transformation\cite{kukreja2026classificationnonfermiliquidsuniversal},
\bqa
d \bar \theta = 
\frac{ \omegaB^{1/2z} }{\Delta_\theta(\omegaB)}
d \theta.
\eqa
$\omega^{-1/2z} d \bar \theta$ measures the angular separation in units of angle-dependent patch size $\Delta_\theta(\omegaB)$.
Like $\theta$, $\bar \theta$ is a compact variable defined in $( -\bar \theta_{max}, \bar \theta_{max}]$,
where 
$\bar \theta_{max} =
\gamma \sqrt{\KFAVdim}$ with an $O(1)$ constant $\gamma$ that depends on $v_{F,\theta}$ and the shape of the Fermi surface.
In this coordinate,
we obtain
$\bT_{\theta_1, \theta_2} = \frac{1}{\omegaB^{1/2z}} 
\frac{d \bar \theta_1}{d \theta_1} 
T^{(r)} \left(  \frac{\bar \theta_1 - \bar \theta_2}{ \omegaB^{1/2z}} \right)$.
Here, $\bar \theta_i=\bar \theta(\theta_i)$.
Since $T^{(r)} \left(  \frac{\bar \theta_1 - \bar \theta_2}{ \omegaB^{1/2z}} \right)$
is sharply peaked at $\bar \theta_1=\bar \theta_2$ and 
$\bar \theta(\theta)$ is a smooth function of $\theta$,
$\frac{d \bar \theta_1}{d \theta_1}$ can be replaced with $\frac{d \bar \theta_2}{d \theta_2}$ to the leading order 
in $\omegaB/\KFAVdim$.
The rotational symmetry is broken by cold spots, if present.
However, the size of cold regions decreases at low energies\cite{kukreja2026classificationnonfermiliquidsuniversal} and their contributions to the eigenvalues of $\bT$ are suppressed by positive powers of $\omegaB/\KFAVdim$.
Up to corrections suppressed by 
$\omegaB/\KFAVdim$,
the $n$-th normalized eigenfunction of $\bT$ is 
\bqa
U_{n,\theta} &=& 
\sqrt{\frac{\pi}{ \bar \theta_{max}}\frac{d\bar \theta }{d\theta}}
e^{i \frac{n \pi \bar \theta}{\bar \theta_{max}}}
\label{eq:eigen}
\eqa
with eigenvalue
\bqa
\bT_n(\omegaB;\KFAVdim) &= &
\int \frac{d x}{2\pi}~
e^{-i \frac{n \pi 
\omegaB^{\frac{1}{2z}}
 }{\bar \theta_{max}} x }
T^{(r)} \left( x \right).
\label{eq:eigen}
\eqa
The eigenvalue is simply the Fourier transformation of $T^{(r)}(x)$ in the $\bar \theta$ space.
The eigenfunctions in the original $\theta$ space are 
the sinusoidal functions
stretched and compressed according to the sizes of angle-dependent local patches\cite{kukreja2026classificationnonfermiliquidsuniversal}.
It is noted that the eigenvalue $\mathcal T_n^{(r)}(\omegaB;\KFAVdim)$ depends on $n$, $\omegaB$ and $\KFAVdim$ through a specific combination, 
\bqa
\frac{n \pi 
\omegaB^{\frac{1}{2z}}
 }{\bar \theta_{max}}  
\sim 
q_\parallel/(\KFAVdim/n).
\label{eq:expy}
\eqa
This is nothing but the ratio between the momentum carried by the critical boson $q_\parallel$
in \eq{eq:qparallel}
and the pitch of the Cooper pair wavefunction, $2 \pi \KFAVdim /n$.

\section{Emergent particle-hole symmetry}
\label{app:PHSymmetry}

In this appendix, we show that \eq{action_low_energy} 
with the four-fermion interaction projected to the pairing channel with zero center of mass momentum is invariant under the combination of particle-hole and time reversal transformations in the large $\KFAVdim$ limit.
This acts as the symmetry for 
\eq{eq:dimensionless4pt} in the large
$\KFAVdim$ limit because only the four-fermion coupling of $\vec q=0$ contributes to it.
The low-energy theory is written as 
\begin{align}
S
&=
\int 
d \mathbf k \,
\psiB^\dagger_j(k_0,\delta,\theta
)
\Bigl[
i k_0
+
v_{F,\theta}
\delta
\Bigr]
\psiB^j(k_0,\delta,\theta)
\nonumber 
+
\frac12
\int 
d \mathbf q\,
{\bf q}^2
|\phi_a(q_0,q,\varphi)|^2
\nonumber \\ &
+
\frac{1}{\sqrt 2}
\int 
d \mathbf k d \mathbf q \,
~
e_{\theta}(\varphi)
\phiB_a( q_0,q,\varphi)
\psiB^\dagger_i( k_0+q_0,\deltaq,\thetaq)
T^{a,i}_{j}
\psiB^j(k_0,\delta,\theta)
\nonumber
\\
&
+
\int 
d \mathbf k  d \mathbf k' d \mathbf q \,
V^{i'j'}_{ij} (\theta',\theta;\vec q)
\psiB_{i'}^{\dagger}\left(\frac{q_0}{2}-k_0',\Delta(\delta',\theta'+\pi,\vec q/2),\Theta(\theta'+\pi,\vec q /2)\right)
\psiB_{j'}^{\dagger}\left(\frac{q_0}{2}+k_0',\Delta(\delta',\theta',\vec q/2),\Theta(\theta',\vec q /2)\right)
\nn & ~~ 
\times 
\psiB^j\left(\frac{q_0}{2}+k_0,\Delta(\delta,\theta,\vec q/2),\Theta(\theta,\vec q /2)\right)
\psiB^i\left(\frac{q_0}{2}-k_0,\Delta(\delta,\theta+\pi,\vec q/2),\Theta(\theta+\pi,\vec q /2)\right) \nn
& + \frac{u}{4} \int  d \mathbf k  d \mathbf k' d \mathbf q \, 
\phiB_a( - \mathbf q-\mathbf k')
\phiB_a(\mathbf k')
\phiB_b(\mathbf k)
\phiB_b(\mathbf q-\mathbf k). 
\label{app:action_low_energy_tangential}
\end{align}
Here, $(\deltaq, \thetasq)_F$ represents the fermionic polar coordinate of the fermionic momentum 
shifted from $(\delta,\theta)_F$ after absorbing a boson with momentum $\vec q=q(\cos\varphi, \sin \varphi)$, where
\begin{equation}
    \begin{aligned}
        \deltaq = \delta+\mathscr{F}_{\varphi,\theta}q+\mathscr G_{\varphi,\theta}\frac{q}{\KFthetadim}, 
        ~~~~~\thetasq = \theta+\mathscr A_{\varphi_\theta}\frac{q}{\KFthetadim}
    \end{aligned}
    \label{eq:DeltaTheta}
\end{equation}
with 
$\mathscr F_{\varphi,\theta} = \cos(\varphi-\theta)-\sin(\varphi-\theta)\frac{\mathbf K_{F,\theta}^\prime}{\KFthetadim}$,
$\mathscr{G}_{\varphi,\theta} =  
      \frac{1}{2}\left( 
        \sin^2(\varphi -\theta )\left[1-\frac{{\bf K}^{\prime\prime}_{F,\theta}}{\KFthetadim}\right]+\sin\left(2\left(\varphi-\theta\right)\right)\frac{{\bf K}^{\prime}_{F,\theta}}{\KFthetadim}\right)$,
$\mathscr{A}_{\varphi,\theta} =  \sin(\varphi-\theta)$.
%
Under the particle-hole transformation defined by 
\begin{equation}
    \begin{aligned}
\psi^{j}(k_0,\delta,\theta) = \epsilon^{ji}\tilde \psi^{\dagger}_i(-k_0,-\delta,\theta+\pi),
~~
\phiB_a(q_0,q,\varphi) 
= -
\tilde 
\phiB_a(q_0,q,\varphi),
\label{app:PH_Transformation}
    \end{aligned}
\end{equation}
the action
becomes
\begin{align}
 S
&=
\int 
d \mathbf k \,
\tilde \psiB^\dagger_j(k_0,\delta,\theta
)
\Bigl[
i k_0
+
v_{F,\theta}
\delta
\Bigr]
\tilde \psiB^j(k_0,\delta,\theta)
\nonumber 
+
\frac12
\int 
d \mathbf q\,
{\bf q}^2
|\tilde \phi_a(q_0,q,\varphi)|^2
\nonumber \\ &
+
\frac{1}{\sqrt 2}
\int 
d \mathbf k d \mathbf q \,
~
 e_{\theta}(\varphi)
\tilde \phiB_a( q_0,q,\varphi)
\tilde \psiB^\dagger_i( k_0+q_0,-\delta,\theta+\pi)
T^{a,i}_{j}
\tilde \psiB^j(k_0,-\deltaq,\thetasq+\pi)
\nonumber
\\
&
+
\int 
d \mathbf k  d \mathbf k' d \mathbf q \,
\tilde V^{i'j'}_{ij} (\theta',\theta;\vec q)
\tilde \psiB_{i'}^{\dagger}\left(\frac{q_0}{2}-k_0',\bar \Delta(\delta',\theta'+\pi,\vec q/2),\Theta(\theta'+\pi,\vec q /2)\right)
\tilde \psiB_{j'}^{\dagger}\left(\frac{q_0}{2}+k_0',\bar \Delta(\delta',\theta',\vec q/2),\Theta(\theta',\vec q /2)\right)
\nn & ~~ 
\times 
\tilde \psiB^j\left(\frac{q_0}{2}+k_0,\bar \Delta(\delta,\theta,\vec q/2),\Theta(\theta,\vec q /2)\right)
\tilde\psiB^i\left(\frac{q_0}{2}-k_0,\bar \Delta(\delta,\theta+\pi,\vec q/2),\Theta(\theta+\pi,\vec q /2)\right) \nn
& + \frac{u}{4} \int  d \mathbf k  d \mathbf k' d \mathbf q \, 
\tilde \phiB_a( - \mathbf q-\mathbf k')
\tilde\phiB_a(\mathbf k')
\tilde\phiB_b(\mathbf k)
\tilde\phiB_b(\mathbf q-\mathbf k), 
\label{app:action_low_energy_tangential_transformed}
\end{align}
where 
$\tilde V^{i'j'}_{ij} (\theta',\theta;\vec q) = V^{i'j'}_{ij}(\theta + \pi,\theta' + \pi;-\vec q)$
and 
$\bar \Delta(\delta, \theta, \vec q) = \delta-\mathscr{F}_{\varphi,\theta}q-\mathscr G_{\varphi,\theta}\frac{q}{\KFthetadim}$.
Unlike the usual particle-hole transformation used in relativistic field theories,
$\psi^i(k_0,\delta,\theta)$
and
$\tilde \psi^i(k_0,\delta,\theta)$
do not carry the same spatial momentum for $\delta \neq 0$ because 
\eq{app:PH_Transformation} maps
a particle above the Fermi surface to a hole below the Fermi surface on the opposite side.
However, the discrepancy, which is of order $\delta$, is sub-leading compared to the Fermi momentum in the large $\KFAVdim$ limit.
The subsequent time-reversal transformation gives
\begin{align}
 S
&=
\int 
d \mathbf k \,
\tilde\psiB^\dagger_j(k_0,\delta,\theta
)
\Bigl[
i k_0
+
v_{F,\theta}
\delta
\Bigr]
\tilde \psiB^j(k_0,\delta,\theta)
\nonumber 
+
\frac12
\int 
d \mathbf q\,
{\bf q}^2
|\tilde \phi_a(q_0,q,\varphi)|^2
\nonumber \\ &
+
\frac{1}{\sqrt 2}
\int 
d \mathbf k d \mathbf q \,
~
e_{\theta}(\varphi)
\tilde\phiB_a( q_0,q,\varphi)
\tilde\psiB^\dagger_i( k_0+q_0,
\bar \Delta(\delta,\theta,\vec q)
,\thetasq)
T^{a,i}_{j}
\psiB^j(k_0,\delta,\theta)
\nonumber
\\
&
+
\int 
d \mathbf k  d \mathbf k' d \mathbf q \,
V^{i'j'}_{ij} (\theta',\theta;\vec q)
\tilde \psiB_{i'}^{\dagger}\left(\frac{q_0}{2}-k_0',\bar \Delta(\delta',\theta'+\pi,\vec q/2),\Theta(\theta'+\pi,\vec q /2)\right)
\tilde\psiB_{j'}^{\dagger}\left(\frac{q_0}{2}+k_0',\bar \Delta(\delta',\theta',\vec q/2),\Theta(\theta',\vec q /2)\right)
\nn & ~~ 
\times 
\tilde\psiB^j\left(\frac{q_0}{2}+k_0,\bar\Delta(\delta,\theta,\vec q/2),\Theta(\theta,\vec q /2)\right)
\tilde \psiB^i\left(\frac{q_0}{2}-k_0,\bar \Delta(\delta,\theta+\pi,\vec q/2),\Theta(\theta+\pi,\vec q /2)\right) \nn
& + \frac{u}{4} \int  d \mathbf k  d \mathbf k' d \mathbf q \, 
\tilde\phiB_a( - \mathbf q-\mathbf k')
\tilde\phiB_a(\mathbf k')
\tilde \phiB_b(\mathbf k)
\tilde \phiB_b(\mathbf q-\mathbf k).
\label{app:action_PH_TR}
\end{align}
\eq{app:action_PH_TR} is identical to the original action except that $\deltaq$ in \eq{eq:DeltaTheta}  
is replaced with
$\bar \Delta(\delta,\theta,\vec q)$.
This difference is not important for the four-fermion coupling because
$\bar \Delta(\delta,\theta,\vec q=0)= \Delta(\delta,\theta,\vec q=0)$,
and
only $V^{i'j'}_{ij} (\theta',\theta;\vec q=0)$
contributes to the correlation function of Cooper pairs in the large $\KFAVdim$ limit.
In the fermion-boson coupling, $\vec q$ is almost tangential to the Fermi surface at low energies and $\varphi \approx \varphi_\theta$,
where $\varphi_\theta$ is the direction of the vector tangential to the Fermi surface at angle $\theta$.
As a result,
$\bar \Delta(\delta, \theta, \vec q) 
\approx \delta+q\Delta\varphi F_\theta-G_\theta \frac{q^2}{\KFthetadim}$, where
$\Delta \varphi = \varphi-\varphi_\theta$ is of the order of 
$\sqrt{\delta/\KFAVdim} < \sqrt{\Lambda/\KFAVdim}$ 
for typical momentum $q$ that scatters fermions within the shell of thickness $\delta$ near the Fermi surface,
$F_{\theta} = \sin( \varphi_\theta-\theta )+\cos( \varphi_\theta-\theta )\frac{\KFthetadim^{\prime}}{\KFthetadim }$
and
$G_{\theta} = \frac{1}{2}\left(\sin^2(\varphi_\theta -\theta )\left[1-\frac{\KFthetadim^{\prime\prime}}{\KFthetadim }\right]+ \sin\left(2\left(\varphi_\theta-\theta\right)\right) \frac{\KFthetadim^{\prime}}{\KFthetadim }\right)$.
The change of variable 
$\varphi \rightarrow -\varphi+2 \varphi_\theta + \frac{2G_\theta}{F_\theta}\frac{q}{\KFthetadim}$
in the fermion-boson coupling 
maps $\bar \Delta(\delta,\theta,\varphi)$ back to $\deltaq \approx \delta -q\Delta\varphi F_\theta+G_\theta \frac{q^2}{\KFthetadim}$ at the expense of shifting the angular variable in the bosonic field as
$\tilde \phi\left(q_0,q,\varphi \right) \rightarrow
\tilde \phi \left(
q_0,q,
\varphi+\frac{2G_\theta}{F_\theta}
\frac{q}{\KFthetadim}
-2\Delta \varphi \right)$. 
Since $\Delta \varphi$ and $q/\KFAVdim$ are small in the large $\KFAVdim$ limit,
we can use
$\tilde \phi\left(q_0,q,\varphi+\frac{2G_\theta}{F_\theta}\frac{q}{\KFthetadim}-2\Delta\varphi\right) \approx \tilde \phi(q_0,q,\varphi)$.
This shows that the theory with the four-fermion coupling projected to the $\vec q=0$ sector is invariant under the particle-hole transformation combined with time reversal 
to the leading order in the large $\KFAVdim$ limit.

\section{Completeness of the flow of diagonal couplings}
\label{app:diagonal_completeness}

\begin{figure}[h]
\includegraphics[width=0.4\textwidth]{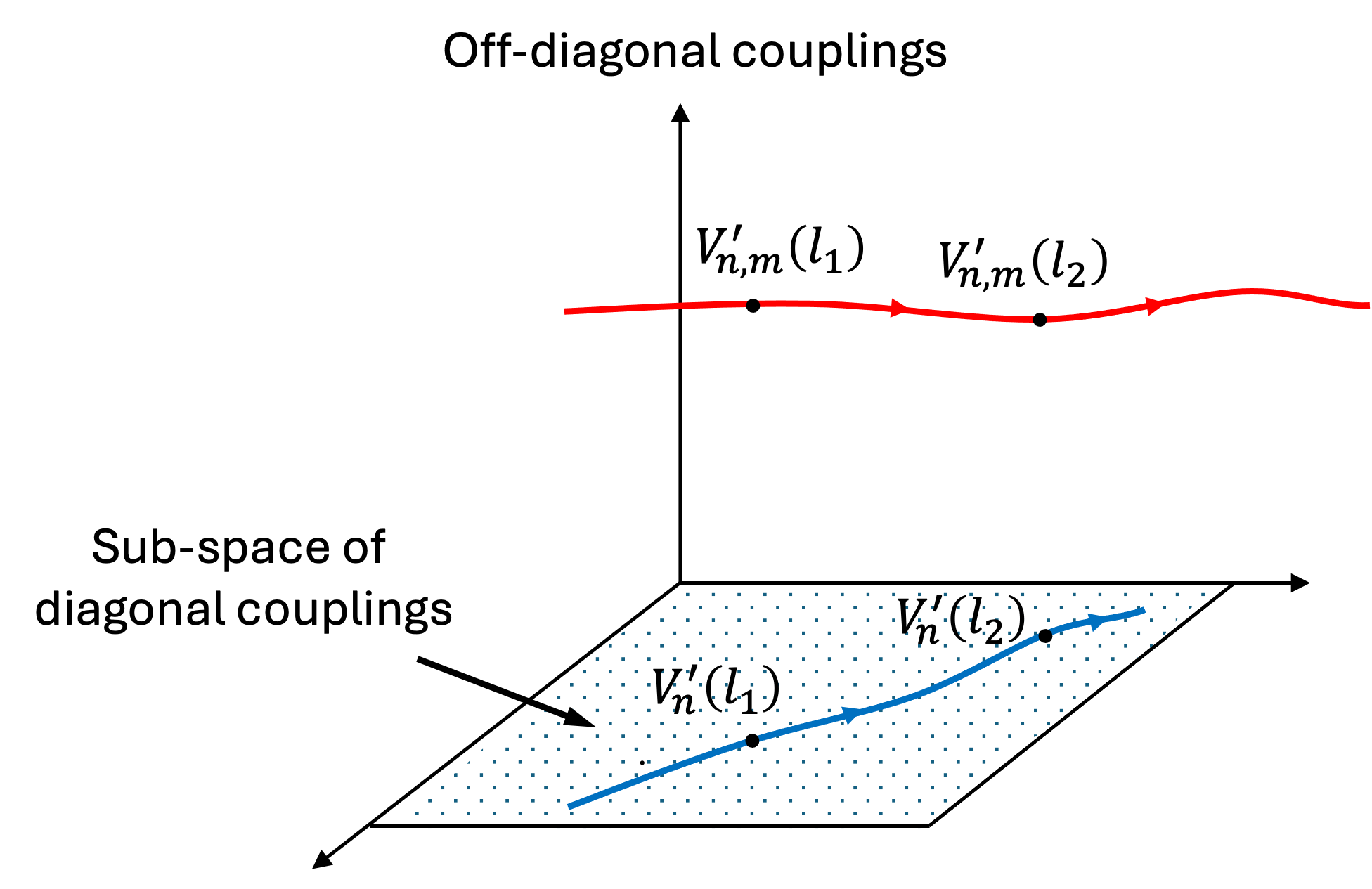}
\caption{
In general, the RG flow maps a coupling matrix with off-diagonal elements to renormalized coupling matrix with off-diagonal elements.
However, this map can be fully constructed from the map restricted on diagonal coupling matrix.
}
\label{fig:diagonal_complete}
\end{figure}

To see this, we first note that
the RG flow is completely specified by the $2\times 2$ matrix 
$u^{(n)}(l_2;l_1)$ for each $n$. 
Not all elements of the matrix are independent because
\bqa
u_{11}^{(n)}(l_2;l_1)
u_{22}^{(n)}(l_2;l_1)
-
u_{12}^{(n)}(l_2;l_1)
u_{21}^{(n)}(l_2;l_1) =1,
\label{eq:detu}
\eqa
which follows from the fact that the exponent in \eq{eq:uabn} is traceless.
The three independent elements of 
$u^{(n)}(l_2;l_1)$ can be fixed from the solutions for 
three sets of diagonal couplings. 
Suppose $\av_n^{[i]}(l_1)$  with $i=1,2,3$ represents three choices of diagonal couplings in the angular momentum channel $n$ at scale $l_1$, 
and $\av_n^{[i]}(l_2)$ are the renormalized couplings at $l_2 > l_1$ that arise from \eq{app:RiccatiSolution0}.
Then, $u_{ab}^{(n)}(l_2;l_1)$ satisfies
\bqa
&&
\left(
\begin{array}{cccc}
\av_n^{[1]}(l_2) ~&~ \av_n^{[1]}(l_2) \av_n^{[1]}(l_1) ~&~ -1 ~&~  -\av_n^{[1]}(l_1)  \\
\av_n^{[2]}(l_2) ~&~ \av_n^{[2]}(l_2) \av_n^{[2]}(l_1) ~&~ -1 ~&~  -\av_n^{[2]}(l_1) \\
\av_n^{[3]}(l_2) ~&~ \av_n^{[3]}(l_2) \av_n^{[3]}(l_1) ~&~ -1 ~&~  -\av_n^{[3]}(l_1) 
\end{array}
\right) 
\left(
\begin{array}{c}
u_{11}^{(n)}(l_2;l_1) \\
u_{12}^{(n)}(l_2;l_1) \\
u_{21}^{(n)}(l_2;l_1) \\
u_{22}^{(n)}(l_2;l_1) 
\end{array}
\right)=0.
\label{eq:threechoices}
\eqa
Because 
$u^{(n)}(l_3;l_1)
=
u^{(n)}(l_3;l_2) 
u^{(n)}(l_2;l_1)$,
it is enough to find 
$u^{(n)}(l_2;l_1)$
for a small step size $l_2-l_1$.
For a small $l_2-l_1$, one can always choose  
$\av_n^{[i]}(l_1)$ 
such that 
the $3 \times 4$ matrix 
in \eq{eq:threechoices}
is non-degenerate.
We can freely choose either
$\av_n^{[i]}(l_1)$
or 
$\av_n^{[i]}(l_2)$
for each $i$.
For example,
we choose
$\av_n^{[1]}(l_1)=0$,
$\av_n^{[2]}(l_1)=\infty$
and
$\av_n^{[3]}(l_2)= -\infty$.
For a small but non-zero $l_2-l_1$,
$\av_n^{[1]}(l_2)$,
$\av_n^{[2]}(l_2)$
and
$\av_n^{[3]}(l_1)$ are finite,
and
the matrix becomes
$\left(
\begin{array}{cccc}
\av_n^{[1]}(l_2) ~&~ 0  ~&~ -1 ~&~  0  \\
0 ~&~ \av_n^{[2]}(l_2)  ~&~ 0 ~&~  - 1 \\
-1 ~&~ -\av_n^{[3]}(l_1) ~&~ 0 ~&~  
0 
\end{array}
\right) 
$ up to a rescaling of each row vector.
This matrix is non-degenerate,
and
Eqs. \eqref{eq:detu}
and 
\eqref{eq:threechoices} completely determine 
$u_{ab}^{(n)}(l_2;l_1)$.
Therefore, the RG trajectories in the subspace of the diagonal couplings fix $u^{(n)}(l_2;l_1)$, which then fix the flow of general couplings, as illustrated in \fig{fig:diagonal_complete}.

\section{
Superconducting transition temperature
}
\label{sec:app_exp}

In this section, we compute $T_{c,n}$ analytically in the limits of large and small $n$. This allows us to extract its superuniversal features that are entirely controlled by the asymptotic coefficients $C_{\pm\infty}$ and superuniversality class. The results from this appendix are used in the main text to distinguish all seven classes on the basis of their universal phase diagrams.

\subsection*{Solutions of the beta functional with constant coefficients}

The computation of $T_{c,n}$ involves finding the solutions to \eq{eq:general_fp_eq_log}, which is complicated by the fact that the coefficients are $y$-dependent.
However, the superuniversal features of $T_{c,n}$, which are independent of the details of the underlying theory, are mainly controlled by the solutions in the asymptotic regions.
We define the large (small) $y$ asymptotic regions to be 
$y > w_L$
($y < w_S$)
in which
$C_y\approx C_{\infty}$
($C_y\approx C_{-\infty}$).
In the intermediate region between $w_S$ and $w_L$, $q_\parallel(l) \sim 2\pi \KFAVdim/n$ and $C_y$ changes significantly, causing active crossovers.
We note that the crossover occurs in the same window of $w_S< y^{(n)}(l) < w_L$ for all $n$, which means that it occurs at lower energy scales for larger $n$.
Therefore, the $y$ coordinate is partitioned into three distinct intervals,
\begin{center}
\begin{tikzpicture}
  \draw[thick,->] (0,0) -- (9,0) node[right] {$y$};
  \node at (1.67,0.55) {$q_{\parallel}(l) \ll \KFAVdim/n$};
  \node at (4.63,0.55) {$q_{\parallel}(l) \sim \bm{k}_{\mathrm{F}}/n$};
  \node at (7.46,0.55) {$q_{\parallel}(l) \gg \bm{k}_{\mathrm{F}}/n$};
  \draw[thick] (3.16,0.12) -- (3.16,-0.12) node[below] {$w_S$};
  \draw[thick] (6.04,0.12)  -- (6.04,-0.12)  node[below] {$w_L$};
\end{tikzpicture}
\end{center}
with the left and right intervals extending to $-\infty$ and $\infty$, respectively. 
In the following discussions, we focus on extracting the behaviours of $T_{c,n}$ by computing $y^*_{-\infty}$, the $y$ coordinate at which the coupling diverges to $-\infty$,
when $y^*_{-\infty} < w_S$ and when $y^*_{-\infty} > w_L$. 
These are particularly interesting and useful because the results are applicable to all non-Fermi liquids, irrespective of the details of the intermediate zone.

As a preliminary step, we first consider the PFP that arises when the discriminant is completely independent of $y$.
For $C_{y} = C \equiv  (\kappa,h,s)$, the general solution of  \eq{eq:general_fp_eq_log} is given by 
\bqa
\aV_y = -\frac{1}{4 \kappa} \begin{cases}
    \sqrt{\eta_P} \tanh\Big[\frac{1}{2}\sqrt{\eta_P} (y-y_i) - \text{arctanh}\Big(\frac{4\kappa \aV_{y_i} + 4h}{\sqrt{\eta_P}}\Big) \Big]+ 4h \quad &\text{for } \eta_P > 0,\\[0.5ex]
    2\Big[(y-y_i) - \frac{1}{2}\frac{1}{\kappa \aV_{y_i} + h} \Big]^{-1}+4h &\text{for }\eta_P = 0 , \\[0.5ex]
    -\sqrt{-\eta_P } \tan\Big[ \frac{1}{2}\sqrt{-\eta_P} (y-y_i) + \arctan \Big(\frac{4\kappa \aV_{y_i} + 4h}{\sqrt{-\eta_P}} \Big)  \Big] +4h&\text{for }\eta_P <0,
\end{cases}
\label{eq:tv_constdisc}
\eqa
with $\eta_P \equiv 16(h^2 + \kappa s)$. From this, it follows that the PFP that starts at $(y_i, \aV_{y_i})$ diverges to $-\infty$ at
\bqa
y_{-\infty}^*-y_i = \begin{cases}
\frac{2}{\sqrt{\eta_P}} \text{arcoth}\Big(\frac{4\kappa \aV_{y_i} + 4h}{\sqrt{\eta_P}} \Big) \quad &\text{for } \eta_P > 0 , \,\, \aV_{y_i} < \frac{-4h-\sqrt{\eta_P}}{4\kappa}, \\[0.5ex]
\frac{1}{2}\frac{1}{\kappa \aV_{y_i} + h} &\text{for } \eta_P = 0,\,\, \aV_{y_i} < -\frac{h}{\kappa}, \\[0.5ex]
-\frac{2}{\sqrt{-\eta_P} } \Big[\frac{\pi}{2} + \arctan\Big(\frac{4\kappa \aV_{y_i} + 4h}{\sqrt{-\eta_P}}  \Big)  \Big] &\text{for } \eta_P < 0, 
\end{cases}.
\label{eq:ysc_constdisc}
\eqa


\subsection*{Calculation of superuniversal behaviour of $T_{c,n}$}

First, we compute $T_{c,n}$ in the small angular momentum limit 
with $ y^{(n)}(0) < w_S$.
Using \eq{eq:ysc_constdisc} with $\eta_P  = \eta_{P,-\infty}$, 
we obtain $\log(T_{c,n}/\Lambda) = 2z(y^*_{-\infty}-y^{(n)}(0))$ given by 
\bqa
\log\Big(\frac{T_{c,n}}{\Lambda} \Big) = -2z \begin{cases}
\frac{2}{\sqrt{\eta_{P,-\infty}}} \text{arcoth}\Big[1 + \frac{4\kappa_{-\infty}}{\sqrt{\eta_{P,-\infty}}}\Big(\aV^\circ_{-\infty}-\av_n(0) \Big) \Big] \quad &\text{for } \eta_{P,-\infty} > 0 ,\, \, \av_n(0) < \aV^\circ_{-\infty}; \\[0.5ex]
\frac{1}{2\kappa_{-\infty}\big(\aV^\halfominus_{-\infty}- \av_n(0)\big)} &\text{for } \eta_{P,-\infty} = 0,\,\,  \av_n(0) < \aV^\halfominus_{-\infty}; \\[0.5ex]
\frac{2}{\sqrt{-\eta_{P,-\infty} }} \Big\{\frac{\pi}{2} + \arctan\Big[\frac{4\kappa_{-\infty}}{\sqrt{-\eta_{P,-\infty}}}\Big(\av_n(0) + \frac{h_{-\infty}}{\kappa_{-\infty}}\Big)  \Big]  \Big\} &\text{for } \eta_{P,-\infty} < 0 
\end{cases}.
\label{eq:Tcn_smalln}
\eqa

Next, we consider the case in which $y^{(n)}(0) > w_L$. 
Even if $y^{(n)}(0)$ is in the large $y$ asymptotic region, the coupling does not necessarily diverge in the large $y$ asymptotic region.
Therefore, we focus on two limiting cases that exhibit superuniversal behaviours. 
The first is when the coupling does indeed diverge in the large $y$ asymptotic region: $y^*_{-\infty} > w_L$. 
This happens only when $\av_{n}(0)$ is sufficiently attractive. 
Setting $y^*_{-\infty} = w_L$ defines the crossover coupling $\Vcrl$ below which $y^*_{-\infty} > w_L$,
\bqa
\Vcrl = -\frac{1}{4\kappa_\infty} \Big\{ \sqrt{\eta_{P,\infty} }\coth\Big[\frac{1}{2}\sqrt{\eta_{P,\infty}}(y^{(n)}(0) - w_L) \Big]  + 4h_{\infty}\Big\}.
\label{eq:V*_L}
\eqa
In the $y^{(n)}(0) \to \infty$ limit, $\Vcrl \to \aV^\circ_{\infty}$, as expected.
In this case, $T_{c,n}$ is obtained from \eq{eq:Tcn_smalln} by replacing $\eta_{P,-\infty}$ with $ \eta_{P,\infty}$  and $\aV^\circ_{-\infty} $ with $ \aV^\circ_{\infty}$, 
\bqa
\log\Big(\frac{T_{c,n}}{\Lambda} \Big) = -\frac{4z}{\sqrt{\eta_{P,\infty}}} \text{arcoth}\Big[1 + \frac{4\kappa_{\infty}}{\sqrt{\eta_{P,\infty}}}\Big(\aV^\circ_{\infty}-\av_n(0) \Big) \Big] \quad &\text{for }  \av_n(0) < \Vcrl.\label{eq:Tcn_L}
\eqa
The second case we consider is when the coupling diverges at $y^*_{-\infty} < w_S$ for $y^{(n)}(0) > w_L$. 
This occurs if the bare coupling is close to the separatrix.
To compute $T_{c,n}$ in this regime analytically, one requires \textit{some} details about the intermediate region in $y$. 
However, for our current purpose of extracting the superuniversal form, we focus only on the fate of couplings that are close to the separatrix PFP prior to entering the intermediate region around $y^{(n)}(l) = w_L$.

\begin{figure}[h]
\centering
\begin{subfigure}{0.45 \textwidth}
\includegraphics[width=\textwidth]{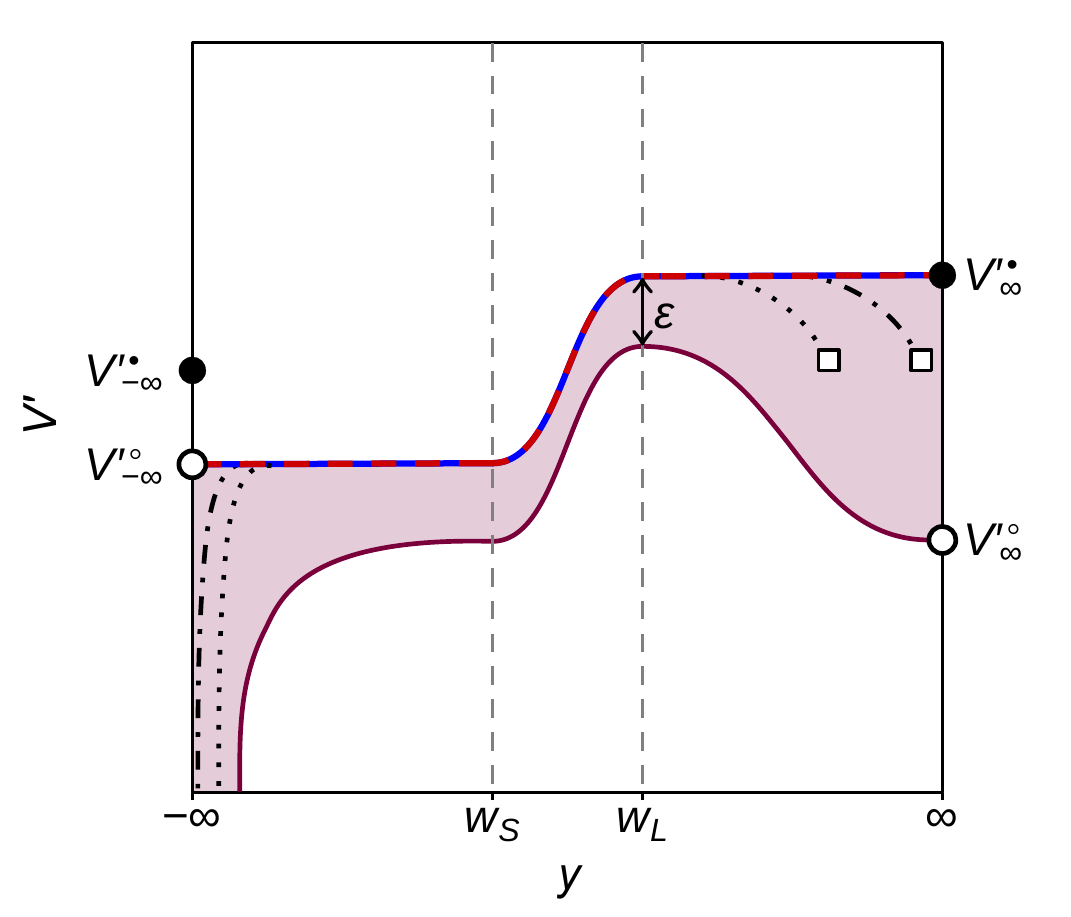}
\caption{}
\label{fig:ab_calc}
 \end{subfigure}
\begin{subfigure}{0.45 \textwidth}
\includegraphics[width=\textwidth]{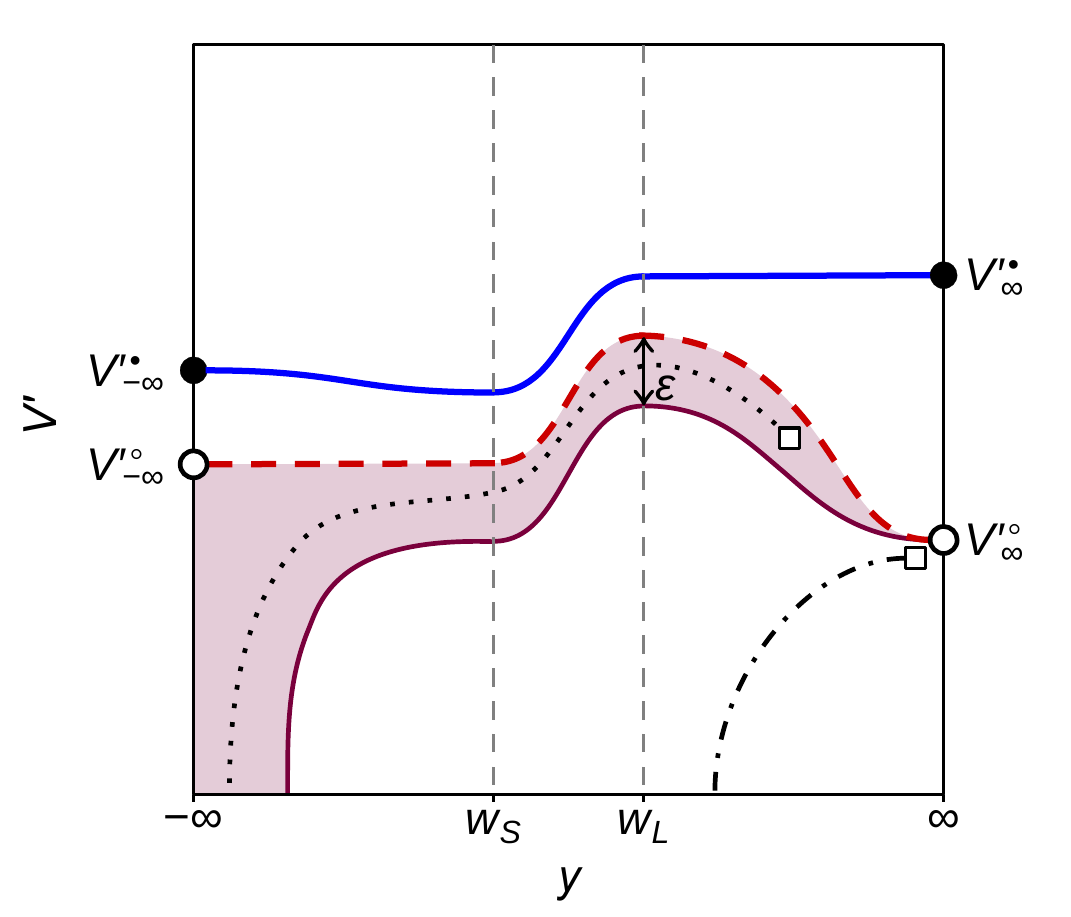}
\caption{}
\label{fig:a_calc}
 \end{subfigure}
\caption{
(a) In class AB, the separatrix connects $\aV^\bullet_\infty$ and $\aV^\circ_{-\infty}$.
Bare couplings placed below the separatrix diverges to $-\infty$ at low energies, but $T_{c,n}$ is suppressed  for large $n$ as the coupling first get attracted close to  the separatrix  in the large $y$ asymptotic region before it grows more attractive in the small $y$ asymptotic region.
The shaded region denotes the collection of PFPs that approaches the separatrix within a small distance $\varepsilon$ at $y=w_L$.
As $y^{(n)}(0)$ is increased, the shaded region fills the whole range of bare coupling between $\aV^\bullet_\infty$ and $\aV^\circ_\infty$.
A similar figure applies to class ABC as well.
(b) In class A, the separatrix connects two unstable asymptotic fixed points.
Consequently, the bare coupling placed below the separatrix begins to grow more attractive immediately.
Hence, $T_{c,n}$ is not suppressed at large $n$,
and the shaded region shrinks at large $y^{(n)}(0)$.
 A similar figure holds for class AC.
}
    \label{fig:ab_and_a_calc}
\end{figure}

First, let us consider classes AB and ABC. 
We recall that, in classes AB and ABC, the separatrix PFP saturates to the stable asymptotic fixed point at large $y > w_L$. 
To facilitate the discussion, we consider the PFP that reaches $y=w_L$ with coupling 
$\aV^S_{w_L} -  \varepsilon \frac{\sqrt{\eta_{P,\infty}}}{4\kappa_{\infty}}$,
where 
$\aV^S_y$ is the separatrix PFP and $\varepsilon$ is a small constant.
The renormalized coupling
reaches between
$\aV^S_{w_L} -   \varepsilon \frac{\sqrt{\eta_{P,\infty}}}{4\kappa_{\infty}} $
and
$\aV^S_{w_L}$ at $w_L$
for the bare coupling between 
$\Vcrs$ and the separatrix at $y^{(n)}(0)$,
where
\bqa
\Vcrs \approx -\frac{1}{4\kappa_\infty} \Big\{ \sqrt{\eta_{P,\infty} }\tanh\Big[\frac{1}{2}\sqrt{\eta_{P,\infty}}(y^{(n)}(0)-w_L) - \text{arctanh}(1-\varepsilon) \Big]   + 4h_{\infty}\Big\}.
\label{eq:V*_S}
\eqa
For any $2 > \varepsilon>0$,
\eq{eq:V*_S} reaches 
$\aV^\circ_{\infty}$ in the large $y^{(n)}(0)$ limit.
In other words, 
all couplings between
$\aV^\circ_{\infty}$ 
and
$\aV^\bullet_{\infty}$ 
in the large $y^{(n)}(0)$ limit
flow arbitrarily close to the separatrix at $w_L$.
This is illustrated in \fig{fig:ab_calc} for class AB, but a similar figure holds for class ABC.

For a generic bare coupling $\av_n(0)>\Vcrl$ with $n$ satisfying $y^{(n)}(0) > w_L$, the renormalized coupling at $w_L$ is finite and given by
\bqa
\aV_{w_L} \approx -\frac{1}{4\kappa_\infty} \bigg\{ \sqrt{\eta_{P,\infty}} \tanh \bigg[ \frac{1}{2}\sqrt{\eta_{P,\infty}} (w_L - y^{(n)}(0)) - \text{arctanh}\bigg(\frac{4\kappa_{\infty} \av_n(0) + 4h_\infty}{\sqrt{\eta_{P,\infty}}}\bigg) \bigg] + 4h_{\infty}\bigg\}.
\eqa
For large $y^{(n)}(0)$, the deviation from the separatrix is small:
$\aV_{w_L} \approx \aV^S_{w_L} + \delta \aV_{w_L}$,
where
\bqa
\delta \aV_{w_L} = -\frac{1}{2\kappa_\infty} \sqrt{\eta_{P,\infty}} e^{-\sqrt{\eta_{P,\infty}}(y^{(n)}(0) - w_L)}\Bigg(\frac{\aV^\bullet_{\infty}-\av_n(0)}{\av_n(0) - \aV^\circ_{\infty}}\Bigg). 
\label{eq:dtvwl}
\eqa
One can track the evolution of the small deviation from the separatrix between $w_L$ and $w_S$ using the linearized flow equation, $-\partial_y \delta \aV_y  =  -4 (\kappa_y\aV^S_y + h_y)\delta \aV_y +\order{\delta\aV^2}$. At $y=w_S$, the deviation from the separatrix becomes
\bqa
\delta \aV_{w_S} \approx A_{w_S;w_L}^S\delta \aV_{w_L},
\label{eq:tvs_tvl}
\eqa
where 
$A_{w_S;w_L}^S = \exp \big\{ -4\int_{w_S}^{w_L} \text{d}y \, (\kappa_y \aV^S_y + h_y)\big\}$.
As the coupling enters the small-$y$ asymptotic region, it obtains the value 
\bqa
\aV_{w_S} \approx \begin{cases}
\aV^\circ_{-\infty}+ \delta \aV_{w_S} \quad &\text{for Class AB},\\
   \aV^\halfominus_{-\infty} + \delta \aV_{w_S} \quad &\text{for Class ABC}.
\end{cases}\label{eq:smally_tv}
\eqa
The subsequent flow in $y<w_S$ can be readily obtained by imposing the initial condition $\aV_{y_i} = \aV_{w_S}$ from \eq{eq:smally_tv} at  $y_i = w_S$ in \eq{eq:ysc_constdisc}. 
From \eq{eq:tvs_tvl} and \eq{eq:dtvwl}, we conclude that the PFP that starts at $(y^{(n)}(0), \av_n(0))$ diverges to $-\infty$ at
\bqa
y_{-\infty}^* = w_S - \begin{cases}
\frac{2}{\sqrt{\eta_{P,-\infty}}} \text{arcoth}\Big[1 +  2A_{w_S;w_L}^S\frac{\sqrt{\eta_{P,\infty}}}{\sqrt{\eta_{P,-\infty}}}e^{-\sqrt{\eta_{P,\infty}} (y^{(n)}(0)-w_L)}\Big(\frac{\aV^\bullet_{\infty}-\av_n(0)}{\av_n(0) - \aV^\circ_{\infty}}\Big) \Big] \quad &\text{for Class AB} ,\\[0.5ex]
\frac{1}{A_{w_S;w_L}^S} \frac{e^{\sqrt{\eta_{P,\infty}} (y^{(n)}(0)-w_L) }}{\sqrt{\eta_{P,\infty}}}\Big(\frac{\av_n(0) - \aV^\circ_{\infty}}{\aV^\bullet_{\infty}-\av_n(0)}\Big) &\text{for Class ABC} ,
\end{cases}
\label{eq:ysc_ababc}
\eqa
$T_{c,n}$ is then given in terms of $n$ and $\av_n(0)$ by
\bqa
\frac{1}{2z}\log(\frac{T_{c,n}}{\Lambda})  = \frac{1}{2z} \log(\frac{T_0}{\Lambda}) - \log(n) -\begin{cases} 
\frac{2}{\sqrt{\eta_{P,-\infty}}} \text{arcoth}\Big[1 + \frac{2}{\mathcal{C}}n^{-\sqrt{\eta_{P,\infty}}}\Big(\frac{\aV^\bullet_{\infty}-\av_n(0)}{\av_n(0) - \aV^\circ_{\infty}}\Big) \Big]  \quad &\text{for Class AB}, \\[0.5ex]
\frac{\mathcal{D}}{2z}n^{\sqrt{\eta_{P,\infty}}}\Big(\frac{\av_n(0) - \aV^\circ_{\infty}}{\aV^\bullet_{\infty}-\av_n(0)}\Big) &\text{for Class ABC},
\end{cases}
\eqa
where 
$\mathcal{D} \equiv \frac{2ze^{-\sqrt{\eta_{P,\infty}}w_L}}{\sqrt{\eta_{P,\infty}} A_{w_S;w_L}^S}\Big(\frac{q_\parallel(0)}{\KFAVdim}\Big)^{\sqrt{\eta_{P,\infty}}}$,
$\mathcal{C} \equiv \frac{\sqrt{\eta_{P,-\infty}}}{2z}\mathcal{D}$
and
$T_0 \equiv \Lambda e^{2zw_S}\Big(\frac{\KFAVdim}{q_\parallel(0)}\Big)^{2z} $.

Let us now consider the same case of $y^{(n)}(0) > w_L$ and $y^*_{-\infty} < w_S$ for classes A and AC. 
In these classes, the separatrix approaches the unstable asymptotic fixed point in the large-$y$ region. 
Consequently, the bare coupling placed below the separatrix at large $y^{(n)}(0)$ immediately grows more attractive under the RG flow.
In order for the renormalized coupling to stay close to the separatrix between 
$\aV^S_{w_L}$ 
and
$\aV^S_{w_L} -   \varepsilon \frac{\sqrt{\eta_{P,\infty}}}{4\kappa_{\infty}} $
at $w_L$,
the bare coupling at $y^{(n)}(0)$ must lie between 
$\aV^S_{y^{(n)}(0)}$ 
and
\bqa
\av_{S,n}^\star(\varepsilon,\Delta') \approx -\frac{1}{4\kappa_\infty} \bigg\{ \sqrt{\eta_{P,\infty} }\tanh\bigg[\frac{1}{2}\sqrt{\eta_{P,\infty}}\big(y^{(n)}(0)-w_L\big) - \text{arctanh}(1-\Delta' - \varepsilon) \bigg]   + 4h_{\infty}\bigg\},
\eqa
where $\Delta'$ is the gap between the separatrix PFP and the stable asymptotic fixed point at $y= w_L$.
In the large $y^{(n)}(0)$ limit, the 
window of bare couplings $\delta \av^\star_{S,n}(\varepsilon,\Delta') \equiv \aV^S_{y^{(n)}(0)} - \av^\star_{S,n}(\varepsilon,\Delta')$  for $\Delta' > 0 $ decays exponentially as
\bqa
\delta \av^{\star}_{S,n}(\varepsilon,\Delta')  
\approx \frac{\varepsilon \sqrt{\eta_{P,\infty}}}{\kappa_{\infty}\Delta'(\Delta' + \varepsilon)} e^{-\sqrt{\eta_{P,\infty}} (y^{(n)}(0) - w_L)}.
\eqa
This is shown schematically for class A in \fig{fig:a_calc}, which may be contrasted with \fig{fig:ab_calc} for class AB.
Therefore, in classes A and AC, 
$T_{c,n}$ is governed entirely by \eq{eq:Tcn_L} in the large angular momentum limit.


\section*{Acknowledgement}

This research was supported by the Natural Sciences 
and Engineering Research Council of
Canada. Research at the Perimeter Institute is supported in part by the
Government of Canada through Industry Canada, and by the Province of
Ontario through the Ministry of Research and Information.

\bibliographystyle{unsrtnat}

\bibliography{references}

\begin{thebibliography}{82}
\providecommand{\natexlab}[1]{#1}
\providecommand{\url}[1]{\texttt{#1}}
\expandafter\ifx\csname urlstyle\endcsname\relax
  \providecommand{\doi}[1]{doi: #1}\else
  \providecommand{\doi}{doi: \begingroup \urlstyle{rm}\Url}\fi

\bibitem[Landau(1957{\natexlab{a}})]{LANDAU}
L.~Landau.
\newblock The theory of a fermi liquid.
\newblock \emph{Sov. Phys. JETP}, 3:\penalty0 920, 1957{\natexlab{a}}.
\newblock URL \url{http://www.jetp.ac.ru/cgi-bin/dn/e_003_06_0920.pdf}.

\bibitem[Landau(1957{\natexlab{b}})]{LANDAU2}
L.~Landau.
\newblock Oscillations in a fermi liquid.
\newblock \emph{Sov. Phys. JETP}, 5:\penalty0 101, 1957{\natexlab{b}}.
\newblock URL \url{http://www.jetp.ac.ru/cgi-bin/dn/e_005_01_0101.pdf}.

\bibitem[Hertz(1976)]{HERTZ}
John~A. Hertz.
\newblock Quantum critical phenomena.
\newblock \emph{Phys. Rev. B}, 14:\penalty0 1165--1184, Aug 1976.
\newblock \doi{10.1103/PhysRevB.14.1165}.
\newblock URL \url{http://link.aps.org/doi/10.1103/PhysRevB.14.1165}.

\bibitem[Millis(1993)]{MILLIS}
A.~J. Millis.
\newblock Effect of a nonzero temperature on quantum critical points in itinerant fermion systems.
\newblock \emph{Phys. Rev. B}, 48:\penalty0 7183--7196, Sep 1993.
\newblock \doi{10.1103/PhysRevB.48.7183}.
\newblock URL \url{http://link.aps.org/doi/10.1103/PhysRevB.48.7183}.

\bibitem[Varma et~al.(1989)Varma, Littlewood, Schmitt-Rink, Abrahams, and Ruckenstein]{VARMALI}
C.~M. Varma, P.~B. Littlewood, S.~Schmitt-Rink, E.~Abrahams, and A.~E. Ruckenstein.
\newblock Phenomenology of the normal state of cu-o high-temperature superconductors.
\newblock \emph{Phys. Rev. Lett.}, 63:\penalty0 1996--1999, Oct 1989.
\newblock \doi{10.1103/PhysRevLett.63.1996}.
\newblock URL \url{https://link.aps.org/doi/10.1103/PhysRevLett.63.1996}.

\bibitem[Polchinski(1994)]{POLCHINSKI2}
Joseph Polchinski.
\newblock Low-energy dynamics of the spinon-gauge system.
\newblock \emph{Nuclear Physics B}, 422\penalty0 (3):\penalty0 617--633, 1994.
\newblock \doi{10.1016/0550-3213(94)90449-9}.
\newblock URL \url{https://www.sciencedirect.com/science/article/pii/0550321394904499?via%3Dihub}.

\bibitem[Lee(1989)]{PLEE1}
Patrick~A. Lee.
\newblock Gauge field, aharonov-bohm flux, and high-${T}_{c}$ superconductivity.
\newblock \emph{Phys. Rev. Lett.}, 63:\penalty0 680--683, Aug 1989.
\newblock \doi{10.1103/PhysRevLett.63.680}.
\newblock URL \url{http://link.aps.org/doi/10.1103/PhysRevLett.63.680}.

\bibitem[Altshuler et~al.(1994)Altshuler, Ioffe, and Millis]{ALTSHULER}
B.~L. Altshuler, L.~B. Ioffe, and A.~J. Millis.
\newblock Low-energy properties of fermions with singular interactions.
\newblock \emph{Phys. Rev. B}, 50:\penalty0 14048--14064, Nov 1994.
\newblock \doi{10.1103/PhysRevB.50.14048}.
\newblock URL \url{http://link.aps.org/doi/10.1103/PhysRevB.50.14048}.

\bibitem[Kim et~al.(1994)Kim, Furusaki, Wen, and Lee]{YBKIM}
Yong~Baek Kim, Akira Furusaki, Xiao-Gang Wen, and Patrick~A. Lee.
\newblock Gauge-invariant response functions of fermions coupled to a gauge field.
\newblock \emph{Phys. Rev. B}, 50:\penalty0 17917--17932, Dec 1994.
\newblock \doi{10.1103/PhysRevB.50.17917}.
\newblock URL \url{http://link.aps.org/doi/10.1103/PhysRevB.50.17917}.

\bibitem[Abanov and Chubukov(2004)]{ABANOV2}
Ar. Abanov and A.~Chubukov.
\newblock Anomalous scaling at the quantum critical point in itinerant antiferromagnets.
\newblock \emph{Phys. Rev. Lett.}, 93:\penalty0 255702, Dec 2004.
\newblock \doi{10.1103/PhysRevLett.93.255702}.
\newblock URL \url{http://link.aps.org/doi/10.1103/PhysRevLett.93.255702}.

\bibitem[Lee(2009)]{SSLEE}
Sung-Sik Lee.
\newblock Low-energy effective theory of fermi surface coupled with u(1) gauge field in $2+1$ dimensions.
\newblock \emph{Phys. Rev. B}, 80:\penalty0 165102, Oct 2009.
\newblock \doi{10.1103/PhysRevB.80.165102}.
\newblock URL \url{http://link.aps.org/doi/10.1103/PhysRevB.80.165102}.

\bibitem[Metlitski and Sachdev(2010{\natexlab{a}})]{MAX0}
Max~A. Metlitski and Subir Sachdev.
\newblock Quantum phase transitions of metals in two spatial dimensions. i. ising-nematic order.
\newblock \emph{Phys. Rev. B}, 82:\penalty0 075127, Aug 2010{\natexlab{a}}.
\newblock \doi{10.1103/PhysRevB.82.075127}.
\newblock URL \url{https://link.aps.org/doi/10.1103/PhysRevB.82.075127}.

\bibitem[Metlitski and Sachdev(2010{\natexlab{b}})]{MAX2}
Max~A. Metlitski and Subir Sachdev.
\newblock Quantum phase transitions of metals in two spatial dimensions. ii. spin density wave order.
\newblock \emph{Phys. Rev. B}, 82:\penalty0 075128, Aug 2010{\natexlab{b}}.
\newblock \doi{10.1103/PhysRevB.82.075128}.
\newblock URL \url{http://link.aps.org/doi/10.1103/PhysRevB.82.075128}.

\bibitem[Nayak and Wilczek(1994)]{NAYAK2}
Chetan Nayak and Frank Wilczek.
\newblock Renormalization group approach to low temperature properties of a non-fermi liquid metal.
\newblock \emph{Nuclear Physics B}, 430\penalty0 (3):\penalty0 534 -- 562, 1994.
\newblock ISSN 0550-3213.
\newblock \doi{https://doi.org/10.1016/0550-3213(94)90158-9}.
\newblock URL \url{http://www.sciencedirect.com/science/article/pii/0550321394901589}.

\bibitem[Schofield(1999)]{SCHOFIELD}
A.~J. Schofield.
\newblock Non-fermi liquids.
\newblock \emph{Contemporary Physics}, 40\penalty0 (2):\penalty0 95--115, 1999.
\newblock \doi{10.1080/001075199181602}.
\newblock URL \url{https://doi.org/10.1080/001075199181602}.

\bibitem[Dalidovich and Lee(2013)]{DENNIS}
Denis Dalidovich and Sung-Sik Lee.
\newblock Perturbative non-fermi liquids from dimensional regularization.
\newblock \emph{Phys. Rev. B}, 88:\penalty0 245106, Dec 2013.
\newblock \doi{10.1103/PhysRevB.88.245106}.
\newblock URL \url{http://link.aps.org/doi/10.1103/PhysRevB.88.245106}.

\bibitem[Senthil(2008)]{SENTHIL}
T.~Senthil.
\newblock Critical fermi surfaces and non-fermi liquid metals.
\newblock \emph{Phys. Rev. B}, 78:\penalty0 035103, Jul 2008.
\newblock \doi{10.1103/PhysRevB.78.035103}.
\newblock URL \url{http://link.aps.org/doi/10.1103/PhysRevB.78.035103}.

\bibitem[Mross et~al.(2010)Mross, McGreevy, Liu, and Senthil]{MROSS}
David~F. Mross, John McGreevy, Hong Liu, and T.~Senthil.
\newblock Controlled expansion for certain non-fermi-liquid metals.
\newblock \emph{Phys. Rev. B}, 82:\penalty0 045121, Jul 2010.
\newblock \doi{10.1103/PhysRevB.82.045121}.
\newblock URL \url{http://link.aps.org/doi/10.1103/PhysRevB.82.045121}.

\bibitem[Fitzpatrick et~al.(2013)Fitzpatrick, Kachru, Kaplan, and Raghu]{FITZPATRICK}
A.~Liam Fitzpatrick, Shamit Kachru, Jared Kaplan, and S.~Raghu.
\newblock Non-fermi-liquid fixed point in a wilsonian theory of quantum critical metals.
\newblock \emph{Phys. Rev. B}, 88:\penalty0 125116, Sep 2013.
\newblock \doi{10.1103/PhysRevB.88.125116}.
\newblock URL \url{http://link.aps.org/doi/10.1103/PhysRevB.88.125116}.

\bibitem[Sur and Lee(2014)]{SHOUVIK2}
Shouvik Sur and Sung-Sik Lee.
\newblock Chiral non-fermi liquids.
\newblock \emph{Phys. Rev. B}, 90:\penalty0 045121, Jul 2014.
\newblock \doi{10.1103/PhysRevB.90.045121}.
\newblock URL \url{http://link.aps.org/doi/10.1103/PhysRevB.90.045121}.

\bibitem[Schlief et~al.(2017)Schlief, Lunts, and Lee]{SCHLIEF}
Andres Schlief, Peter Lunts, and Sung-Sik Lee.
\newblock Exact critical exponents for the antiferromagnetic quantum critical metal in two dimensions.
\newblock \emph{Phys. Rev. X}, 7:\penalty0 021010, Apr 2017.
\newblock \doi{10.1103/PhysRevX.7.021010}.
\newblock URL \url{https://link.aps.org/doi/10.1103/PhysRevX.7.021010}.

\bibitem[Else et~al.(2021)Else, Thorngren, and Senthil]{PhysRevX.11.021005}
Dominic~V. Else, Ryan Thorngren, and T.~Senthil.
\newblock Non-fermi liquids as ersatz fermi liquids: General constraints on compressible metals.
\newblock \emph{Phys. Rev. X}, 11:\penalty0 021005, Apr 2021.
\newblock \doi{10.1103/PhysRevX.11.021005}.
\newblock URL \url{https://link.aps.org/doi/10.1103/PhysRevX.11.021005}.

\bibitem[Debbeler and Metzner(2023)]{PhysRevB.107.165152}
Lukas Debbeler and Walter Metzner.
\newblock Non-fermi liquid behavior at flat hot spots from quantum critical fluctuations at the onset of charge- or spin-density wave order.
\newblock \emph{Phys. Rev. B}, 107:\penalty0 165152, Apr 2023.
\newblock \doi{10.1103/PhysRevB.107.165152}.
\newblock URL \url{https://link.aps.org/doi/10.1103/PhysRevB.107.165152}.

\bibitem[Berg et~al.(2019)Berg, Lederer, Schattner, and Trebst]{annurev:/content/journals/10.1146/annurev-conmatphys-031218-013339}
Erez Berg, Samuel Lederer, Yoni Schattner, and Simon Trebst.
\newblock Monte carlo studies of quantum critical metals.
\newblock \emph{Annual Review of Condensed Matter Physics}, 10\penalty0 (Volume 10, 2019):\penalty0 63--84, 2019.
\newblock ISSN 1947-5462.
\newblock \doi{https://doi.org/10.1146/annurev-conmatphys-031218-013339}.
\newblock URL \url{https://www.annualreviews.org/content/journals/10.1146/annurev-conmatphys-031218-013339}.

\bibitem[Wilson(1971{\natexlab{a}})]{WILSONRG1}
Kenneth~G. Wilson.
\newblock Renormalization group and critical phenomena. i. renormalization group and the kadanoff scaling picture.
\newblock \emph{Phys. Rev. B}, 4:\penalty0 3174--3183, Nov 1971{\natexlab{a}}.
\newblock \doi{10.1103/PhysRevB.4.3174}.
\newblock URL \url{https://link.aps.org/doi/10.1103/PhysRevB.4.3174}.

\bibitem[Wilson(1971{\natexlab{b}})]{WILSONRG2}
Kenneth~G. Wilson.
\newblock Renormalization group and critical phenomena. ii. phase-space cell analysis of critical behavior.
\newblock \emph{Phys. Rev. B}, 4:\penalty0 3184--3205, Nov 1971{\natexlab{b}}.
\newblock \doi{10.1103/PhysRevB.4.3184}.
\newblock URL \url{https://link.aps.org/doi/10.1103/PhysRevB.4.3184}.

\bibitem[Borges et~al.(2023)Borges, Borissov, Singh, Schlief, and Lee]{BORGES2023169221}
Francisco Borges, Anton Borissov, Ashutosh Singh, Andr{\'e}s Schlief, and Sung-Sik Lee.
\newblock Field-theoretic functional renormalization group formalism for non-fermi liquids and its application to the antiferromagnetic quantum critical metal in two dimensions.
\newblock \emph{Annals of Physics}, 450:\penalty0 169221, 2023.
\newblock ISSN 0003-4916.
\newblock \doi{https://doi.org/10.1016/j.aop.2023.169221}.
\newblock URL \url{https://www.sciencedirect.com/science/article/pii/S0003491623000064}.

\bibitem[Kukreja et~al.(2024)Kukreja, Besharat, and Lee]{PhysRevB.110.155142}
Shubham Kukreja, Afshin Besharat, and Sung-Sik Lee.
\newblock Projective fixed points for non-fermi liquids: A case study of the ising-nematic quantum critical metal.
\newblock \emph{Phys. Rev. B}, 110:\penalty0 155142, Oct 2024.
\newblock \doi{10.1103/PhysRevB.110.155142}.
\newblock URL \url{https://link.aps.org/doi/10.1103/PhysRevB.110.155142}.

\bibitem[Kukreja et~al.(2026)Kukreja, Willerton, and Lee]{kukreja2026classificationnonfermiliquidsuniversal}
Shubham Kukreja, Dawson~M. Willerton, and Sung-Sik Lee.
\newblock Classification of non-fermi liquids and universal superconducting fluctuations, 2026.
\newblock URL \url{https://arxiv.org/abs/2601.21047}.

\bibitem[Kim et~al.(1995)Kim, Lee, and Wen]{PhysRevB.52.17275}
Yong~Baek Kim, Patrick~A. Lee, and Xiao-Gang Wen.
\newblock Quantum boltzmann equation of composite fermions interacting with a gauge field.
\newblock \emph{Phys. Rev. B}, 52:\penalty0 17275--17292, Dec 1995.
\newblock \doi{10.1103/PhysRevB.52.17275}.
\newblock URL \url{https://link.aps.org/doi/10.1103/PhysRevB.52.17275}.

\bibitem[Patel et~al.(2023)Patel, Guo, Esterlis, and Sachdev]{doi:10.1126/science.abq6011}
Aavishkar~A. Patel, Haoyu Guo, Ilya Esterlis, and Subir Sachdev.
\newblock Universal theory of strange metals from spatially random interactions.
\newblock \emph{Science}, 381\penalty0 (6659):\penalty0 790--793, 2023.
\newblock \doi{10.1126/science.abq6011}.
\newblock URL \url{https://www.science.org/doi/abs/10.1126/science.abq6011}.

\bibitem[Phillips et~al.(2020)Phillips, Yeo, and Huang]{Phillips:2020aa}
Philip~W. Phillips, Luke Yeo, and Edwin~W. Huang.
\newblock Exact theory for superconductivity in a doped mott insulator.
\newblock \emph{Nature Physics}, 16\penalty0 (12):\penalty0 1175--1180, 2020.
\newblock \doi{10.1038/s41567-020-0988-4}.
\newblock URL \url{https://doi.org/10.1038/s41567-020-0988-4}.

\bibitem[Son(1999)]{PhysRevD.59.094019}
D.~T. Son.
\newblock Superconductivity by long-range color magnetic interaction in high-density quark matter.
\newblock \emph{Phys. Rev. D}, 59:\penalty0 094019, Apr 1999.
\newblock \doi{10.1103/PhysRevD.59.094019}.
\newblock URL \url{https://link.aps.org/doi/10.1103/PhysRevD.59.094019}.

\bibitem[Metlitski et~al.(2015)Metlitski, Mross, Sachdev, and Senthil]{PhysRevB.91.115111}
Max~A. Metlitski, David~F. Mross, Subir Sachdev, and T.~Senthil.
\newblock Cooper pairing in non-fermi liquids.
\newblock \emph{Phys. Rev. B}, 91:\penalty0 115111, Mar 2015.
\newblock \doi{10.1103/PhysRevB.91.115111}.
\newblock URL \url{https://link.aps.org/doi/10.1103/PhysRevB.91.115111}.

\bibitem[Abanov and Chubukov(2020)]{PhysRevB.102.024524}
Artem Abanov and Andrey~V. Chubukov.
\newblock Interplay between superconductivity and non-fermi liquid at a quantum critical point in a metal. i.
\newblock \emph{Phys. Rev. B}, 102:\penalty0 024524, Jul 2020.
\newblock \doi{10.1103/PhysRevB.102.024524}.
\newblock URL \url{https://link.aps.org/doi/10.1103/PhysRevB.102.024524}.

\bibitem[Wu et~al.(2020{\natexlab{a}})Wu, Abanov, Wang, and Chubukov]{PhysRevB.102.024525}
Yi-Ming Wu, Artem Abanov, Yuxuan Wang, and Andrey~V. Chubukov.
\newblock Interplay between superconductivity and non-fermi liquid at a quantum critical point in a metal. ii.
\newblock \emph{Phys. Rev. B}, 102:\penalty0 024525, Jul 2020{\natexlab{a}}.
\newblock \doi{10.1103/PhysRevB.102.024525}.
\newblock URL \url{https://link.aps.org/doi/10.1103/PhysRevB.102.024525}.

\bibitem[Wu et~al.(2020{\natexlab{b}})Wu, Abanov, and Chubukov]{PhysRevB.102.094516}
Yi-Ming Wu, Artem Abanov, and Andrey~V. Chubukov.
\newblock Interplay between superconductivity and non-fermi liquid behavior at a quantum critical point in a metal. iii.
\newblock \emph{Phys. Rev. B}, 102:\penalty0 094516, Sep 2020{\natexlab{b}}.
\newblock \doi{10.1103/PhysRevB.102.094516}.
\newblock URL \url{https://link.aps.org/doi/10.1103/PhysRevB.102.094516}.

\bibitem[Wu et~al.(2021{\natexlab{a}})Wu, Zhang, Abanov, and Chubukov]{PhysRevB.103.024522}
Yi-Ming Wu, Shang-Shun Zhang, Artem Abanov, and Andrey~V. Chubukov.
\newblock Interplay between superconductivity and non-fermi liquid at a quantum critical point in a metal. iv.
\newblock \emph{Phys. Rev. B}, 103:\penalty0 024522, Jan 2021{\natexlab{a}}.
\newblock \doi{10.1103/PhysRevB.103.024522}.
\newblock URL \url{https://link.aps.org/doi/10.1103/PhysRevB.103.024522}.

\bibitem[Wu et~al.(2021{\natexlab{b}})Wu, Zhang, Abanov, and Chubukov]{PhysRevB.103.184508}
Yi-Ming Wu, Shang-Shun Zhang, Artem Abanov, and Andrey~V. Chubukov.
\newblock Interplay between superconductivity and non-fermi liquid behavior at a quantum-critical point in a metal. v.
\newblock \emph{Phys. Rev. B}, 103:\penalty0 184508, May 2021{\natexlab{b}}.
\newblock \doi{10.1103/PhysRevB.103.184508}.
\newblock URL \url{https://link.aps.org/doi/10.1103/PhysRevB.103.184508}.

\bibitem[Zhang et~al.(2021)Zhang, Wu, Abanov, and Chubukov]{PhysRevB.104.144509}
Shang-Shun Zhang, Yi-Ming Wu, Artem Abanov, and Andrey~V. Chubukov.
\newblock Interplay between superconductivity and non-fermi liquid at a quantum critical point in a metal. vi.
\newblock \emph{Phys. Rev. B}, 104:\penalty0 144509, Oct 2021.
\newblock \doi{10.1103/PhysRevB.104.144509}.
\newblock URL \url{https://link.aps.org/doi/10.1103/PhysRevB.104.144509}.

\bibitem[Wang et~al.(2017{\natexlab{a}})Wang, Raghu, and Torroba]{PhysRevB.95.165137}
Huajia Wang, Srinivas Raghu, and Gonzalo Torroba.
\newblock Non-fermi-liquid superconductivity: Eliashberg approach versus the renormalization group.
\newblock \emph{Phys. Rev. B}, 95:\penalty0 165137, Apr 2017{\natexlab{a}}.
\newblock \doi{10.1103/PhysRevB.95.165137}.
\newblock URL \url{https://link.aps.org/doi/10.1103/PhysRevB.95.165137}.

\bibitem[Raghu et~al.(2015)Raghu, Torroba, and Wang]{PhysRevB.92.205104}
S.~Raghu, Gonzalo Torroba, and Huajia Wang.
\newblock Metallic quantum critical points with finite bcs couplings.
\newblock \emph{Phys. Rev. B}, 92:\penalty0 205104, Nov 2015.
\newblock \doi{10.1103/PhysRevB.92.205104}.
\newblock URL \url{https://link.aps.org/doi/10.1103/PhysRevB.92.205104}.

\bibitem[Chubukov and Schmalian(2005)]{CHUBUKOVSC}
Andrey~V. Chubukov and J\"org Schmalian.
\newblock Superconductivity due to massless boson exchange in the strong-coupling limit.
\newblock \emph{Phys. Rev. B}, 72:\penalty0 174520, Nov 2005.
\newblock \doi{10.1103/PhysRevB.72.174520}.
\newblock URL \url{https://link.aps.org/doi/10.1103/PhysRevB.72.174520}.

\bibitem[Abanov et~al.(2003)Abanov, Chubukov, and Schmalian]{ABANOV3}
Ar. Abanov, Andrey~V. Chubukov, and J.~Schmalian.
\newblock Quantum-critical theory of the spin-fermion model and its application to cuprates: Normal state analysis.
\newblock \emph{Adv. Phys.}, 52\penalty0 (3):\penalty0 119--218, 2003.
\newblock \doi{10.1080/0001873021000057123}.
\newblock URL \url{http://dx.doi.org/10.1080/0001873021000057123}.

\bibitem[Wang et~al.(2017{\natexlab{b}})Wang, Schattner, Berg, and Fernandes]{PhysRevB.95.174520}
Xiaoyu Wang, Yoni Schattner, Erez Berg, and Rafael~M. Fernandes.
\newblock Superconductivity mediated by quantum critical antiferromagnetic fluctuations: The rise and fall of hot spots.
\newblock \emph{Phys. Rev. B}, 95:\penalty0 174520, May 2017{\natexlab{b}}.
\newblock \doi{10.1103/PhysRevB.95.174520}.
\newblock URL \url{https://link.aps.org/doi/10.1103/PhysRevB.95.174520}.

\bibitem[Lederer et~al.(2017)Lederer, Schattner, Berg, and Kivelson]{LEDERERREV}
Samuel Lederer, Yoni Schattner, Erez Berg, and Steven~A. Kivelson.
\newblock Superconductivity and non-fermi liquid behavior near a nematic quantum critical point.
\newblock \emph{Proceedings of the National Academy of Sciences}, 114\penalty0 (19):\penalty0 4905--4910, 2017.
\newblock ISSN 0027-8424.
\newblock \doi{10.1073/pnas.1620651114}.
\newblock URL \url{https://www.pnas.org/content/114/19/4905}.

\bibitem[Lederer et~al.(2015)Lederer, Schattner, Berg, and Kivelson]{LEDERER}
S.~Lederer, Y.~Schattner, E.~Berg, and S.~A. Kivelson.
\newblock Enhancement of superconductivity near a nematic quantum critical point.
\newblock \emph{Phys. Rev. Lett.}, 114:\penalty0 097001, Mar 2015.
\newblock \doi{10.1103/PhysRevLett.114.097001}.
\newblock URL \url{https://link.aps.org/doi/10.1103/PhysRevLett.114.097001}.

\bibitem[Fulde and Ferrell(1964)]{PhysRev.135.A550}
Peter Fulde and Richard~A. Ferrell.
\newblock Superconductivity in a strong spin-exchange field.
\newblock \emph{Phys. Rev.}, 135:\penalty0 A550--A563, Aug 1964.
\newblock \doi{10.1103/PhysRev.135.A550}.
\newblock URL \url{https://link.aps.org/doi/10.1103/PhysRev.135.A550}.

\bibitem[LeClair et~al.(2004)LeClair, Mar\'{\i}a~Rom\'an, and Sierra]{PhysRevB.69.020505}
Andr\'e LeClair, Jos\'e Mar\'{\i}a~Rom\'an, and Germ\'an Sierra.
\newblock Russian doll renormalization group and superconductivity.
\newblock \emph{Phys. Rev. B}, 69:\penalty0 020505, Jan 2004.
\newblock \doi{10.1103/PhysRevB.69.020505}.
\newblock URL \url{https://link.aps.org/doi/10.1103/PhysRevB.69.020505}.

\bibitem[Hauck et~al.(2020)Hauck, Klug, Esterlis, and Schmalian]{HAUCK2020168120}
Daniel Hauck, Markus~J. Klug, Ilya Esterlis, and J{\"o}rg Schmalian.
\newblock Eliashberg equations for an electron--phonon version of the sachdev--ye--kitaev model: Pair breaking in non-fermi liquid superconductors.
\newblock \emph{Annals of Physics}, 417:\penalty0 168120, 2020.
\newblock ISSN 0003-4916.
\newblock \doi{https://doi.org/10.1016/j.aop.2020.168120}.
\newblock URL \url{https://www.sciencedirect.com/science/article/pii/S0003491620300531}.
\newblock Eliashberg theory at 60: Strong-coupling superconductivity and beyond.

\bibitem[Li et~al.(2023)Li, Louie, and Lee]{PhysRevB.107.L041103}
Zi-Xiang Li, Steven~G. Louie, and Dung-Hai Lee.
\newblock Emergent superconductivity and non-fermi liquid transport in a doped valence bond solid insulator.
\newblock \emph{Phys. Rev. B}, 107:\penalty0 L041103, Jan 2023.
\newblock \doi{10.1103/PhysRevB.107.L041103}.
\newblock URL \url{https://link.aps.org/doi/10.1103/PhysRevB.107.L041103}.

\bibitem[Wang et~al.(2016)Wang, Abanov, Altshuler, Yuzbashyan, and Chubukov]{Wang_Chubukov_First_Matsubara}
Yuxuan Wang, Artem Abanov, Boris~L. Altshuler, Emil~A. Yuzbashyan, and Andrey~V. Chubukov.
\newblock Superconductivity near a quantum-critical point: The special role of the first matsubara frequency.
\newblock \emph{Phys. Rev. Lett.}, 117:\penalty0 157001, Oct 2016.
\newblock \doi{10.1103/PhysRevLett.117.157001}.
\newblock URL \url{https://link.aps.org/doi/10.1103/PhysRevLett.117.157001}.

\bibitem[Wu et~al.(2019{\natexlab{a}})Wu, Abanov, Wang, and Chubukov]{Wu_Chubukov_Matsubara}
Yi-Ming Wu, Artem Abanov, Yuxuan Wang, and Andrey~V. Chubukov.
\newblock Special role of the first matsubara frequency for superconductivity near a quantum critical point: Nonlinear gap equation below ${T}_{c}$ and spectral properties in real frequencies.
\newblock \emph{Phys. Rev. B}, 99:\penalty0 144512, Apr 2019{\natexlab{a}}.
\newblock \doi{10.1103/PhysRevB.99.144512}.
\newblock URL \url{https://link.aps.org/doi/10.1103/PhysRevB.99.144512}.

\bibitem[Wu et~al.(2019{\natexlab{b}})Wu, Abanov, and Chubukov]{Pairing_Wu_Chubukov}
Yi-Ming Wu, Artem Abanov, and Andrey~V. Chubukov.
\newblock Pairing in quantum critical systems: Transition temperature, pairing gap, and their ratio.
\newblock \emph{Phys. Rev. B}, 99:\penalty0 014502, Jan 2019{\natexlab{b}}.
\newblock \doi{10.1103/PhysRevB.99.014502}.
\newblock URL \url{https://link.aps.org/doi/10.1103/PhysRevB.99.014502}.

\bibitem[Chubukov et~al.(2020)Chubukov, Abanov, Wang, and Wu]{CHUBUKOV2020168142}
Andrey~V. Chubukov, Artem Abanov, Yuxuan Wang, and Yi-Ming Wu.
\newblock The interplay between superconductivity and non-fermi liquid at a quantum-critical point in a metal.
\newblock \emph{Annals of Physics}, 417:\penalty0 168142, 2020.
\newblock ISSN 0003-4916.
\newblock \doi{https://doi.org/10.1016/j.aop.2020.168142}.
\newblock URL \url{https://www.sciencedirect.com/science/article/pii/S0003491620300750}.
\newblock Eliashberg theory at 60: Strong-coupling superconductivity and beyond.

\bibitem[Lee(2008)]{PhysRevB.78.085129}
Sung-Sik Lee.
\newblock Stability of the u(1) spin liquid with a spinon fermi surface in $2+1$ dimensions.
\newblock \emph{Phys. Rev. B}, 78:\penalty0 085129, Aug 2008.
\newblock \doi{10.1103/PhysRevB.78.085129}.
\newblock URL \url{https://link.aps.org/doi/10.1103/PhysRevB.78.085129}.

\bibitem[Bardeen et~al.(1957{\natexlab{a}})Bardeen, Cooper, and Schrieffer]{BCS1}
J.~Bardeen, L.~N. Cooper, and J.~R. Schrieffer.
\newblock Microscopic theory of superconductivity.
\newblock \emph{Phys. Rev.}, 106:\penalty0 162--164, Apr 1957{\natexlab{a}}.
\newblock \doi{10.1103/PhysRev.106.162}.
\newblock URL \url{https://link.aps.org/doi/10.1103/PhysRev.106.162}.

\bibitem[Bardeen et~al.(1957{\natexlab{b}})Bardeen, Cooper, and Schrieffer]{BCS2}
J.~Bardeen, L.~N. Cooper, and J.~R. Schrieffer.
\newblock Theory of superconductivity.
\newblock \emph{Phys. Rev.}, 108:\penalty0 1175--1204, Dec 1957{\natexlab{b}}.
\newblock \doi{10.1103/PhysRev.108.1175}.
\newblock URL \url{https://link.aps.org/doi/10.1103/PhysRev.108.1175}.

\bibitem[Shankar(1994)]{SHANKAR}
R.~Shankar.
\newblock Renormalization-group approach to interacting fermions.
\newblock \emph{Rev. Mod. Phys.}, 66:\penalty0 129--192, Jan 1994.
\newblock \doi{10.1103/RevModPhys.66.129}.
\newblock URL \url{http://link.aps.org/doi/10.1103/RevModPhys.66.129}.

\bibitem[Raghu et~al.(2010)Raghu, Kivelson, and Scalapino]{PhysRevB.81.224505}
S.~Raghu, S.~A. Kivelson, and D.~J. Scalapino.
\newblock Superconductivity in the repulsive hubbard model: An asymptotically exact weak-coupling solution.
\newblock \emph{Phys. Rev. B}, 81:\penalty0 224505, Jun 2010.
\newblock \doi{10.1103/PhysRevB.81.224505}.
\newblock URL \url{https://link.aps.org/doi/10.1103/PhysRevB.81.224505}.

\bibitem[Balatsky(1993)]{Balatsky01101993}
A.~V. Balatsky.
\newblock Superconducting instability in a non-fermi liquid scaling approach.
\newblock \emph{Philosophical Magazine Letters}, 68\penalty0 (4):\penalty0 251--256, 1993.
\newblock \doi{10.1080/09500839308242421}.
\newblock URL \url{https://doi.org/10.1080/09500839308242421}.

\bibitem[Sudb\o{}(1995)]{PhysRevLett.74.2575}
A.~Sudb\o{}.
\newblock Pair susceptibilities and gap equations in non-fermi liquids.
\newblock \emph{Phys. Rev. Lett.}, 74:\penalty0 2575--2578, Mar 1995.
\newblock \doi{10.1103/PhysRevLett.74.2575}.
\newblock URL \url{https://link.aps.org/doi/10.1103/PhysRevLett.74.2575}.

\bibitem[Yin and Chakravarty(1996)]{doi:10.1142/S0217979296000349}
Lan Yin and Sudip Chakravarty.
\newblock Spectral anomaly and high temperature superconductors.
\newblock \emph{International Journal of Modern Physics B}, 10\penalty0 (07):\penalty0 805--845, 1996.
\newblock \doi{10.1142/S0217979296000349}.
\newblock URL \url{https://doi.org/10.1142/S0217979296000349}.

\bibitem[Esterlis and Schmalian(2019)]{PhysRevB.100.115132}
Ilya Esterlis and J\"org Schmalian.
\newblock Cooper pairing of incoherent electrons: An electron-phonon version of the sachdev-ye-kitaev model.
\newblock \emph{Phys. Rev. B}, 100:\penalty0 115132, Sep 2019.
\newblock \doi{10.1103/PhysRevB.100.115132}.
\newblock URL \url{https://link.aps.org/doi/10.1103/PhysRevB.100.115132}.

\bibitem[Ye et~al.(2022)Ye, Lee, and Zou]{PhysRevLett.128.106402}
Weicheng Ye, Sung-Sik Lee, and Liujun Zou.
\newblock Ultraviolet-infrared mixing in marginal fermi liquids.
\newblock \emph{Phys. Rev. Lett.}, 128:\penalty0 106402, Mar 2022.
\newblock \doi{10.1103/PhysRevLett.128.106402}.
\newblock URL \url{https://link.aps.org/doi/10.1103/PhysRevLett.128.106402}.

\bibitem[Bittanti et~al.(1991)Bittanti, Laub, and Willems]{bittanti1991riccati}
Sergio Bittanti, Alan~J. Laub, and Jan~C. Willems, editors.
\newblock \emph{The Riccati Equation}.
\newblock Communications and Control Engineering. Springer-Verlag, Berlin, Heidelberg, 1991.
\newblock ISBN 978-3-540-53099-2.
\newblock \doi{10.1007/978-3-642-58223-3}.

\bibitem[Arora et~al.(2020)Arora, Polski, Zhang, Thomson, Choi, Kim, Lin, Wilson, Xu, Chu, Watanabe, Taniguchi, Alicea, and Nadj-Perge]{Arora2020}
Harpreet~Singh Arora, Robert Polski, Yiran Zhang, Alex Thomson, Youngjoon Choi, Hyunjin Kim, Zhong Lin, Ilham~Zaky Wilson, Xiaodong Xu, Jiun-Haw Chu, Kenji Watanabe, Takashi Taniguchi, Jason Alicea, and Stevan Nadj-Perge.
\newblock Superconductivity in metallic twisted bilayer graphene stabilized by wse2.
\newblock \emph{Nature}, 583:\penalty0 379--384, 2020.
\newblock \doi{10.1038/s41586-020-2473-8}.

\bibitem[Park et~al.(2021)Park, Cao, Watanabe, Taniguchi, and Jarillo-Herrero]{Park2021}
Jeong~Min Park, Yuan Cao, Kenji Watanabe, Takashi Taniguchi, and Pablo Jarillo-Herrero.
\newblock Tunable strongly coupled superconductivity in magic-angle twisted trilayer graphene.
\newblock \emph{Nature}, 590:\penalty0 249--255, 2021.
\newblock \doi{10.1038/s41586-021-03192-0}.

\bibitem[Hao et~al.(2021)Hao, Zimmerman, Ledwith, Khalaf, Najafabadi, Watanabe, Taniguchi, Vishwanath, and Kim]{Hao2021}
Zeyu Hao, A.~M. Zimmerman, Patrick Ledwith, Eslam Khalaf, Danial~Haie Najafabadi, Kenji Watanabe, Takashi Taniguchi, Ashvin Vishwanath, and Philip Kim.
\newblock Electric field--tunable superconductivity in alternating-twist magic-angle trilayer graphene.
\newblock \emph{Science}, 371\penalty0 (6534):\penalty0 1133--1138, 2021.
\newblock \doi{10.1126/science.abg0399}.
\newblock URL \url{https://www.science.org/doi/abs/10.1126/science.abg0399}.

\bibitem[Zhou et~al.(2021)Zhou, Xie, Taniguchi, Watanabe, and Young]{Zhou2021}
Haoxin Zhou, Tian Xie, Takashi Taniguchi, Kenji Watanabe, and Andrea~F. Young.
\newblock Superconductivity in rhombohedral trilayer graphene.
\newblock \emph{Nature}, 598:\penalty0 434--438, 2021.
\newblock \doi{10.1038/s41586-021-03926-0}.

\bibitem[Yankowitz et~al.(2019)Yankowitz, Chen, Polshyn, Zhang, Watanabe, Taniguchi, Graf, Young, and Dean]{Yankowitz2019}
Matthew Yankowitz, Shaowen Chen, Hryhoriy Polshyn, Yuxuan Zhang, K.~Watanabe, T.~Taniguchi, David Graf, Andrea~F. Young, and Cory~R. Dean.
\newblock Tuning superconductivity in twisted bilayer graphene.
\newblock \emph{Science}, 363\penalty0 (6431):\penalty0 1059--1064, 2019.
\newblock \doi{10.1126/science.aav1910}.
\newblock URL \url{https://www.science.org/doi/abs/10.1126/science.aav1910}.

\bibitem[Lu et~al.(2019)Lu, Stepanov, Yang, Xie, Aamir, Das, Urgell, Watanabe, Taniguchi, Zhang, Bachtold, MacDonald, and Efetov]{Lu2019}
Xiaobo Lu, Petr Stepanov, Wei Yang, Ming Xie, Mohammed~Ali Aamir, Ipsita Das, Carles Urgell, Kenji Watanabe, Takashi Taniguchi, Guangyu Zhang, Adrian Bachtold, Allan~H. MacDonald, and Dmitri~K. Efetov.
\newblock Superconductors, orbital magnets and correlated states in magic-angle bilayer graphene.
\newblock \emph{Nature}, 574:\penalty0 653--657, 2019.
\newblock \doi{10.1038/s41586-019-1695-0}.

\bibitem[Saito et~al.(2020)Saito, Ge, Watanabe, Taniguchi, and Young]{Saito2020}
Yu~Saito, Jingyuan Ge, Kenji Watanabe, Takashi Taniguchi, and Andrea~F. Young.
\newblock Independent superconductors and correlated insulators in twisted bilayer graphene.
\newblock \emph{Nature Physics}, 16:\penalty0 926--930, 2020.
\newblock \doi{10.1038/s41567-020-0928-3}.

\bibitem[Stepanov et~al.(2020)Stepanov, Das, Lu, Fahimniya, Watanabe, Taniguchi, Koppens, Lischner, Levitov, and Efetov]{Stepanov2020}
Petr Stepanov, Ipsita Das, Xiaobo Lu, Ali Fahimniya, Kenji Watanabe, Takashi Taniguchi, Frank H.~L. Koppens, Johannes Lischner, Leonid Levitov, and Dmitri~K. Efetov.
\newblock Untying the insulating and superconducting orders in magic-angle graphene.
\newblock \emph{Nature}, 583:\penalty0 375--378, 2020.
\newblock \doi{10.1038/s41586-020-2459-6}.

\bibitem[Ye et~al.(2012)Ye, Zhang, Akashi, Bahramy, Arita, and Iwasa]{Ye2012}
J.~T. Ye, Y.~J. Zhang, R.~Akashi, M.~S. Bahramy, R.~Arita, and Y.~Iwasa.
\newblock Superconducting dome in a gate-tuned band insulator.
\newblock \emph{Science}, 338\penalty0 (6111):\penalty0 1193--1196, 2012.
\newblock \doi{10.1126/science.1228006}.
\newblock URL \url{https://www.science.org/doi/abs/10.1126/science.1228006}.

\bibitem[Costanzo et~al.(2016)Costanzo, Jo, Berger, and Morpurgo]{Costanzo2016}
Davide Costanzo, Sanghyun Jo, Helmuth Berger, and Alberto~F. Morpurgo.
\newblock Gate-induced superconductivity in atomically thin mos2 crystals.
\newblock \emph{Nature Nanotechnology}, 11:\penalty0 339--344, 2016.
\newblock \doi{10.1038/nnano.2015.314}.

\bibitem[Mandal and Lee(2015)]{IPSITA}
Ipsita Mandal and Sung-Sik Lee.
\newblock Ultraviolet/infrared mixing in non-fermi liquids.
\newblock \emph{Phys. Rev. B}, 92:\penalty0 035141, Jul 2015.
\newblock \doi{10.1103/PhysRevB.92.035141}.
\newblock URL \url{http://link.aps.org/doi/10.1103/PhysRevB.92.035141}.

\bibitem[Halperin et~al.(1993)Halperin, Lee, and Read]{HALPERINHALF}
B.~I. Halperin, Patrick~A. Lee, and Nicholas Read.
\newblock Theory of the half-filled landau level.
\newblock \emph{Phys. Rev. B}, 47:\penalty0 7312--7343, Mar 1993.
\newblock \doi{10.1103/PhysRevB.47.7312}.
\newblock URL \url{https://link.aps.org/doi/10.1103/PhysRevB.47.7312}.

\bibitem[Zhan et~al.(2025)Zhan, Zhang, Zhang, Liu, Nie, Chen, Jiao, Komijani, Smidman, Steglich, Coleman, and Yuan]{bp6l-46z7}
Jin Zhan, Yongjun Zhang, Jiawen Zhang, Yu~Liu, Zhiyong Nie, Yuxin Chen, Lin Jiao, Yashar Komijani, Michael Smidman, Frank Steglich, Piers Coleman, and Huiqiu Yuan.
\newblock Critical fluctuations and conserved dynamics in a strange ferromagnetic metal.
\newblock \emph{Phys. Rev. Lett.}, 135:\penalty0 266504, Dec 2025.
\newblock \doi{10.1103/bp6l-46z7}.
\newblock URL \url{https://link.aps.org/doi/10.1103/bp6l-46z7}.

\bibitem[Li et~al.(2021)Li, Jiang, Li, Zhang, Kang, Zhu, Watanabe, Taniguchi, Chowdhury, Fu, Shan, and Mak]{Li:2021aa}
Tingxin Li, Shengwei Jiang, Lizhong Li, Yang Zhang, Kaifei Kang, Jiacheng Zhu, Kenji Watanabe, Takashi Taniguchi, Debanjan Chowdhury, Liang Fu, Jie Shan, and Kin~Fai Mak.
\newblock Continuous mott transition in semiconductor moir{\'e}superlattices.
\newblock \emph{Nature}, 597\penalty0 (7876):\penalty0 350--354, 2021.
\newblock \doi{10.1038/s41586-021-03853-0}.
\newblock URL \url{https://doi.org/10.1038/s41586-021-03853-0}.

\bibitem[Podolsky et~al.(2009)Podolsky, Paramekanti, Kim, and Senthil]{PhysRevLett.102.186401}
Daniel Podolsky, Arun Paramekanti, Yong~Baek Kim, and T.~Senthil.
\newblock Mott transition between a spin-liquid insulator and a metal in three dimensions.
\newblock \emph{Phys. Rev. Lett.}, 102:\penalty0 186401, May 2009.
\newblock \doi{10.1103/PhysRevLett.102.186401}.
\newblock URL \url{https://link.aps.org/doi/10.1103/PhysRevLett.102.186401}.

\bibitem[{Darius Shi}(2026)]{2026arXiv260827453D}
Zhengyan {Darius Shi}.
\newblock {Metallogenic quantum criticality: Fermi surface nucleation at transitions between gapped phases}.
\newblock \emph{arXiv e-prints}, art. arXiv:2608.27453, August 2026.
\newblock \doi{10.48550/arXiv.2608.27453}.

\end{thebibliography}

\end{document}